\documentclass[dvips,twoside, a4paper,12pt]{preprint}

\usepackage{times}
\usepackage{theorem}
\usepackage{epsfig}

\usepackage{mhequ}

\def\R{{\bf R}}

\def\Ref#1{(\ref{#1})}
\def\vir{{\rm,}}
\def\vers{\rightarrow}

\def\ve{\varepsilon}
\def\vp{\varphi}

\def\dd{\partial}
\def\fract#1#2{\textstyle{\frac{#1}{#2}}}
\def\bea{\begin{equa}}
\def\eea{\end{equa}}
\def\bes{\begin{equs}}
\def\ees{\end{equs}}
\def\dimf{f}

\def\TT{^{\rm\small T}}
\def\JJ{J}

\theoremstyle{plain}
\theorembodyfont{\it}
\newtheorem{Lemma}{Lemma}
\theorembodyfont{\rm}

\newtheorem{Example}[Lemma]{Example}
\newtheorem{Remark}[Lemma]{Remark}

\usepackage{sub_JP}

\newcommand{\ddp}{\delta p}
\newcommand{\ddq}{\delta q}
\newcommand{\ddxi}{\delta \xi}

\newcommand{\NN}{{\mathcal N}}

\newcommand{\nn}{{\bf n}}

\def\fld#1#2{\bigg(\!\!\begin{array}{c}\textstyle{#1}\cr\textstyle{#2}\end{array}\!\!\bigg)}

\def\D{{\rm D}}
\def\T{{\rm T}}
\def\OO{{\cal O}}
\let\epsilon=\varepsilon
\def\proclaim#1#2{\hfill\break\noindent{\bf #1:~} {\it #2}\hfill\break}

\def\EE#1{^{(#1)}}

\newenvironment{mylist}%
{\begin{list}{}{\setlength{\itemsep}{0pt}\setlength{\topsep}{\parsep}%
\setlength{\labelsep}{1ex}\setlength{\leftmargin}{\labelwidth}}}{\end{list}}

\def\DD{{\rm D}}
\def\DDD#1#2#3{\left.\D#1^{#2}\right\vert_{#3}}
\def\DPhi#1#2{\DDD{\Phi}{#1}{#2}}
\def\DF#1#2{\DDD{F}{#1}{#2}}
\def\DC#1{\DDD{C}{}{#1}}
\def\L{{\rm L}}
\def\text{\leavevmode}
\def\LP{{\rm LP}}
\def\P{{\rm P}}
\def\T{{\rm T}}
\def\A{A}
\def\XX{\mu}

\begin{document}

\title{Lyapunov Modes in Hard-Disk Systems}
\author{Jean-Pierre Eckmann${}^{1,2}$, Christina Forster${}^3$,\\
\\
Harald A. Posch${}^3$, Emmanuel Zabey${}^1$}
\institute{%
${}^1$\,D\'epartement de Physique Th\'eorique,\\$\,{}^2$\,Section de Math\'ematiques,\\
\hphantom{${}^1$}\,Universit\'e de Gen\`eve, Switzerland\\
${}^3$\,Institute for Experimental Physics, 
University of Vienna, Austria}
\maketitle

\begin{abstract}We consider simulations of a 2-dimensional gas of hard disks in a
rectangular container and study the Lyapunov spectrum near the
vanishing Lyapunov exponents. To this spectrum are associated
``eigen-directions'', called Lyapunov modes. We carefully analyze
these modes and show how they are naturally associated with vector
fields over the container. We also show that the Lyapunov exponents,
and the coupled dynamics of the modes (where it exists) follow linear laws,
whose coefficients only depend on the density of the gas, but not on
aspect ratio and very little on the boundary conditions.
\end{abstract}

%\bigskip

%\begin{keywords}
%\end{keywords}

%\bigskip
%\bigskip
%\bigskip

%
%  Introduction
%  ============
%
\section{Introduction}
\label{s:intro}

In this paper, we study the Lyapunov spectra of two-dimensional hard-disk systems 
and, in particular, the associated ``Lyapunov modes'' \cite{MPH98,PH00}.  Recently, 
this topic has received considerable attention
\cite{EG00,MM01,MM03,TM03,WB04,FHPH04}, and a lot of progress has been
made in the understanding of the issues involved.
In the present work we synthesize and expand the results found
earlier.
In particular, we completely classify
these modes and give a simple interpretation of their dynamics, in
particular for systems with arbitrary aspect ratio. We
furthermore present new simulations in support of this classification.\\

The Lyapunov {\em exponents} describe the rates of exponential growth, or decay, of
infinitesimal phase-space perturbations, and are taken to be ordered according to
$\lambda\EE1>\lambda\EE2> \cdots >\lambda\EE\ell$.\footnote{$\ell$ is at most
the dimension of the phase space. Note that $\lambda \EE{j}$ stands
for {\em different} Lyapunov exponents, while we will use $\lambda _i$
  when we consider them with multiplicity.} Because of the Hamiltonian nature of the problem, 
they come in conjugate pairs, 
$$
\lambda\EE j = -\lambda\EE{\ell-j+1}~.
$$
As is well-known, 0 is always a Lyapunov exponent for such systems and therefore, as a 
consequence, $\ell $ is odd.
At any point $\xi$ in phase space, the tangent space $TX(\xi)$
decomposes into a sum 
$$
TX(\xi)=E\EE{1}(\xi)\oplus\cdots\oplus E\EE{\ell}(\xi)~,
$$
where $E\EE{j}(\xi)$  
is the (linear) space of those perturbations of the initial
condition
$\xi$ whose growth rate is
$\lambda\EE{j}$ for the forward dynamics, and $-\lambda \EE{j}$ for the
time reversed dynamics. This decomposition is called the Oseledec splitting.\\ 

We say that the Lyapunov exponent $\lambda \EE{j}$ is $d$-fold
degenerate if $\dim E\EE {j}(\xi)= d$.
It should be noted that, when the Lyapunov exponents are $d$-fold
degenerate, 
only the subspace corresponding to all $d$ of them is well defined\footnote{In this, and
many other aspects, the theory of Lyapunov exponents is very similar
to that of matrices.}. The main idea of our full classification of Lyapunov exponents and their
modes is based on this simple observation.\\

The Lyapunov {\em modes} are defined as
follows: at time $t=0$, we take $n$ orthogonal tangent vectors at $\xi$ and, by applying to them 
the tangent-space dynamics\footnote{See Section \ref{s:tandyn} for details.} for a 
long-enough time $t$, map them onto $n$ vectors which, generally, are not orthogonal 
but still span an $n$-dimensional subspace $S_n(t)$. If, instead of $n$, we consider 
only $n-1$ vectors, they similarly span an $n-1$ dimensional subspace $S_{n-1}(t)$, such that
$S_{n-1}(t)\subset S_{n}(t)$. 
The {\bf Lyapunov mode} is a unit vector in $S_{n}(t)$ which is
orthogonal to the space $S_{n-1}(t)$.
We will give a precise definition in the next section, explain the
algorithmic aspects in Sect.~\ref{s:defmodes} and relate the modes
to Oseledec's subspaces in Sect.~\ref{s:ergod}\\

\begin{figure}[ht]
\begin{center}
\epsfig{file=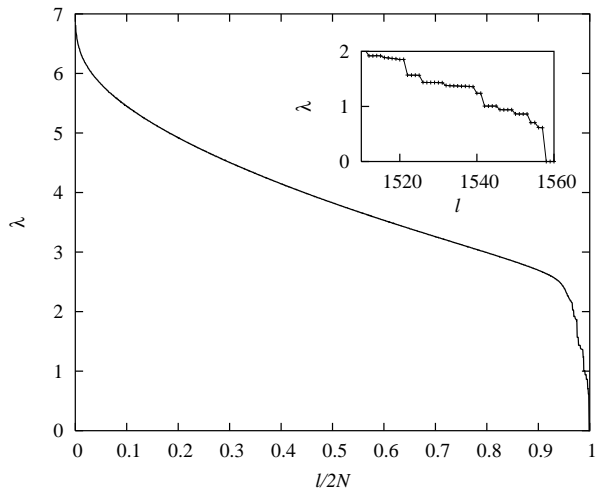, width=6.67cm, angle=-90}
\caption{Lyapunov spectrum for $N =780$ hard disks at a density
$\rho\equiv N/(L_x L_y) = 0.8$ in a 
rectangular periodic box with an aspect ratio $L_y/L_x = 0.867$. 
The insert provides a magnified view of the mode regime. $l$ is the 
Lyapunov index numbering the exponents.}
\label{f:lyap_spect}
\end{center}
\end{figure}

The study of Lyapunov modes \cite{PH00,FHPH04} has revealed interesting spatial
structures which we will define later but which come in two types: {\em localized structures} 
associated with the large positive and negative Lyapunov exponents, and {\em smooth delocalized 
structures} of wave-like type for exponents close to zero \cite{MPH98,FHPH04}. 
The exponents associated with the latter are degenerate and give rise to 
a step-like appearance of the Lyapunov spectrum as is shown, for example, 
in Fig.~\ref{f:lyap_spect}.\\
%%%%%%%%%%%%%%%%%%%%%%%%%%%%%%%%%%%%%%%%%%%%%%%%%%%%%%%%%%%%%%%%%%%%%%%%%%%%%%%
%%%%%%%%%%%%%%%%%%%%%%%%%%%%%%%%%%%%%%%%%%%%%%%%%%%%%%%%%%%%%%%%%%%%%%%%%%%%%%%

The discovery of these structures has led to several studies \cite{EG00,MM01,MM03,TM03,WB04}
which go some way in explaining their origin and their dynamics.
In this paper, we show how they are related to the symmetries of the container in which 
the particles move (including the boundary conditions).  At the same time we obtain a
classification of the degeneracies of the Lyapunov exponents near zero. This classification 
allows to view the so-called ``mode dynamics'' from a new geometrical perspective.\\ 

The paper starts with a summary of results, passes through a precise
definition of the Lyapunov modes, and then describes them as vector
fields.
In Sect.~\ref{s:classif} these vector fields are classified, and their
dynamics is studied in Sect.~\ref{s:dynamics}. The last sections deal
with the density dependence and with possible hydrodynamic aspects.

%%%%%%%%%%%%%%%%%%%%%%%%%%%%%%%%%%%%%%%%%%%%%%%%%%%%%%%%%%%%%%%%%%%%%%
\section{Notation and summary of results}
\label{s:tandyn}
%%%%%%%%%%%%%%%%%%%%%%%%%%%%%%%%%%%%%%%%%%%%%%%%%%%%%%%%%%%%%%%%%%%%%%

We consider a system of $N$ hard disks of diameter $\sigma$ and mass $m$
moving in a two-dimensional rectangular container with sides of lengths $L_x$ and
$L_y$. At this point, we do not need to specify 
the boundary conditions.\footnote{We will consider reflecting 
and periodic boundary conditions.} The phase space of such a system is
$$
X\,=\, \R^{2N}\times ([0,L_x]\times[0,L_y])^N ~,
$$
and a phase point $\xi$ in $X$ is
$$
\xi\,=\,(p,q)\,=\,(p_1,\dots,p_N,q_1,\dots,q_N)~.
$$
When necessary, we will write $q_i=(q_{i,x},q_{i,y})$ to distinguish the two position
components of particle $i$, and similarly for the momenta. 
The dynamics of the system is that of free flight, interrupted by elastic binary 
collisions.  If $\xi_0$ is the state of the system at time 0, then $\xi_t=\Phi^t(\xi_0)$ 
is the state at time $t$, where $\Phi^t: X\to X$ defines the flow.\\

Apart from issues of differentiability, which will be addressed when we describe 
the numerical implementation in Section \ref{s:def}, we call $\D \Phi^t$ 
the tangent flow. Informally speaking, it is a $4N\times 4N$ matrix of partial 
derivatives and can be thought of as the first-order term (in $\epsilon $)
in an expansion of the perturbed flow,
$$
\Phi^t(\xi_0 +\epsilon \delta\xi)\,=\,
\Phi^t(\xi_0)+\epsilon \,\DPhi{t}{\xi_0}\cdot \delta\xi +\OO(\epsilon^2
)~.
$$
The vector $\delta\xi$ lies in the tangent space $T X$ 
(at $\xi_0$) to the manifold $X$; in our case  $T X(\xi_0)=\R^{4N}$.
We next invoke the multiplicative ergodic theorem of Oseledec
\cite{O68,ER85}. To do so, we need ergodicity of the Liouville
measure. 
Without further knowledge, we assume this for our case.

\proclaim{Theorem}{There exist an integer $\ell$ and numbers $\lambda \EE{j},
  j=1,\dots,\ell$ such that for almost every $\xi\in X$ (with respect to the 
  Liouville measure) the tangent space splits into
$$
T X(\xi)\,=\,E\EE1(\xi) \oplus E\EE2(\xi) \oplus \cdots\oplus
E\EE\ell(\xi)~,
$$
with the property:
$$
\lim_{t\to\pm\infty } \frac{1}{|t|}\log \|\DPhi{t}{\xi}
 \delta\xi\| = \pm\lambda\EE{j} 
$$
for $\delta\xi\in E\EE{j}(\xi)$.
The spaces $E\EE{j}(\xi)$ are covariant:  $\DPhi{t}{\xi} E\EE{j}(\xi)=E\EE{j}(\Phi^t(\xi))$.}

\noindent
The dimension $d\EE{j}$ of $E\EE{j}(\xi)$ is called multiplicity
of the exponent $\lambda\EE{j}$. In general the $E\EE{j}$ are not orthogonal to
each other.\\

We also use the notation
\begin{equ}
\lambda_1\geq\lambda_2\geq\ldots\geq\lambda _i\geq\ldots\geq\lambda_{4N}
\end{equ}
to denote the Lyapunov exponents repeated with multiplicities, 
where the index is referred to as the Lyapunov index.  This notation is 
more adequate for describing numerical methods of measurement,
in which tangent-space dynamics is probed by a set of vectors.
With these notations the relation between the $\lambda \EE{j}$ and the
$\lambda _i$ is given by
$$
\lambda \EE{j}
=\lambda_{\dimf\EE{j-1}+1}=\dots= \lambda_{f\EE{j}}~,
$$
where $\dimf\EE{j}=d\EE{1}+\cdots+ d\EE{j}$.
Choosing $j\in \{1,\dots,\ell\}$ and letting $F\EE{j}=E\EE{1}\oplus \cdots\oplus
E\EE{j}$, we define the $d\EE{j}$ Lyapunov modes associated with
$\lambda \EE{j}$ as any orthogonal spanning set of the space
\begin{equ}[e:mj]
M\EE{j}\equiv(F\EE{j-1})^\bot
\cap F\EE{j}~.
\end{equ}
A finer decomposition of this spanning set will be obtained when we
describe the algorithm used, see Sect.~\ref{s:defmodes}.
The subspaces $M\EE{j}$ are very similar to the subspaces $E\EE{j}$: they have
the same dimension and also satisfy $F\EE{j}=M\EE{1}\oplus \cdots\oplus
M\EE{j}$. However, they are not identical, because of the orthogonality
constraint in \Ref{e:mj}. This will be explained in Sect.~\ref{s:ergod}.\\

In this summary, we focus on rectangular boxes with {\bf periodic} boundary conditions.
Narrow  systems ($L_y<2\sigma$) or  systems with reflecting boundaries
have a very similar Lyapunov spectrum, but some exponents found in the periodic case
are either absent, or appear with smaller multiplicities. We shall
treat such systems in Sect.~\ref{s:differ} but concentrate, until then,
on the ``general'' periodic case. However, it should be noted that
systems with reflecting boundaries give important information on the
relation between the vanishing and the small Lyapunov exponents 
(see Ex.~\ref{ex:Dirichlet} below).\\

The Lyapunov exponents near zero are found to be proportional to the 
wave numbers of the system,
\begin{equ}[e:k]
k_{(n_x,n_y)}=\sqrt{\big({\fract{2\pi}{L_x}}\,n_x\big)^2
+\big({\fract{2\pi}{L_y}}\,n_y\big)^2}~\vir\quad n_x,n_y=0,1,\ldots\;\;.
\end{equ}\\
\proclaim{Remark}{This result, as well as all results mentioned below,
  are to be understood in the limit of an infinite number of disks, at fixed density. In
  particular, we omit higher order terms (in $k$) in most of our statements. }

The Lyapunov exponents have the following properties:

\proclaim{Lyapunov Spectrum}{For gases of hard disks,
the Lyapunov exponents near zero 
are fully determined by two (positive) constants, $c_{\L}$ and $c_{\T}$. For small-enough
$\nn=(n_x,n_y)$, these exponents lie on two straight
lines.\footnote{For both the longitudinal 
and transverse modes the linear $k$ dependence of $\lambda$ is only the first term of 
an expansion in powers of $k$ \cite{FHPH04}, $\lambda = c k + c_2 k^2 \ldots$. For
positive $\lambda$, $c_2$ is positive, but small.
For a second order calculation in $k$, see \cite{WB04}. }
\begin{mylist}
\item[1)]
Transverse branch {\rm(\T)}\,:~$\lambda=\pm c_{\T}\,k_\nn$, with multiplicity 4 (2 if either
$n_x$ or $n_y$ is zero)
\item[2)]
Longitudinal branch {\rm(\L)}\,:~$\lambda = \pm c_{\L}\,k_\nn$, with multiplicity 8 (4 if either
$n_x$ or $n_y$ is zero)
\end{mylist}
}
\noindent
The multiplicities of both branches will be explained by 
simple geometric observations in Sect.~\ref{s:lpt}. While 
the linear laws resemble the square roots of the eigenvalues
of a Laplacian in the box, we have no explanation beyond those already given
in Ref.~\cite{EG00}. (The square root is related to the symplectic
nature of the problem.) 
Additional degeneracies arise in square systems and may accidentally 
appear also for specific aspect ratios ${{L_y}/{L_x}}$.\\

\begin{figure}
\begin{center}
\epsfig{file=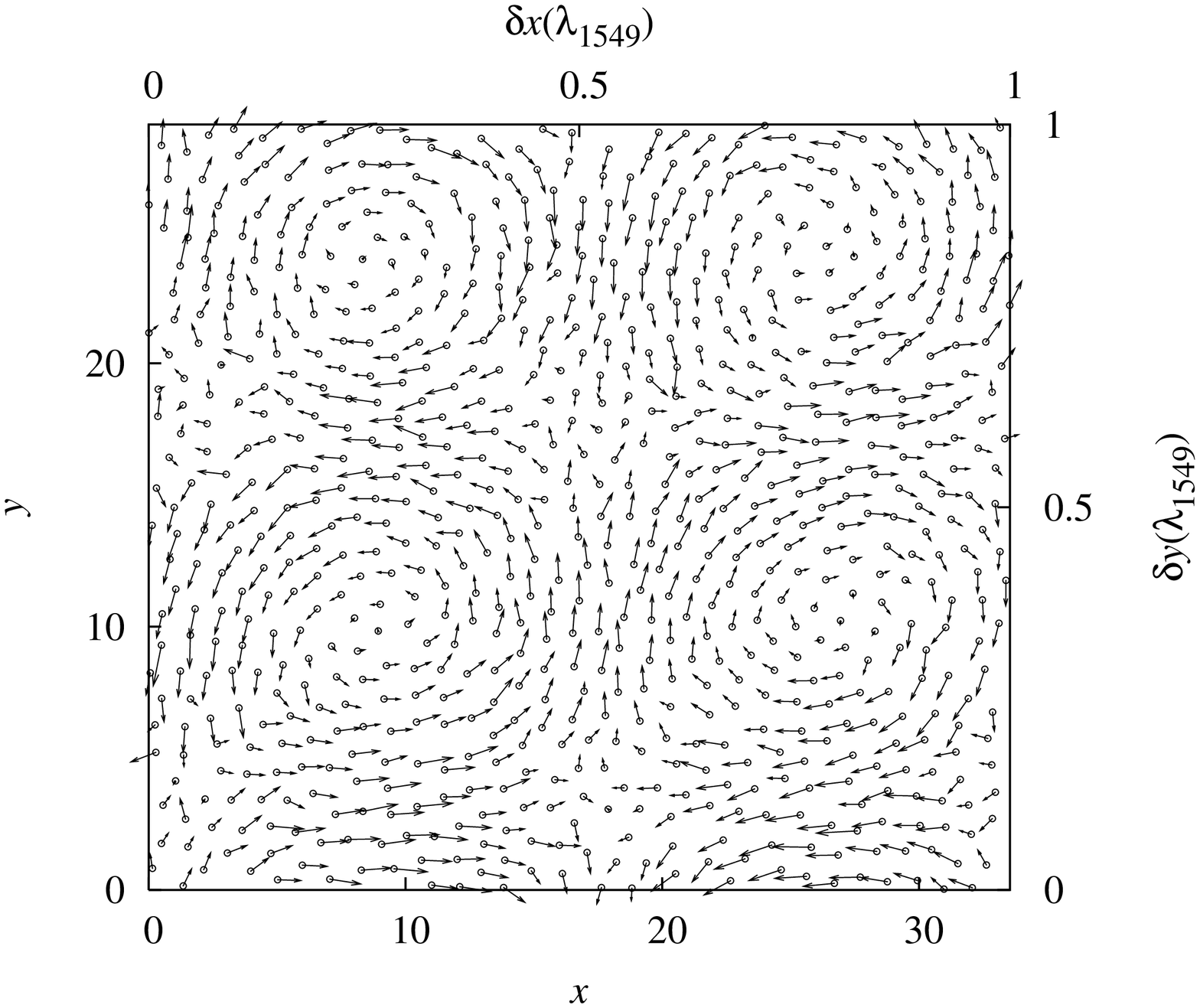, width=5cm, angle=-90}
\epsfig{file=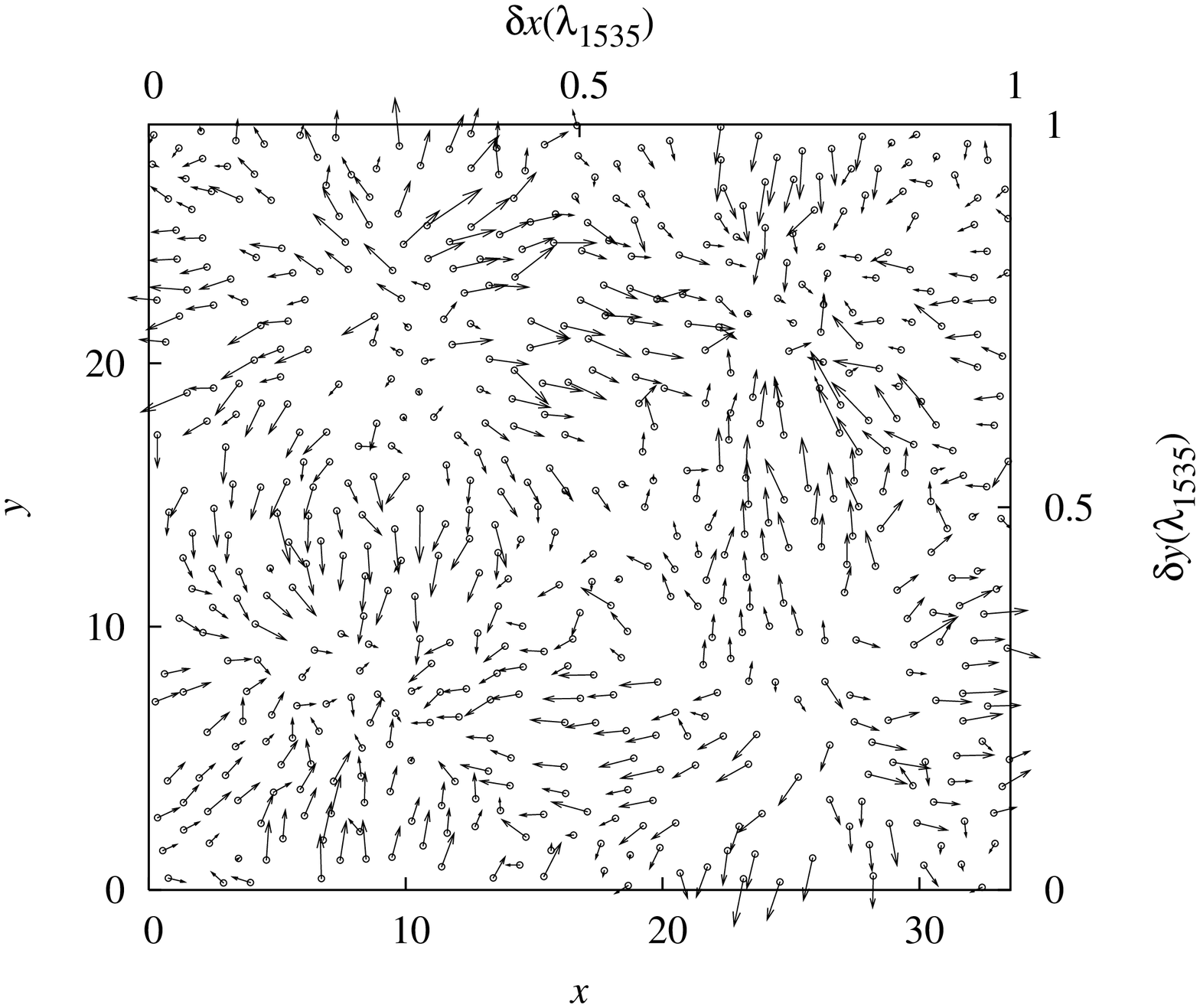, width=5cm, angle=-90}
\caption{Modes for the 780-disk system 
with periodic boundaries characterized in Fig.~\ref{f:lyap_spect}.
Left: Transverse mode \T(1,1) belonging to $\lambda_{1549}$.
Right: Longitudinal mode for $\lambda_{1535}$ belonging to an {\LP} pair 
{\LP}(1,1).}
\label{f:mode_example}
\end{center}
\end{figure}

We next explain how to visualize a mode \cite{MP02,FHPH04}.
Fix $\xi=(p,q)\in X,$ and let $\delta\xi$ be a Lyapunov mode of $TX(\xi)$.
The vector $\delta\xi=(\delta p,\delta q)$ has $4N$ components, $2N$ associated 
with the momenta and $2N$ with the positions.  Consider, for example, the $q$ components,
$\delta q_1,\dots,\delta q_{N}$, where each $\delta q_j$ is in $\R^2$ (corresponding to
the infinitesimal $x$ and $y$ displacements of $q_j$, $j=1,\dots,N$). 
By drawing the perturbation vectors $\delta q_j$ at the positions $q_j$ of 
the particles in the box, one obtains a field of vectors 
as is shown in Fig.~\ref{f:mode_example}.  For dense-enough fluids we obtain a vector 
field in every point of the box by interpolating between the particles.\\

%%%%%%%%%%%%%%%%%%%%%%%%%%%%%%%%%%%%%%%%%%%%%%%%%%%%%%%%%%%%%%%%%%%%%%%%%%%%%%
%%%%%%%%%%%%%%%%%%%%%%%%%%%%%%%%%%%%%%%%%%%%%%%%%%%%%%%%%%%%%%%%%%%%%%%%%%%%%%

It has been observed
that for those Lyapunov exponents close to zero, this vector field is
well approximated by trigonometric functions of the spatial
coordinates $x$ and $y$ \cite{MPH98,FHPH04}. In particular, the number of nodes $(n_x,n_y)$
of the vector field determines a wave number $k_\nn$. We then say that
$\delta\xi$ is a mode of wave number $k_\nn$. Our second main result is:

\proclaim{Mode Classification}{The subspaces $M\EE{j}(\xi)$ (defined in
  \Ref{e:mj}) belonging to
Lyapunov exponents $\lambda \EE{j}$
close to zero fall into two categories: 
\begin{mylist}
\item[1)]
Transverse branch: modes associated with a Lyapunov exponent 
$\pm c_{\T}\,k_\nn$ are divergence-free periodic fields of wave number $k_\nn$.
\item[2)]
Longitudinal branch: The modes associated with
the Lyapunov exponent $\pm c_{\L}\,k_\nn$ are of two types:
\begin{mylist}
\item[(i)] 
Half of them are irrotational periodic vector fields of wave number $k_\nn$
 (called \L-modes); 
\item[(ii)]
The others are scalar modulations with wave number $k_\nn$ 
of the momentum field,
(called P or momentum modes).
\end{mylist}
\end{mylist}
}
The L and the P-modes turn out to be paired:
the P-mode has components $p_i A(q_i)$ and the corresponding L-mode
has components
$\nabla A(q_i)$, where $A$ is a scalar function, see \Ref{e:PLT}.
In the following, we refer to an \L-mode 
and its corresponding \P-mode as an {\bf {\LP} pair}.
We will give more details on this relation below.
Results in this direction have been obtained in
Refs.~\cite{MM01,TM03,WB04}. \\

An interesting question, which is related to some hydrodynamic aspects 
of the fluid,
is an apparent propagation of the L and P-modes in physical 
space. It is a consequence of the motion of the tangent vectors
in the subspaces $M\EE{j}(\xi_t)$, to which we refer as {\em mode dynamics}. 
It will be described in detail in Section \ref{s:dynamics}.
We note that 
this motion is not only determined by the tangent flow, but also by the 
re-orthonormalization process part of the algorithm for the simulation.  
Mode dynamics was observed early on 
\cite{PH00,MP02}, but attempts to compute the propagation velocities are still 
scarce and, at present, only work for low densities \cite{MM03,WB04}.
Here, we give a more precise definition and provide numerical results.

\proclaim{Mode Dynamics}{In an $M\EE{j}$ space of Longitudinal and \P-modes,
the mode dynamics couples {\LP} pairs. When restricted
to the two-dimensional subspace spanned by a given {\LP} pair,
it reduces to a rotation at constant angular velocity $\omega_\nn$ 
which is proportional to the wave number $k_\nn$, 
\begin{equ}
\omega_\nn=v\,k_\nn~\vir
\end{equ}
where $v$ has the dimension of a velocity.  }

In the remainder of the paper we state these results more precisely 
and give details about how they are obtained. They are of two types: First, 
a theoretical description of the modes, which is based on the symmetries of the
system. Second, a detailed account of the numerical algorithms
necessary to substantiate our claims (the difficulty being the
decomposition of the $d\EE{j}$ dimensional spaces $M\EE{j}(\xi)$).

%%%%%%%%%%%%%%%%%%%%%%%%%%%%%%%%%%%%%%%%%%%%%%%%%%%%%%%%%%%%%%%%%%%%%%
\section{Tangent-space dynamics}
\label{s:def}
%%%%%%%%%%%%%%%%%%%%%%%%%%%%%%%%%%%%%%%%%%%%%%%%%%%%%%%%%%%%%%%%%%%%%%

The dynamics of hard disks consists of phases of free flight
interrupted by instantaneous elastic
collisions. We denote the map for free flights of duration $\tau$ by
$F^{\tau}$, and the collision map by $C$. Then, the evolution of an
initial state, $\xi_0$, 
is given by
\begin{equ}[e:flow]
\xi_t=F^{\tau_n}\circ\cdots\circ C\circ F^{\tau_2}\circ C\circ F^{\tau_1}~\xi_0~\vir
\end{equ}
where $\tau_1,\ldots,\tau_n$ are the time intervals between successive
collisions.  The {\em tangent-space dynamics} of an infinitesimal perturbation $\ddxi_0$ 
of  $\xi_0$ is given by the tangent map of the flow \Ref{e:flow} :
\begin{equ}
\ddxi_t=\DF{\tau_n}{\xi_n^+}\cdots\DC{\xi_2^-}\cdot 
\DF{\tau_2}{\xi_1^+}
\cdot\DC{\xi_1^-}\cdot\DF{\tau_1}{\xi_0}~\ddxi_0~\vir
\end{equ}
where $\xi_k^-$ denotes the state just before the $k^{\rm th}$ collision,
and $\xi_k^+=C(\xi_k^-)$ is 
the state immediately after.\footnote{We disregard problems of
differentiability which appear for the (rare) tangent collisions.} Here, 
$\DF\tau{\xi_k}$ and $\DC{\xi_k}$ are 
$4N\times4N$ symplectic matrices. For the sake of simplicity,
the flow is also called $\Phi^t$: namely $\Phi^t(\xi_0)=\xi_t$, and
we often write $\ddxi_t=\DPhi{t}{\xi_0}\cdot\ddxi_0$. By convention,
if $t$ is a collision time, $\ddxi_t$ denotes a tangent vector 
immediately before that collision.

%%%%%%%%%%%%%%%%%%%%%%%%%%%%%%%%%%%%%%%%%%%%%%%%%%%%%%%%%%%%%%%%%%%%%%%
%
\subsection{Numerical procedure and Lyapunov modes}
\label{s:defmodes}
%%%%%%%%%%%%%%%%%%%%%%%%%%%%%%%%%%%%%%%%%%%%%%%%%%%%%%%%%%%%%%%%%%%%%%%

    Extensive numerical simulations are used to
establish the classification and dynamics of the modes. Here, we
briefly summarize our algorithm described in detail in Refs.~\cite{DPH96,PH00} 
and provide a precise definition of the Lyapunov modes we observe. 
It is well known that the exponential growth rate of a typical
$k$-dimensional volume element is given by $\lambda_1+\cdots+\lambda_k$, 
\begin{equ}[e:volgrowth]
\lambda_1+\cdots+\lambda_k=
\lim_{t\vers\infty}\frac1t\log\|\ddxi^1_t\wedge\cdots\wedge \ddxi^k_t\|.
\end{equ}
If the tangent vectors $\ddxi^1,\ldots,\ddxi^k$ are linearly independent
at time zero, they will remain so, because the tangent flow is time reversible.  
Orthogonality is not preserved by the tangent flow, because, generally, 
$\D\Phi^t$ is not an orthogonal matrix. However, \Ref{e:volgrowth} still holds if 
the vectors $\ddxi^1_t,\ldots,\ddxi^k_t$ are replaced by a (Gram-Schmidt) 
orthogonalized 
set of vectors $\delta\eta^1_t,\ldots,\delta\eta^k_t$.\footnote{$\delta\eta^j_t=
\delta\xi^j_t-\sum_{i=1}^{j-1}(\delta\xi^j_t\cdot\delta\xi^i_t)/(\|\delta\xi^i_t\|^2)
\delta\xi^i_t $} It takes thus the simpler form 
\begin{equ}[e:volgrowth2]
\lambda_1+\cdots+\lambda_k=
\lim_{t\vers\infty}\frac1t\sum_{i=1}^k \log\|\delta\eta^i_t\|~.
\end{equ}
Since \Ref{e:volgrowth2} holds for every $k\leq 4N$, the exponent $\lambda_k$
turns out to be equal to the growth rate of the $k^{\rm th}$ orthogonalized 
vector $\delta\eta^k_t$. \\

The numerical method of Benettin {\em et al.}\cite{B80} and Shimada
{\em et al.}\cite{S79} -- and indeed any algorithm -- is based on this
construction, although its 
basic objects are not the $\delta\eta^k_t$ vectors. As time increases,
they all would get exponentially close to the most-unstable direction, become
numerically indistinguishable, and diverge. Instead of orthogonalizing {\em once} at
time $t$, the tangent dynamics is applied to a set of tangent vectors,
which are periodically replaced by an orthogonalized set,
that we denote by $\delta\gamma^1_t,\ldots,\delta\gamma^k_t$ . The modified dynamics
is therefore that of an orthogonal frame \cite{DPH96,PH00}. One assumes that
\Ref{e:volgrowth2} is still valid if $\delta\eta^k_t$ is replaced by
$\delta\gamma^k_t$.
The $k^{\rm th}$ Lyapunov mode (or Lyapunov vector) at time $t$ is, by definition, the 
vector $\delta\gamma^k_t$, and it is associated with the exponent $\lambda_k$.\\

Our study starts with the observation that we consider a system
with non-trivial Lyapunov exponents.
To guarantee the convergence of the numerical
algorithm, additional
properties of the dynamical system are needed: Namely, that the system
has well-defined local stable and unstable subspaces associated with
every Lyapunov exponent (close to zero). Results in this direction
have been obtained for hard-disk systems in \cite{C83,W90,BLPS92}.
A stronger property, hyperbolicity, has been recently
proved for hard disk systems with randomly
chosen masses in \cite{SS99,S02,S04}.
We assume here that these results hold for our system 
as well.  As already advocated in Ref.~\cite{ER85}, what matters from a
physicist's point of view is that the numerical studies behave as if 
this were true.
Under the above assumptions, the $k^{\rm th}$ mode will align with 
the corresponding
unstable subspace.
In other words,
measured modes will be orthogonal spanning sets of the $M\EE{j}$
subspaces defined in Eq.~\Ref{e:mj}.\\

The algorithm we use in our numerical work \cite{DPH96,PH00} is based on  
the principles just outlined. 
We restrict our considerations to
hard-disk systems without external interaction.
Since there is no potential energy, the dynamics is the same
at any (total) energy, up to a rescaling of time.
The natural unit of time of the system 
is $((m \sigma^2
N)/K)^{1/2}$.
Throughout, reduced units are used, for which the
particle mass $m$, the disk diameter $\sigma$, and the kinetic energy per particle, $K/N$,
are unity. 
The density, defined by $\rho = N/V$ and the aspect ratio, defined by
$A = L_y/L_x$, are the only relevant macroscopic parameters. 
Here, $V = L_x L_y$
is the area of the (rectangular) simulation box whose sides are $L_x$ and $L_y$ 
in the $x$ and $y$ directions, respectively. All our numerical examples are for 
densities $\rho \le 0.8$ characteristic of dense or 
dilute (if $\rho < 0.1)$ hard disk gases. The results are insensitive to the
time between successive Gram-Schmidt re-orthonormalization steps.

%%%%%%%%%%%%%%%%%%%%%%%%%%%%%%%%%%%%%%%%%%%%%%%%%%%%%%%%%%%%%%%%%%%%%%%
%
\subsection{Modes and Oseledec subspaces}
\label{s:ergod}
%%%%%%%%%%%%%%%%%%%%%%%%%%%%%%%%%%%%%%%%%%%%%%%%%%%%%%%%%%%%%%%%%%%%%%%

We have remarked in Sect.~\ref{s:tandyn} that the spaces $E\EE{j}$ 
of Oseledec' theorem 
are in general not identical to the spaces of the modes $M\EE{j}$ defined in 
the Eq.~\Ref{e:mj}. Here, we explain this difference.\\

The covariant subspaces $E\EE1,\dots,E\EE\ell$ of the Oseledec
splitting are obtained as follows~\cite{ER85}: the multiplicative ergodic theorem 
(for reversible systems) states that the matrices 
$$
\Lambda_\pm(\xi)\equiv\lim_{t\vers\pm\infty}
(\DDD\Phi{\pm t}\xi\TT\DDD\Phi{\pm t}\xi)^{\fract{1}{2|t|}}
$$
exist with probability one. The eigenvalues of $\Lambda_+$
are $\exp(\lambda\EE1)>\dots> \exp(\lambda\EE\ell)$ and the eigenvalues
of $\Lambda_-$ are $\exp(-\lambda\EE\ell)>\dots> \exp(-\lambda\EE{1})$. 
Since both $\Lambda_+$ and $\Lambda_-$ are symmetric, their eigenspaces
define two orthogonal decompositions of the tangent space
$$
TX=U_\pm\EE1\oplus\dots\oplus U_\pm\EE\ell~{\rm,}
$$
where $U_\pm\EE j$ is the eigenspace of $\Lambda_\pm$ associated to
$\exp(\pm\lambda\EE j)$. Note that the $U_\pm\EE j$ subspaces
are pairwise orthogonal but in general {\em not} 
covariant. However, for $j\in\{1,\dots,\ell\}$, the 
subspaces
$$
U_+\EE j\oplus\dots\oplus U_+\EE\ell\quad {\rm and}\quad
U_-\EE 1\oplus\dots\oplus U_-\EE j
$$
are covariant. They are, respectively, the subspace of the 
$\ell-j+1$ most stable directions of $\Lambda_+$ and 
the subspace of the $j$ most unstable directions of $\Lambda_-$ (or equivalently
its $j$ most stable directions {\em in the past}).
For every $\xi$, the invariant subspaces $E\EE j$ are then given by
$$
E\EE j = 
\big(U_-\EE 1\oplus\dots\oplus U_-\EE j\big)\cap
\big(U_+\EE j\oplus\dots\oplus U_+\EE\ell\big)~.
$$
The $E\EE j$ spaces are covariant but in general {\em not} orthogonal.
One can show that
$$
U_-\EE1\oplus\dots\oplus U_-\EE j=
E\EE1\oplus\dots\oplus E\EE j
\equiv F\EE j
~{\rm,}
$$
where $F\EE j$ was introduced in Sect.~\ref{s:tandyn}.
Using the definition \Ref{e:mj}, one easily verifies that $M\EE j=U_-\EE j$.

\begin{Remark}
In our numerical method, the modes are obtained at the phase points $\Phi^t(\xi_0)$
for large~$t$, {\em i.e.}, at the {\em end} of integrated trajectories. 
It is therefore consistent
that they should contain information about the {\em past} and not the future,
that is about $\Lambda_-$ and not $\Lambda_+$.
\end{Remark}

In general, $U_-\EE j$, $U_+\EE j$ and $E\EE j$ are different.
Hard-disk systems have a special property: the three spaces coincide for 
the Lyapunov exponent $\lambda\EE m=0$, where $m\equiv\fract12(\ell+1)$
is the middle index. 
We shall denote by $\NN$ the covariant
subspace $\NN=E\EE m$, also called the null subspace. 
In Sect.~\ref{s:null}, we give the explicit form of $\NN$. Using
the explicit Jacobians for the particle collisions 
and for the free-streaming motion, as given in \cite{DPH96,DP97,PH00},
one can check that the orthogonal
complement to the null subspace,  $\NN^\bot$,
is also covariant. From this statement, one can
show that $\NN=U_+\EE m=U_-\EE m$. Our measured modes  $M\EE m=U_-\EE m$
therefore span the null space. Since this
argument seems to fail for $j\neq m$, we can only conclude that modes 
of $M\EE j$ are perturbations whose
exponential growth rate is {\em at least} $\lambda\EE j$.

%%%%%%%%%%%%%%%%%%%%%%%%%%%%%%%%%%%%%%%%%%%%%%%%%%%%%%%%%%%%%%%%%%%% 
%
\subsection{Two simplifications}
\label{s:stable}
%%%%%%%%%%%%%%%%%%%%%%%%%%%%%%%%%%%%%%%%%%%%%%%%%%%%%%%%%%%%%%%%%%%% 

Two additional properties greatly simplify the classification of the modes:
the symplecticity of the tangent flow and an observed property ``$\delta q$
is proportional to $\delta p$''. The symplecticity of the tangent flow means that
\begin{equ}[e:symp]
\DDD\Phi{t}{\xi}\TT
J\DDD\Phi{t}{\xi}=J~,
\end{equ}
where
the $4N\times 4N$ matrix $J$ is defined as\footnote{With the reduced units defined above,
positions and velocities are dimensionless.}
\begin{equ}
\JJ:\pmatrix{\delta p\cr\delta q}\mapsto\pmatrix{0&-1\cr1&0}
\pmatrix{\delta p\cr\delta q}=\pmatrix{-\delta q\cr\delta p}~.
\end{equ}
As a consequence of \Ref{e:symp}, both $\Lambda_+$ and $\Lambda_-$ are symplectic
(see {\it e.g.},  \cite{W88})
and, as is well known, Lyapunov exponents of Hamiltonian
systems come in pairs $\lambda\EE{j}$, $\lambda\EE{\ell-j+1}$ $ =-\lambda\EE{j}$
of equal multiplicity.
The matrix $J$ relates the $U_\pm\EE j$
subspaces by
\begin{equ}[e:sympJ]
U_\pm\EE{j}(\xi)=JU_\pm\EE{\ell-j+1}(\xi)~.
\end{equ}
Therefore, it is sufficient to measure the Lyapunov exponents and the modes
for the positive part of the spectrum.
Note that $J$ is {\em not} the (derivative of the) velocity 
reversal $(\delta p,\delta q)\mapsto(-\delta p, \delta q)$ and does not
change the sign of time.\\

\begin{figure}[ht]
\begin{center}
\epsfig{file=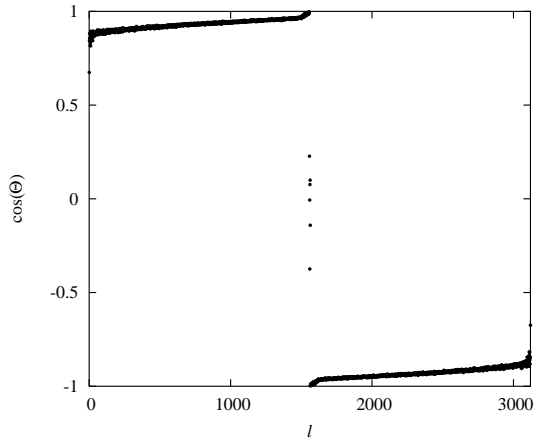, width=6cm, angle=-90}
\caption{ Value of $\cos(\Theta)=(\delta
  q \cdot \delta p)/(|\delta q|\cdot|\delta p|)$, as a function of the
Lyapunov index $l$ for an instantaneous configuration of the system 
characterized in Fig.~\ref{f:lyap_spect}.
Here, $\Theta$ is the angle between the $2N$-dimensional vectors of 
the perturbation components of all particle positions and velocities 
for a Lyapunov vector $\xi_l$.
For the small positive exponents, for which $l < 2N-2 = 1558$, 
this angle vanishes, for the small negative exponents, for which $l > 2N+3 = 1563$, 
it is equal to $\pi$. For the six zero modes, $1558 \le l \le 1563$,
the angle lies between these limiting values.}
\label{f:vel_PBC}
\end{center}
\end{figure}

The simulations of our system exhibit
an additional structure which
considerably simplifies the analysis of Lyapunov modes near zero.
{\em Unstable} perturbations ($\lambda>0$) have the (approximate) form
\begin{equ}
\ddp = C(\xi,\lambda)\,\ddq~\vir\quad C>0~\vir
\end{equ}
while the {\em stable} ($\lambda<0$) perturbations are of the form
\begin{equ}
-C'(\xi,\lambda)\,\ddp = \ddq~\vir\quad C'>0~.
\end{equ}
Here, $C$ and $C'$ are {\em numbers}.
In the left panel of
Fig.~\ref{f:vel_PBC} we demonstrate that the perturbations $\delta q$ and 
$\delta p$ associated with each Lyapunov mode are nearly parallel or anti-parallel
for large $N$. 
The Equation~\Ref{e:sympJ} also suggest $C=C'$, which is well
verified by the simulations.\\

%%%%%%%%%%%%%%%%%%%%%%%%%%%%%%%%%%%%%%%%%%%%%%%%%%%%%%%%%%%%%%%%%%%%%
%%%%%%%%%%%%%%%%%%%%%%%%%%%%%%%%%%%%%%%%%%%%%%%%%%%%%%%%%%%%%%%%%%%%%

{\em We can therefore restrict our classification  to
the $\ddq$ part of the modes corresponding to positive exponents.} If
we can associate every measured exponent $\lambda$ to a given $\ddq$, then
we will know that the exponents $\lambda$ and $-\lambda$ correspond
to the modes $(\ddp,\ddq)$ with $\ddp=C\ddq$ and $\ddp=-C^{-1}\ddq$,
respectively.

%%%%%%%%%%%%%%%%%%%%%%%%%%%%%%%%%%%%%%%%%%%%%%%%%%%%%%%%%%%%%%%%%%%%%%%%%%%%%%
%
\subsection{Tangent vectors as vector fields}
\label{s:vf}%
%%%%%%%%%%%%%%%%%%%%%%%%%%%%%%%%%%%%%%%%%%%%%%%%%%%%%%%%%%%%%%%%%%%%%%%%%%%%%%

\begin{figure}
\begin{center}
\epsfig{file=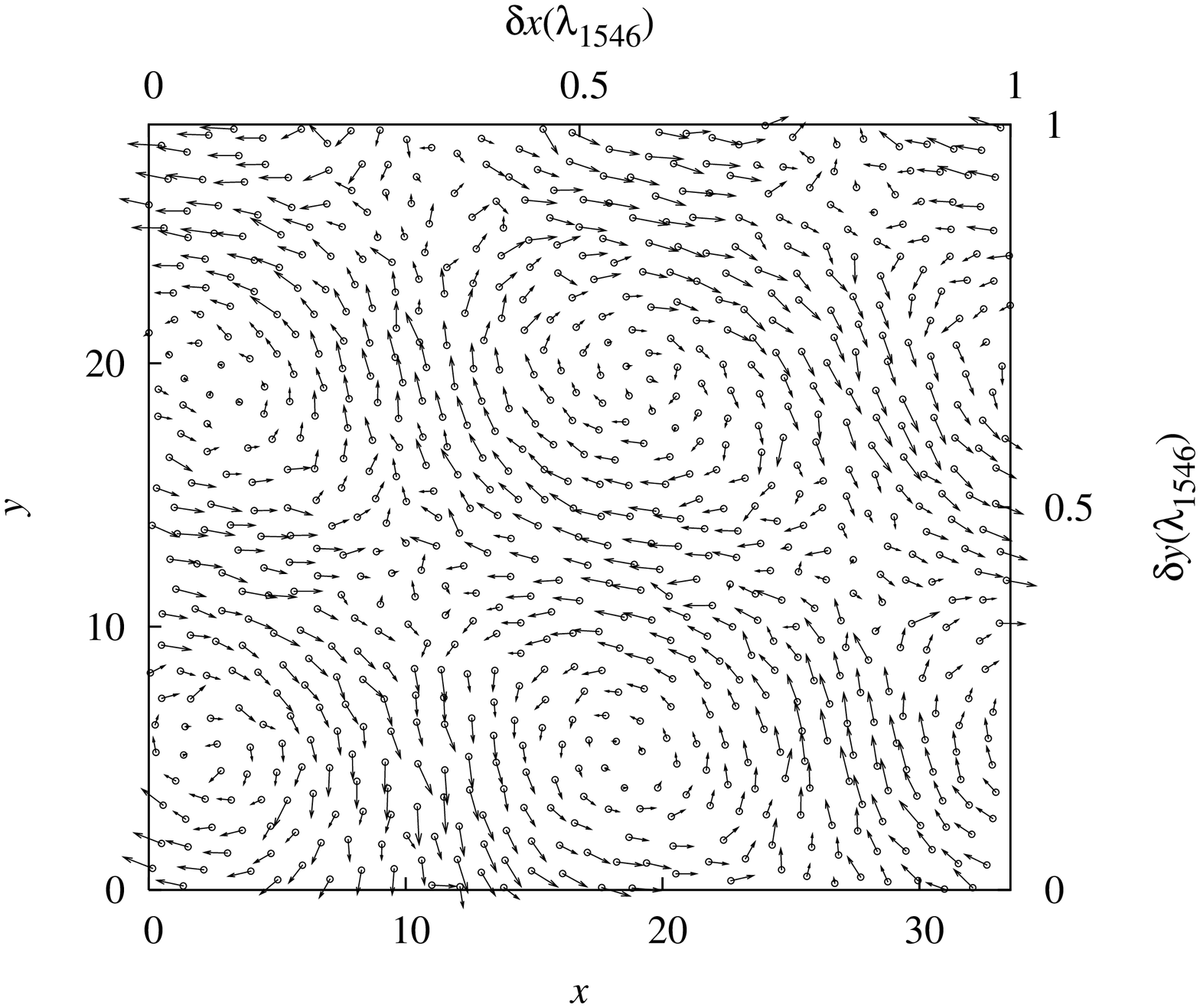, width=5cm, angle=-90}
\epsfig{file=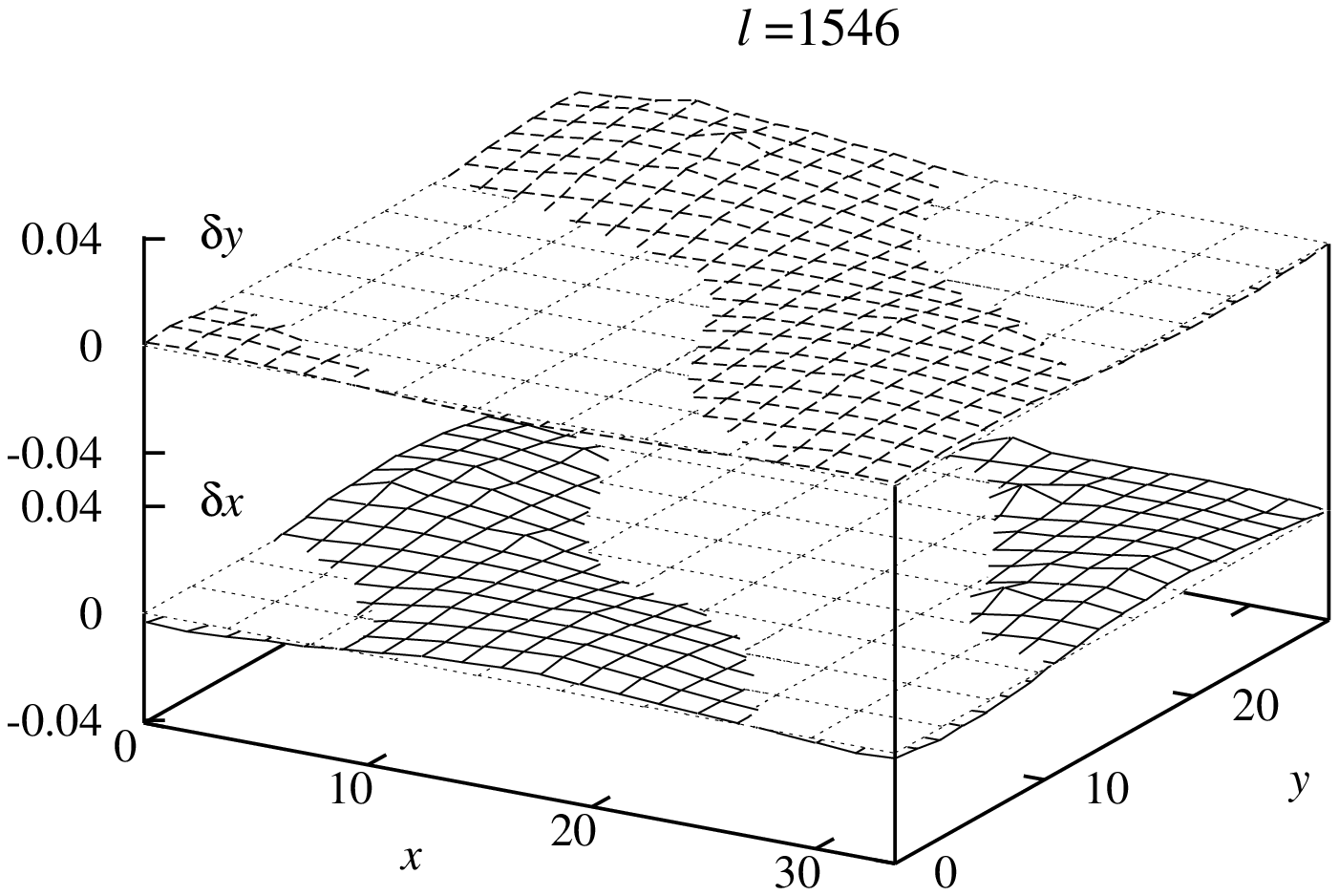, width=6cm, angle=-90}
\caption{The transverse mode \T(1,1) for $\lambda_{1546}$ of Fig.~\ref{f:disp_rel}.
Left: Interpretation as a vector field; Right: Alternative representation as 
periodic spatial 
patterns of the position perturbations $\delta x_i$ and $\delta y_i$ of the 
particles, which emphasizes the wave vector parallel to a diagonal of the
simulation box.}
\label{f:field}
\end{center}
\end{figure}

   The components of a tangent vector 
$\ddxi=(\ddp,\ddq)$ are the perturbation components
of the positions and velocities of 
all particles. As a graphical representation of
such a vector, we show in the left panel of Fig.~\ref{f:field} 
the instantaneous positional perturbations of 
all particles {\em at their positions} in
physical space, where the arrows indicate the directions and strengths. 
It belongs to a Lyapunov exponent 
$\lambda_{1546}$ indicated by the enlarged circle in
Fig.~\ref{f:disp_rel}. 
A qualitatively identical figure is obtained, if 
the positional displacements of all particles
are replaced by their momentum displacements, as explained in Sect.~\ref{s:stable}.
Thus, this figure is a complete
representation of the Lyapunov vector 
$\ddxi=(\ddp,\ddq)$ belonging to $\lambda_{1546}$ in 
Fig.~\ref{f:disp_rel}. The transverse modes for $\lambda_{1549}$ on
the left panel of
Fig.~\ref{f:mode_example}, and for $\lambda_{1548}$ at the bottom left
panel of
Fig.~\ref{f:modes_all}, are other examples of transverse modes belonging to the same 
degenerate exponent. \\

%%%%%%%%%%%%%%%%%%%%%%%%%%%%%%%%%%%%%%%%%%%%%%%%%%%%%%%%%%%%%%%%%%%%%%%
%%%%%%%%%%%%%%%%%%%%%%%%%%%%%%%%%%%%%%%%%%%%%%%%%%%%%%%%%%%%%%%%%%%%%%%

    We interpret the left panel of Fig.~\ref{f:field} as a two-dimensional 
{\em vector field} $\vp$ which -- up to a constant phase -- is well 
described by 
\begin{equ}
\fld{\vp_x(x,y)}{\vp_y(x,y)}=\fld{\phantom-\alpha_1\,\cos(k_x\,x)\,\sin(k_y\,y)}
          {-\alpha_2\sin(k_x\,x)\,\cos(k_y\,y)}~\vir
\end{equ}
where $k_x=\fract{2\pi}{L_x}$ and $k_y=\fract{2\pi}{L_y}$, and $\alpha_1,\alpha_2$
are two constants. For this reason,
we assign the node numbers $(n_x,n_y)=(1,1)$ to this mode.\\

   To be more precise, let $r=(x,y)\in[0,L_x)\times[0,L_y)$.
We say that a two-dimensional smooth vector 
field $\vp=(\vp_x,\vp_y)$ over the position space,
is {\em sampled} by the infinitesimal displacements $\ddq$ of the $N$ disks at 
their reference positions $q$, when
\begin{equ}
\vp(q_j)=\delta q_j~, \text{for all } j=1,\dots,N~.
\end{equ}
Thinking of a tangent vector in terms of an associated
vector field is meaningful if there are ``sufficiently'' many particles to sample the field
on a typical length scale of its variation. Once this condition breaks down for larger exponents, 
as it will for
large-enough $k$, the modes disappear\footnote{and the sampled vector
field is ill-defined.}, and so do the steps in the Lyapunov spectrum.
In the following we use a notation which does not distinguish between tangent vectors 
and their two-dimensional vector fields.

%%%%%%%%%%%%%%%%%%%%%%%%%%%%%%%%%%%%%%%%%%%%%%%%%%%%%%%%%%%%%%%%%%%%%%%%%
%
\section{Observation and description of the modes}
\label{s:classif}
%%%%%%%%%%%%%%%%%%%%%%%%%%%%%%%%%%%%%%%%%%%%%%%%%%%%%%%%%%%%%%%%%%%%%%%%%

In this section, we describe the Lyapunov
spectrum near 0
and the corresponding modes, as they are measured in numerical
experiments. In the two following subsections, we
will explain how these modes can be understood on the basis of
symmetry breaking of ``zero modes''.\\

\begin{figure}
\begin{center}
\epsfig{file=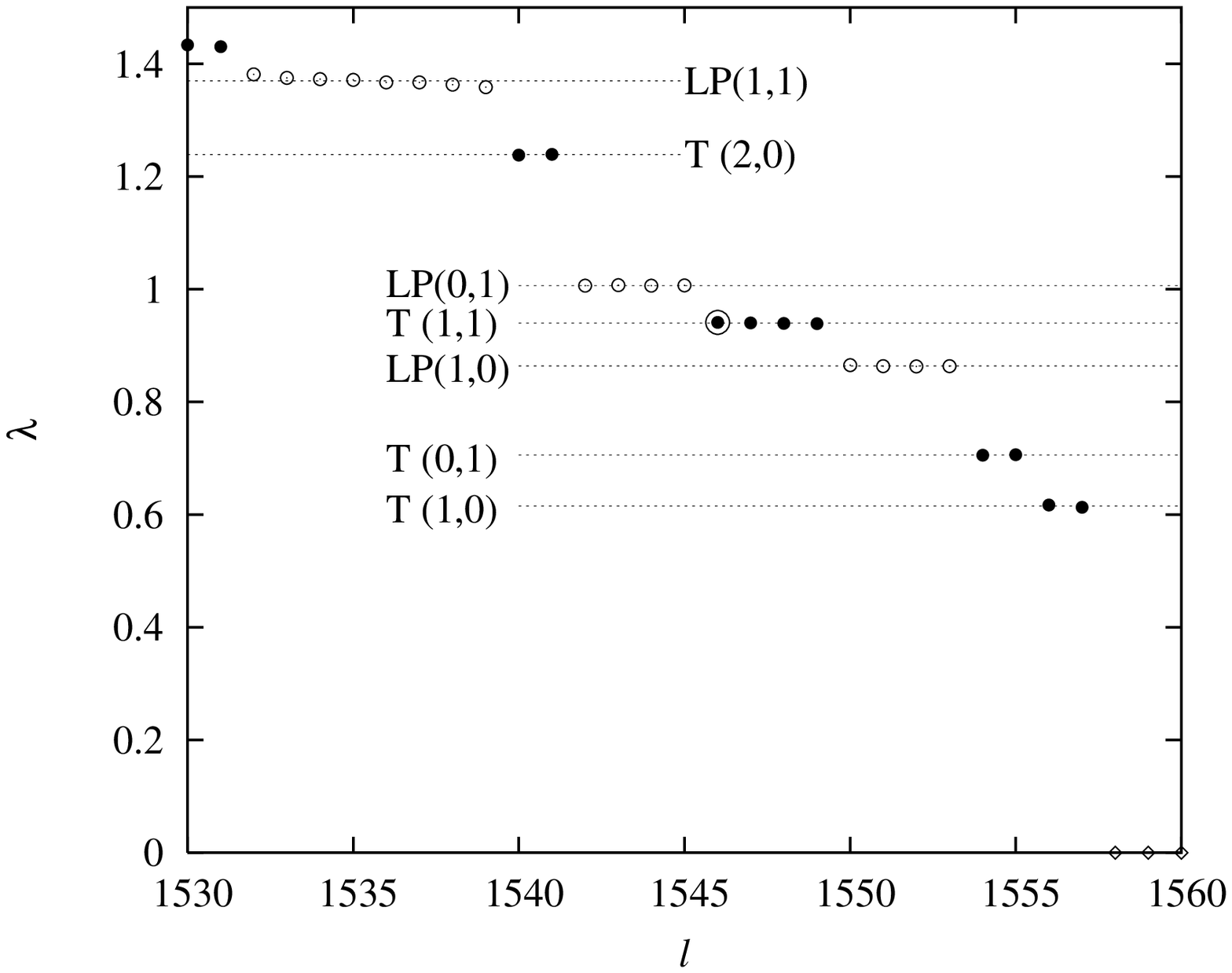, width=6cm, angle=-90}
\hspace{1cm}
\epsfig{file=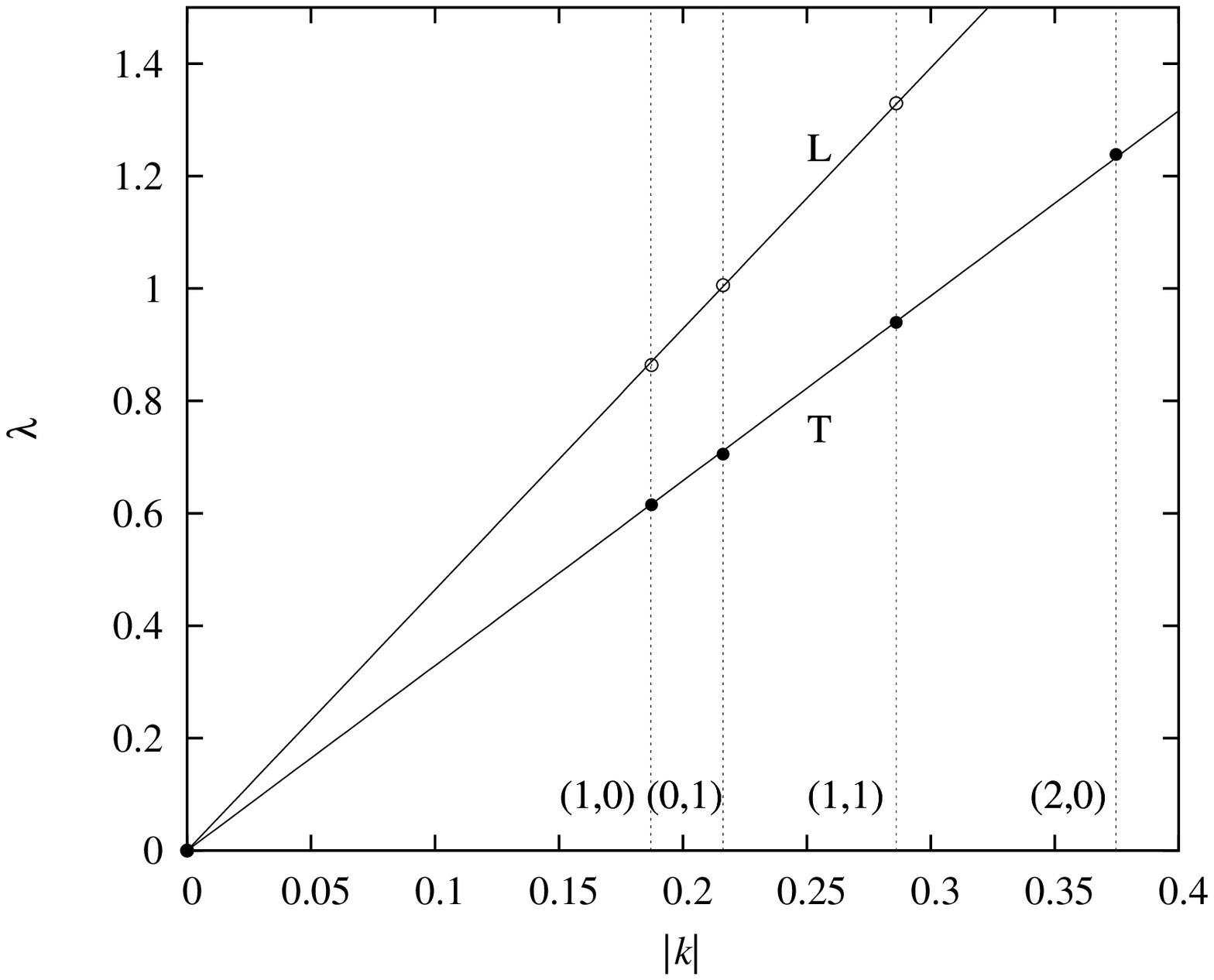, width=6cm,  angle=-90}
\caption{Lyapunov spectrum and ``dispersion relations'' for the 
780-disk system characterized in Fig.~\ref{f:lyap_spect},
and the density $\rho = 0.8.$ Left: Lyapunov
exponents are ordered by size and repeated with multiplicities.
The specially-marked point corresponds to the transverse mode 
shown in  Fig.~\ref{f:field}. Right: Lyapunov 
exponents as a function of their wave number. The respective labels 
$\L$ and $\T$ refer to the longitudinal and transverse branches.}
\label{f:disp_rel}
\end{center}
\end{figure}

We illustrate our assertions with the system already introduced in 
Fig.~\ref{f:lyap_spect}, which contains $N=780$ particles in a rectangular 
periodic box with an aspect ratio $ L_y / L_x =0.867$.
It corresponds to a hard-disk gas with a density  $\rho = 0.8$,
slightly below the fluid-to-solid phase transition density \cite{DPH96}.
The left panel of Fig.~\ref{f:disp_rel} provides a magnified view of
the smallest positive Lyapunov exponents for this system.\footnote{The conjugate 
negative exponents are not shown, see \cite{PH00,FHPH04}.} 
The exponents, ordered by size and repeated with their multiplicities, 
are plotted as a function of their index.
Degenerate exponents with a multiplicity $d\ge2$ appear therefore as
``steps''. To account for the
wave-like appearance of the modes, we associate  with each Lyapunov vector a 
wave number $k_{(n_x,n_y)}$, as in \Ref{e:k},
where the non-negative integers $\nn=(n_x,n_y)$ count the nodes in the respective 
directions.\footnote{In many cases, the number of nodes is easy to determine,
as in Fig.~\ref{f:field}. To facilitate an objective identification of the 
wave vectors modes associated with larger exponents, which are noisier,
Fourier-transforms are used.}\\

%%%%%%%%%%%%%%%%%%%%%%%%%%%%%%%%%%%%%%%%%%%%%%%%%%%%%%%%%%%%%%%%%%%%%%%%%%%%%%%%%%%
%%%%%%%%%%%%%%%%%%%%%%%%%%%%%%%%%%%%%%%%%%%%%%%%%%%%%%%%%%%%%%%%%%%%%%%%%%%%%%%%%%%

When the small Lyapunov exponents are plotted as a function of their corres\-ponding wave number, 
they all lie on two curves, sometimes referred to as
``dispersion relations'' \cite{PH00,FHPH04}. This is demonstrated on
the right panel 
of Fig.~\ref{f:disp_rel}.  
On this plot, degenerate exponents are represented by a single point.
For reasons discussed below, the upper branch is called 
{\em longitudinal} (L), and the lower {\em transverse} (T). 
It is experimentally found that for a given wave number, the multiplicity of
the L branch is twice that of the T branch, as mentioned already in Sect.~\ref{s:tandyn}.
\subsection{Vanishing Lyapunov exponents}
\label{s:null}
%%%%%%%%%%%%%%%%%%%%%%%%%%%%%%%%%%%%%%%%%%%%%%%%%%%%%%%%%%%%%%%%%%%%%%%

We start our description with the six modes associated 
with the six vanishing Lyapunov exponents, commonly referred to as {\em zero
modes}. Four of them are induced by the homogeneity of space, and two
are consequences of the homogeneity of time. They span a
six-dimensional subspace $\NN(\xi)$
of the tangent space, $TX(\xi)$, at any
phase point $\xi$.\footnote{It is important to keep in mind that
$\NN(\xi)$ really depends on $\xi$, see below.}
These 6 zero modes play a
fundamental role in understanding the nature of the modes associated
with Lyapunov exponents close to zero.\\

In order to show which symmetries give rise to the zero modes, 
we list in Table \ref{t:lyapnull} the six corresponding elementary transformations.
This defines the six zero modes $\ddxi_1$ to $\ddxi_6$ in a notation that 
separates the $x$ and $y$ components of $\ddp$ and $\ddq$.
The vectors $\ddxi_1$ and $\ddxi_2$
correspond to a perturbation of the total momentum in the $x$ and $y$
directions, $\ddxi_3$ and $\ddxi_4$ to an (infinitesimal) uniform translation 
of the origin, $\ddxi_5$ to a 
change of energy, and $\ddxi_6$ to a change of the origin of time. \\

\begin{table}[ht]
\begin{equ}
\begin{array}{|l|l|}
\hline & \\[-0.4cm]
{\rm Transformation}&{\rm Generator} \\[0.1cm] \hline & \\[-0.4cm]
(p,q)\mapsto(p_x+\ve1,p_y,q_x,q_y)&\ddxi_1 = (1,0,0,0)\cr
(p,q)\mapsto(p_x,p_y+\ve1,q_x,q_y)&\ddxi_2 = (0,1,0,0)\cr
(p,q)\mapsto(p_x,p_y,q_x+\ve1,q_y)&\ddxi_3 = (0,0,1,0)\cr
(p,q)\mapsto(p_x,p_y,q_x,q_y+\ve1)&\ddxi_4 = (0,0,0,1)\cr
(p,q)\mapsto(p_x+\ve p_x,p_y+\ve p_y,q_x,q_y)&\ddxi_5 = (p_x,p_y,0,0)\cr
(p,q)\mapsto(p_x,p_y,q_x+\ve p_x,q_y+\ve p_y)\quad&\ddxi_6 = (0,0,p_x,p_y)\\[0.1cm] \hline
\end{array}
\medskip
\end{equ}
\caption{Central subspace for the vanishing Lyapunov exponents. 
Notation : $1=(1,1,\ldots,1)$, 
and $0=(0,0,\ldots,0)$. All vectors have $4N$ components.  }
\label{t:lyapnull}
\end{table}

One can explicitly check that the subspace
${\rm Span}\{\delta\xi_1,\ldots,\delta\xi_6\}$
is covariant and that its vectors have a sub-exponential growth (or decay).
It thus coincides with the null space $\NN(\xi)$ of Sect.~\ref{s:ergod}.

\begin{Remark}
The space $\NN(\xi)$ can be further decomposed into three covariant 
subspaces $\NN_x={\rm Span}\{\ddxi_1,\ddxi_3\}$, 
$\NN_y={\rm Span}\{\ddxi_2,\ddxi_4\}$
and $\NN_p={\rm Span}\{\ddxi_5,\ddxi_6\}$, each of which independently satisfies the 
properties listed for $\NN(\xi)$. In systems with reflecting boundaries,
only $\NN_p$ is present, while $\NN_x$ and $\NN_y$ are absent, because they are 
related to  translation invariance.  If only the $x$ direction is periodic 
\cite{TM03}, the space of zero modes is reduced to $\NN_x\oplus \NN_p$.
\end{Remark}

%%%%%%%%%%%%%%%%%%%%%%%%%%%%%%%%%%%%%%%%%%%%%%%%%%%%%%%%%%%%%%%%%%%%%%%
%
\subsection{Longitudinal, Transverse and \P-modes}
\label{s:lpt}
%%%%%%%%%%%%%%%%%%%%%%%%%%%%%%%%%%%%%%%%%%%%%%%%%%%%%%%%%%%%%%%%%%%%%%%

Following our argument of Sect.~\ref{s:stable}, we need only describe 
the $\delta q$ part of the modes. Therefore, we consider the three
transformations (see Table~\ref{t:lyapnull})
\begin{equa}[e:zero]
\ddxi_3:(q_{x,j},q_{y,j})&\mapsto(q_{x,j}+\ve,q_{y,j})\\
\ddxi_4:(q_{x,j},q_{y,j})&\mapsto(q_{x,j},q_{y,j}+\ve)\\
\ddxi_6:(q_{x,j},q_{y,j})&\mapsto(q_{x,j}+\ve p_{x,j},q_{y,j}+\ve p_{y,j})~\vir
\end{equa}
associated to zero modes, and claim the following~:

%%%%%%%%%%%%%%%%%%%%%%%%%%%%%%%%%%%%%%%%%%%%%%%%%%%%%%%%%%%%%%%%%%%%%%%
\proclaim{Modes Classification I}{%
Modes of wave number $k_\nn$ are scalar modulations of \Ref{e:zero}
with wave number $k_\nn$. More precisely, they are obtained by
replacing $\ve$ with $\ve\,\A(q_{x,j},q_{y,j})$ in \Ref{e:zero},
where
the real scalar function $\A$ is of the form
\begin{equ}
\A(x,y) =\sum_{|\ell |=n_x,|m|=n_y}c_{\ell ,m}\,\exp\big(i(\ell \,k_xx+m\,k_yy)\big)~.
\end{equ}
}
%%%%%%%%%%%%%%%%%%%%%%%%%%%%%%%%%%%%%%%%%%%%%%%%%%%%%%%%%%%%%%%%%%%%%%%
%
The space of such modulations has dimension 4 in general and dimension 2 if either $n_x$
or $n_y$ vanishes.

\begin{Example}
We consider $\nn=(1,0)$, $(0,1)$ and $(1,1)$.
We use the notation $c_x=\cos(k_xx)$, $s_x=\cos(k_xx)$, $c_y=\cos(k_yy)$ and
$s_y=\sin(k_yy)$. A basis of the space of functions with wave
number $k_\nn$ is shown in Table \ref{t:Vn}. Of course, this choice
fixes a constant phase for the sine and cosine functions. 
%\footnote{%
%We prefer the multiplicative notation to the additive one of \Ref{e:func}. 
%The latter can, of course, be obtained with identities such as 
%$2\sin x\cos y=\sin(x+y)+sin(x-y)$.}.

\begin{table}[ht]
\begin{equ}
\begin{array}{|c|l|c|}
\hline
\;\nn\;&{\rm Function\ }\A&{\rm dim}\\\hline
(1,0)&c_x,s_x&2\\
(0,1)&c_y,s_y&2\\
(1,1)&c_xc_y,s_xc_y,c_xs_y,s_xs_y&4\\
\hline
\end{array}
\end{equ}
\caption{Functions with wave number $k_\nn$}
\label{t:Vn}
\end{table}
\end{Example}

If the Lyapunov exponent were a function of $k_\nn$ only,
we would expect 12-fold degeneracy in  Fig.~\ref{f:disp_rel},
(resp. 6-fold
if either $n_x$ or $n_y=0$), since each of the three perturbations
of \Ref{e:zero} can be modulated by the four (resp. two) functions of
Table~\ref{t:Vn}. 
However, this degeneracy is broken into an
8+4 (resp. 4+2) structure:

\proclaim{Mode Classification II}{
The Lyapunov vectors of wave number $k_\nn$ have two possible Lyapunov 
exponents: $|\lambda|=c_{\T}\,|k_\nn|$ or $|\lambda|=c_{\L}\,|k_\nn|$, corresponding to the
Transverse or Longitudinal branch of Fig.~\ref{f:disp_rel}.
The modes for these two branches are obtained as follows:
\begin{mylist}
\item[1)] 
Transverse branch: transverse modes are obtained by combining the modulations of 
$\ddxi_3$ and $\ddxi_4$ in a {\em divergence-free} vector field.
We denote by $\T(\nn)$ the space of such vector fields. 
\item[2)] Longitudinal branch:
\begin{mylist}
\item[(i)]
Longitudinal modes are {\em irrotational} vector fields 
one obtains by combining the modulations of $\ddxi_3$ and $\ddxi_4$.
We denote the corresponding space by $\L(\nn)$. 
\item[(ii)]
\P-modes are modulations of $\ddxi_6$, and we denote the corresponding subspace
by $\P(\nn)$.
\end{mylist}
\end{mylist}
The three subspaces $\T(\nn)$, $\L(\nn)$ and $\P(\nn)$ have dimension 4
(or dimension 2 if either $n_x$ or $n_y$ vanishes).
We denote by ${\LP}(\nn)\equiv \L(\nn)\oplus \P(\nn)$ the subspace corresponding
to the Longitudinal branch. It has dimension 8 (4 if either $n_x$ or $n_y$ vanishes).
}

\begin{Remark}
The divergence and curl of a vector field $\vp=(\vp_x,\vp_y)$ are, of course,
\begin{equ}
\nabla\cdot\vp=\dd_x\vp_x+\dd_y\vp_y~\vir\quad
\nabla\wedge\vp=\dd_x\vp_y - \dd_y\vp_x~.
\end{equ}
Since every two-dimensional vector field uniquely decomposes into a sum
of a divergence-free and an irrotational vector field, the spaces $\L(\nn)$ and
$\T(\nn)$ span all possible two-dimensional vector fields of wave number $k_\nn$.
\end{Remark}

There is a simple way to build $\P(\nn)$, $\L(\nn)$ and 
$\T(\nn)$ from the scalar modulations of Table~\ref{t:Vn}. If $\A$ is a
modulation of wave number $k_\nn$, then we have
\begin{equ}[e:PLT]
p\A\in \P(\nn)~\vir\quad\nabla\A\in \L(\nn)~\vir\quad\nabla\wedge\A\in \T(\nn)~\vir
\end{equ}
where by definition
\begin{equ}
\nabla\A=\fld{\dd_x\A}{\dd_y\A}~\vir\quad\nabla\wedge\A=\fld{\dd_y\A}{-\dd_x\A}~.
\end{equ}
This construction is also useful because it naturally defines what
we shall call {\bf {LP} pairs}, by which we denote a field of $\L(\nn)$ and a field
of $\P(\nn)$ originating from the same scalar modulation, as in \Ref{e:PLT}.
We show below that {\LP} pairs play an important role. Indeed,
when only some of the modes are present because of the boundary conditions,
{\LP} pairs are never broken: both fields are present, or both are missing
(see Sect.~\ref{s:differ}). We shall also see in Sect.~\ref{s:dynamics} that the dynamics
of the modes mostly takes place between {\LP} pairs.

\begin{figure}
\begin{center}
\epsfig{file=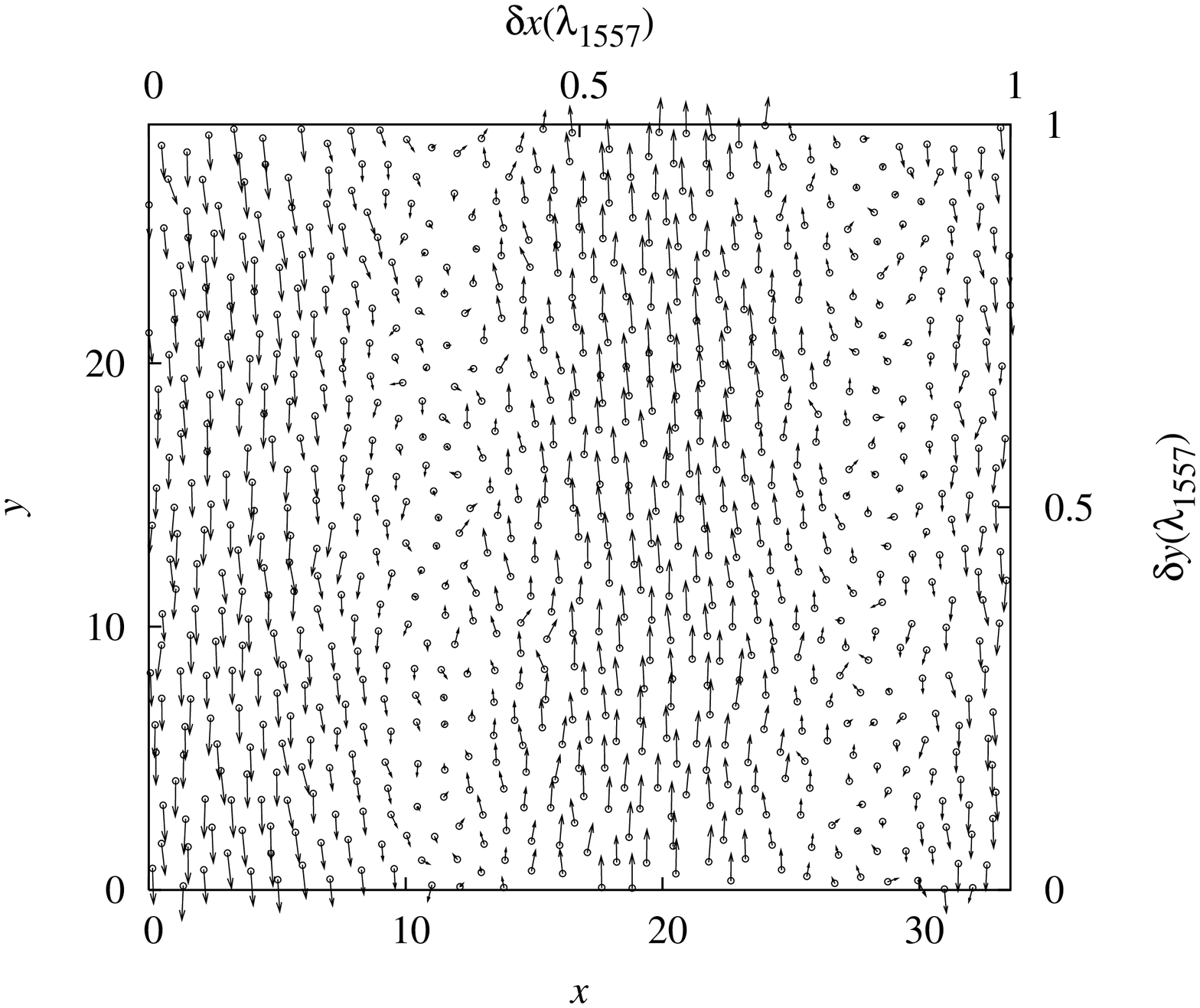, width=5cm, angle=-90}
\epsfig{file=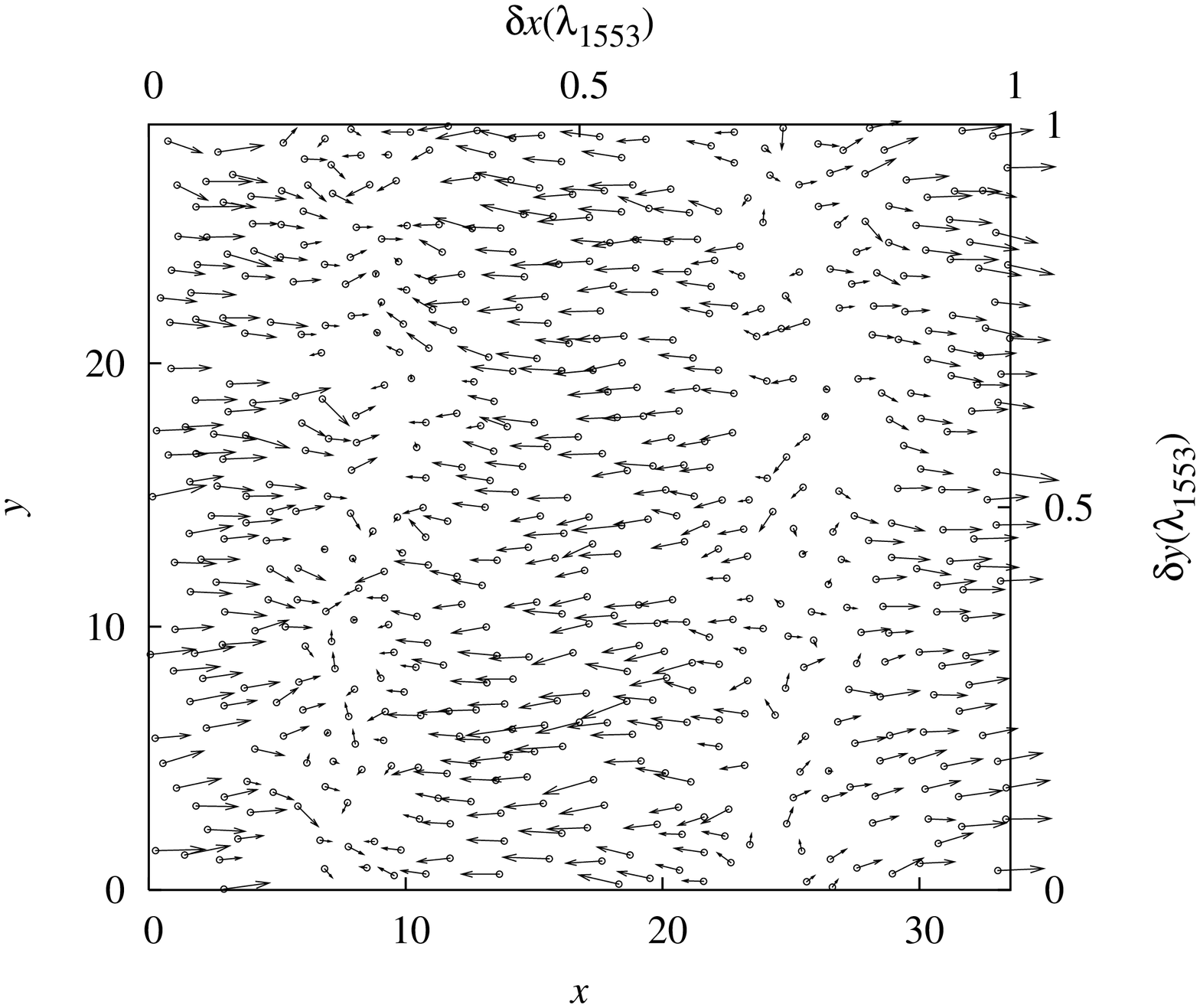, width=5cm, angle=-90}
\epsfig{file=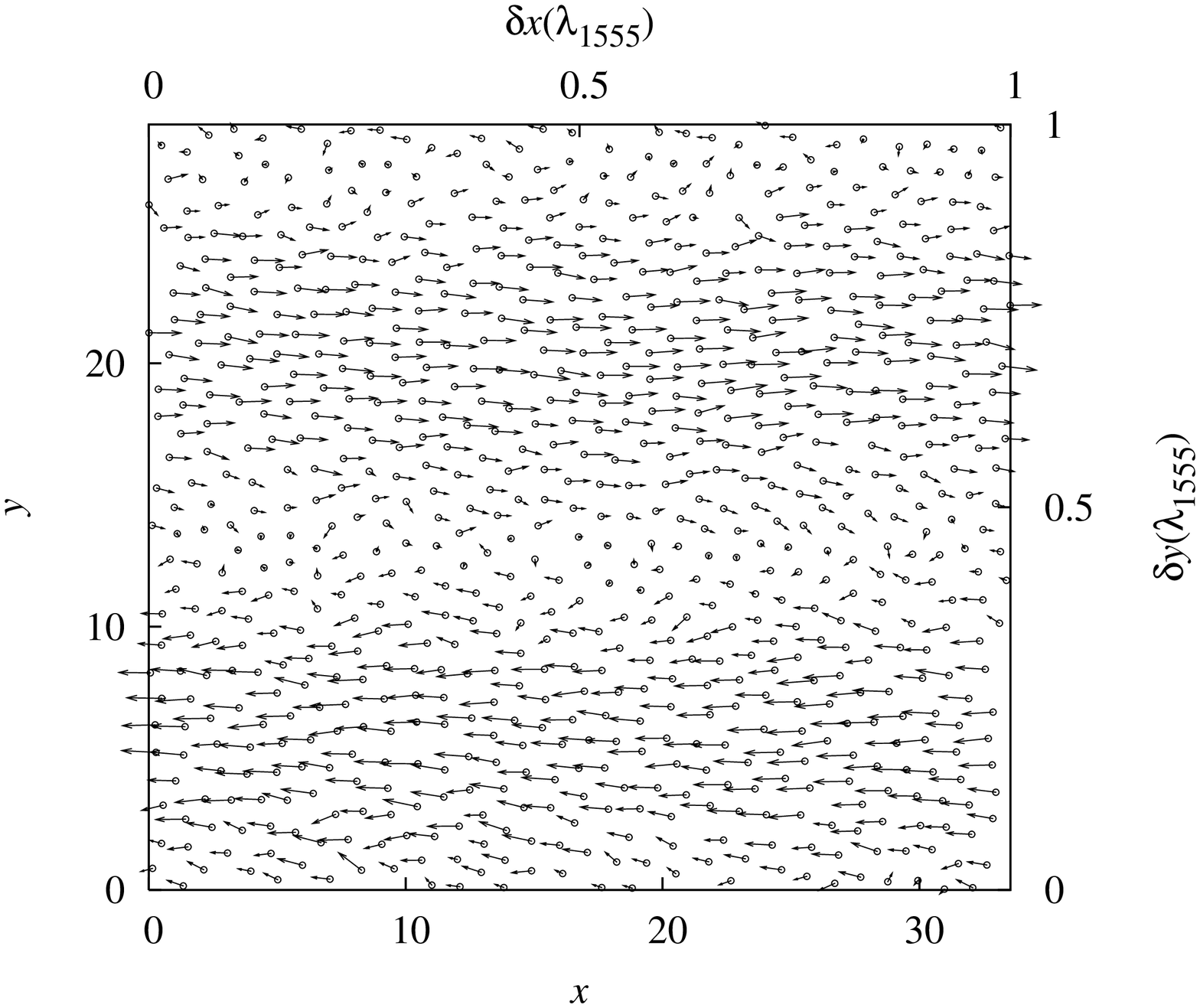, width=5cm, angle=-90}
\epsfig{file=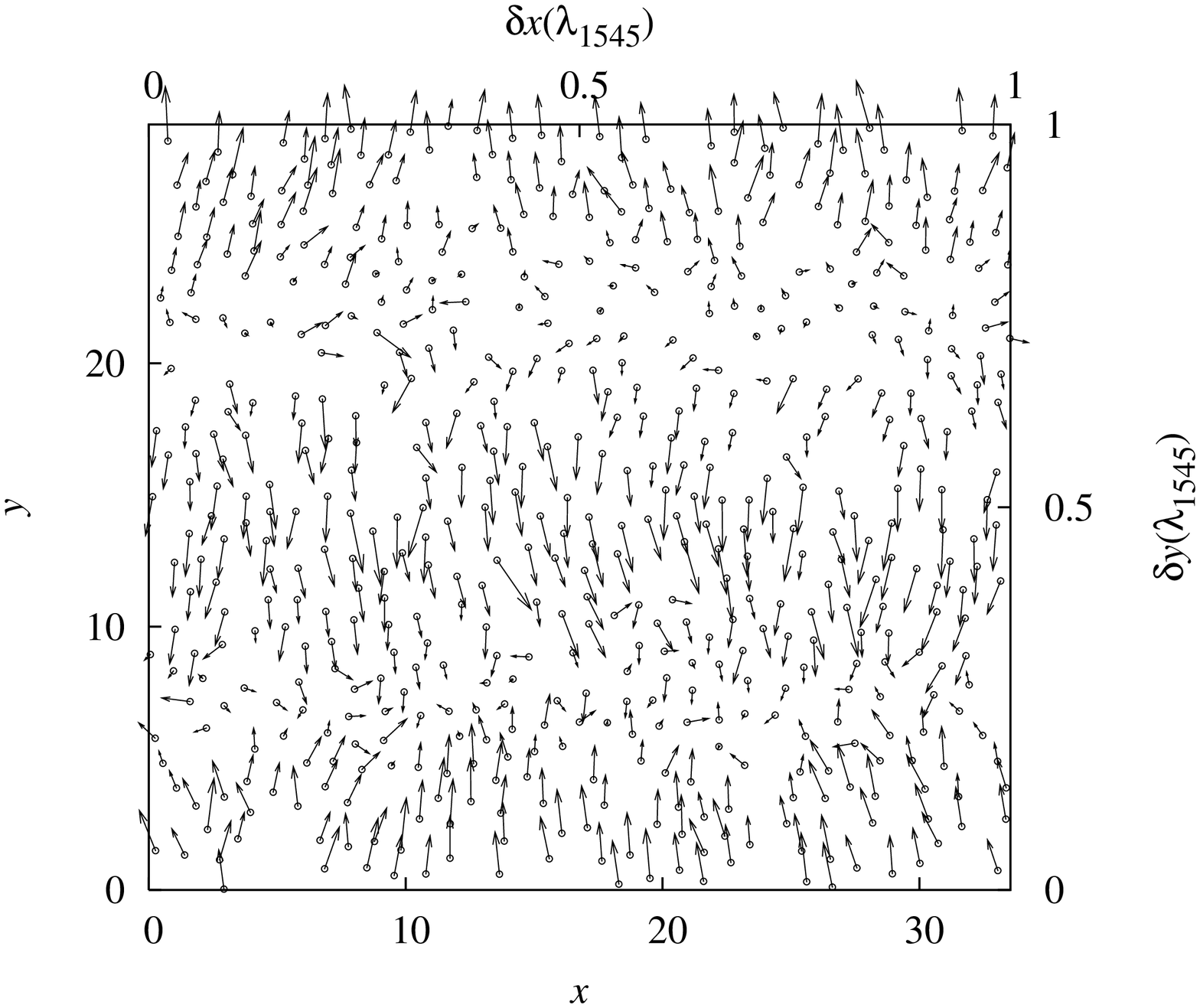, width=5cm, angle=-90}
\epsfig{file=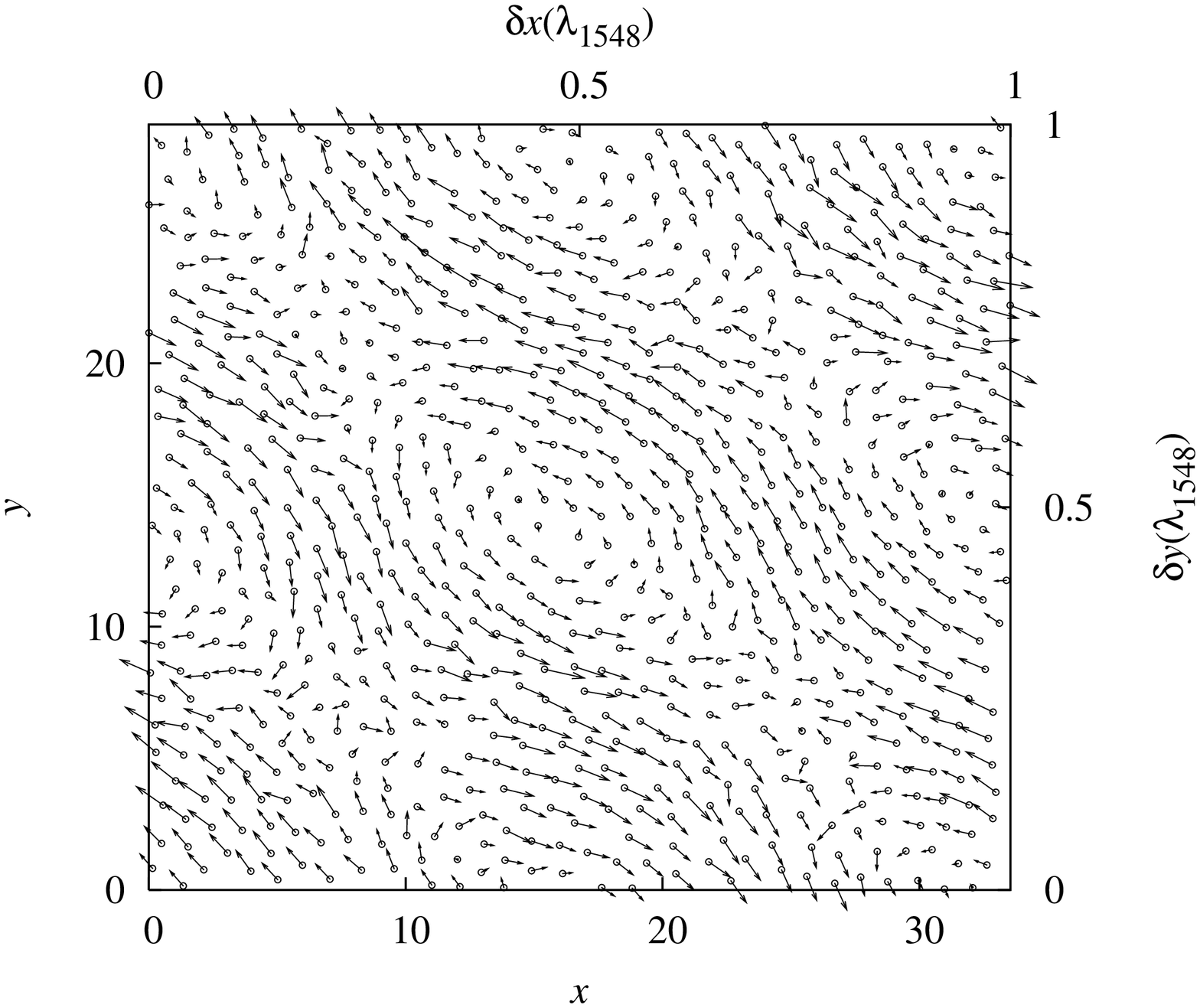, width=5cm, angle=-90}
\epsfig{file=figures/Mode_arrow_PBC_LP11.ps, width=5cm, angle=-90}
\caption{Snapshots of Lyapunov modes for the periodic 780-disk system of
Sect.~\ref{s:classif}.
Left, from top to bottom: Transverse modes \T(1,0), \T(0,1),
and \T(1,1) belonging to $\lambda_{1557}, \lambda_{1555},$ and 
$\lambda_{1548}$, respectively, of Fig.~\ref{f:disp_rel}. Right, from top to bottom: 
Vector fields for longitudinal modes, which belong to the {\LP} pairs {\LP}(1,0), {\LP}(0,1), 
and {\LP}(1,1), and which are associated with the respective
exponents $\lambda_{1553}, \lambda_{1545}$ and $\lambda_{1535}$, 
of Fig.~\ref{f:disp_rel}.  }
\label{f:modes_all}
\end{center}
\end{figure}

\begin{Example}
For the three modes of lowest wave number, Table \ref{t:Wn} lists
a basis of $\T(\nn)$, $\L(\nn)$ and $\P(\nn)$. 
Fields are given in a non-normalized form to keep notation short.
Corresponding $\L(\nn)$ fields and $\P(\nn)$ fields are {\LP} pairs.
Fig.~\ref{f:modes_all} provides examples for \T{}  and \L-modes.

\newcommand{\phiTx}{{\fract1{k_x}}\,}
\newcommand{\phiTy}{{\fract1{k_y}}\,}
\newcommand{\phiLx}{\phiTy}
\newcommand{\phiLy}{\phiTx}

\begin{table}[ht]
\begin{equ}
\begin{array}{|c||l||l|l|}
\hline&&&\\[-0.25cm]
\nn&{\rm Basis\ of\ }\T(\nn)&{\rm Basis\ of\ }\L(\nn)&{\rm Basis\ of\ }\P(\nn)\\[0.15cm]
\hline&&&\\[-0.3cm]
(1,0)&\fld0{c_x},\fld0{s_x}&\fld{c_x}0,\fld{s_x}0&
\fld{p_x}{p_y}\,s_x,\fld{p_x}{p_y}\,c_x\\
&&&\\[-0.3cm]
(0,1)&\fld{c_y}0,\fld{s_y}0&\fld0{c_y},\fld0{s_y}&
\fld{p_x}{p_y}\,s_y,\fld{p_x}{p_y}\,c_y\\
&&&\\[-0.3cm]
(1,1)&\fld{\phiTx c_xs_y}{-\phiTy s_xc_y},\fld{\phiTx s_xc_y}{-\phiTy c_xs_y}
&\fld{\phiLx c_xs_y}{\phiLy s_xc_y},\fld{\phiLx s_xc_y}{\phiLy c_xs_y}&
\fld{p_x}{p_y}\,s_xs_y,\fld{p_x}{p_y}\,c_xc_y\\
&&&\\[-0.3cm]
&\fld{\phiTx s_xs_y}{\phiTy c_xc_y},\fld{\phiTx c_xc_y}{\phiTy s_xs_y}
&\fld{\phiLx s_xs_y}{-\phiLy c_xc_y},\fld{\phiLx c_xc_y}{-\phiLy s_xs_y}&
\fld{p_x}{p_y}\,c_xs_y,\fld{p_x}{p_y}\,s_xc_y\\[-0.3cm]
&&&\\\hline
\end{array}
\end{equ}
\caption{Decomposition of $\nn$-modes for rectangular systems with periodic boundaries.}
\label{t:Wn}
\end{table}

\end{Example}

Modes of $\L(\nn)$ and $\T(\nn)$ are wavelike perturbations of the 
position space, and, therefore, are similar to the modes that appear in
hydrodynamics 
(see Sect.~\ref{s:hydro}). In particular, when either $n_x$ or $n_y$ vanish, the 
fields of $\L(\nn)$ and $\T(\nn)$ are, respectively, 
longitudinal and transverse to the wave vector. This observation
is the reason for the names of the two branches in Fig.~\ref{f:disp_rel}.
To stay in line with this now-accepted terminology,
we keep it also for the case $n_x\cdot n_y\ne 0$.
\P-modes are more complex than the other modes, because
they depend not only on the positions of the perturbed
particles but also on their velocities. \\

At this point we are able  to characterize the spectrum and its
multiplicities with only two constants, $c_{\L}$ and $c_{\T}$, which have the
physical dimension of {\em velocities}. 
We shall see in Sect.~\ref{s:density} that these velocities 
depend on the density of the system, but are insensitive to
the system size, the boundary type, and the aspect ratio.
Therefore, it is tempting to think of  $c_{\L}$ and $c_{\T}$  as thermodynamic velocities.
However, as discussed in Sect.~\ref{s:density}, no obvious interpretation
could be found so far.
%
%%%%%%%%%%%%%%%%%%%%%%%%%%%%%%%%%%%%%%%%%%%%%%%%%%%%%%%%%%%%%%%%%%%%%%%%%
%%%%%%%%%%%%%%%%%%%%%%%%%%%%%%%%%%%%%%%%%%%%%%%%%%%%%%%%%%%%%%%%%%%%%%%%%

%%%%%%%%%%%%%%%%%%%%%%%%%%%%%%%%%%%%%%%%%%%%%%%%%%%%%%%%%%%%%%%%%%%%%%%%%
%
\subsection{Other aspect ratios and boundary conditions}
\label{s:differ}
%
%%%%%%%%%%%%%%%%%%%%%%%%%%%%%%%%%%%%%%%%%%%%%%%%%%%%%%%%%%%%%%%%%%%%%%%%%

We summarize here some observations which concern different boundary
conditions and degeneracies.
\begin{mylist}

\item[i)]
The Lyapunov spectrum is more degenerate for square systems, for which
$k_{(n_x,n_y)}$ = $k_{(n_y,n_x)}$. In this case the 
multiplicities are doubled with respect to the general case, for which $n_x\neq n_y$.
Other ``accidental'' degeneracies may occur: for instance,
parameters can be found for which $c_{\T}\,|k_{(1,1)}| = c_{\L}\,|k_{(1,0)}|$.

\item[ii)] Systems with {\bf reflecting} boundaries \cite{TM03} develop only a subset of 
the modes encountered so far, which can be found by the following
simple, and obvious, rules:
\begin{itemize}
\item
The fundamental wave vectors are $k_x=\frac\pi{L_x}$ and $k_y=\frac\pi{L_y}$ 
(not $2\pi$). Here, $L_x$ and $L_y$ are the {\em effective} box sizes,
obtained from the actual side lengths by subtracting one particle diameter $\sigma$.  
\item
The fields of $\T(\nn)$ and $\L(\nn)$ have to satisfy Dirichlet conditions, namely 
to be tangent to the boundary:
\begin{equ}
\vp_x(0,y)=\vp_x(L_x,y)=0\quad{\rm and}\quad\vp_y(x,0)=\vp_y(x,L_y)=0~.
\end{equ}
If expressed in terms of sines and cosines, this means that $\vp_x$ may
contain $\sin(k_xx)$ but not $\cos(k_xx)$, and so on.
\item 
As for periodic boundary conditions,
when an L-mode is present, its paired P-mode is always present.
\end{itemize}

\item[iii)] 
Hybrid systems with a reflecting boundaries in one direction and periodic boundaries 
along the other behave as expected \cite{TM03}: the fundamental wave vectors are chosen
according to the boundary type, and the Dirichlet conditions are only applied
to one component of the field.

\item[iv)]
Narrow systems, for instance those with $\sigma<L_y<2\sigma$ and
$L_x\gg\sigma$,\footnote{Recall that $\sigma$ is the diameter of the disks.}
only develop modes with $n_y=0$ \cite{FMP04}. This follows, since a vector field varying
along the $y$ axis cannot be sampled by a single particle. Therefore, the Lyapunov 
spectrum of such a system is  greatly simplified, since the modes are restricted to
$\L(n_x,0)$ and $\T(n_x,0)$. 

\end{mylist}

\begin{Example}
\label{ex:Dirichlet}
Table \ref{t:WnDirichlet} shows which modes of Table \ref{t:Wn} satisfy
the Dirichlet condition and are thus present in a system
with reflecting boundaries.\footnote{For such a choice of basis vectors 
the origin of the coordinate system is at the bottom-left corner of the
simulation box.}
We stress that such a system has only {\em two}
vanishing Lyapunov exponents which are associated with $\ddxi_5$ and $\ddxi_6$ of 
Table \ref{t:lyapnull}. {\it Therefore, modes appear which 
are {\bf not} modulations of zero modes of the system.} The crucial
observation here is that even if one of the fundamental symmetries is
broken by the boundary condition, the modulation, as defined in
Eq.~(\ref{e:PLT}), may still satisfy that boundary condition. For
example, if we have reflecting boundaries on the walls $\{x=0\}$ and
$\{x=L_x\}$, then any perturbation $A(x,y)= \sin(m k_x x) B(y)$ will
be acceptable, whereas, of course, $A(x,y)= \cos(m k_x x) B(y)$ would not.

\begin{table}[ht]
\begin{equ}
\begin{array}{|c||l||l|l|}
\hline&&&\\[-0.25cm]
\nn&\T(\nn)&\L(\nn)&\P(\nn)\\[0.15cm]
\hline&&&\\[-0.3cm]
(1,0)&{\rm none}&\fld{s_x}0&\fld{p_x}{p_y}\,c_x\\
&&&\\[-0.3cm]
(0,1)&{\rm none}&\fld0{s_y}&\fld{p_x}{p_y}\,c_y\\
&&&\\[-0.3cm]
(1,1)&\fld{s_xc_y}{-c_xs_y}&\fld{s_xc_y}{c_xs_y}&\fld{p_x}{p_y}\,c_xc_y\\[-0.2cm]
&&&\\\hline
\end{array}
\end{equ}
\caption{Mode decomposition for rectangular systems with reflecting boundaries.}
\label{t:WnDirichlet}
\end{table}
\end{Example}

Systems with reflecting boundaries and narrow systems are easier to
study numerically,
because the multiplicities of the L, P and T spaces are smaller than in the
periodic rectangular case. In particular, the {\LP}(1,1) space has only dimension 2
when the boundaries are reflecting. For that reason, we illustrate 
some of the issues below also with the ``reflecting-wall version'' of the
780-disk system introduced in Fig.~\ref{f:LP_dynamics_refl}.

%%%%%%%%%%%%%%%%%%%%%%%%%%%%%%%%%%%%%%%%%%%%%%%%%%%%%%%%%%%%%%%%%%%%%%%%%
%
\subsection{How to measure P-modes}
\label{s:mesp}
%
%%%%%%%%%%%%%%%%%%%%%%%%%%%%%%%%%%%%%%%%%%%%%%%%%%%%%%%%%%%%%%%%%%%%%%%%%

Contrary to $\L(\nn)$ and $\T(\nn)$, the tangent 
subspace  $\P(\nn)$ {\em depends on the state that it
perturbs}. Moreover,  modes of $\P(\nn)$ are not really vector fields,
because the velocities of the particles in a typical configuration do not depend
smoothly on position.
In order to ``see'' a \P-mode, we face two problems:
\begin{mylist}

\item[i)]
A typical measured vector of an ${\LP}(\nn)$ space is a {\em superposition}
of vectors of $\L(\nn)$ and $\P(\nn)$.

\item[ii)]
Even when a mode of $\P(\nn)$ is isolated, it is not smooth and does not
``look like'' a vector field over the box.

\end{mylist}
We explain our solution to the first problem with the simplest possible example: the
${\LP}(1,0)$ space for a rectangular system with {\bf reflecting boundaries}
(see Sect.~\ref{s:differ}). This space
has dimension two and is defined by the two normalized spanning vectors
\begin{equ}[e:fldLP]
\fld{\vp^{\L}_x}{\vp^{\L}_y}=\frac1{z_1}\,\fld{s_x}{0}~\vir\quad
\fld{\vp^{\P}_x}{\vp^{\P}_y}=\frac1{z_2}\,\fld{p_x}{p_y}\,c_x~,\quad k_x=\frac\pi{L_x}~\vir
\end{equ}
where $z_1$ and $z_2$ are normalization constants.
At any time $t$,
we have two measured modes, $\psi^1$ and $\psi^2$ (vectors with $2N$ components, 
since only the $\ddq$ part is observed), whose span is (numerically very close to) ${\LP}(1,0)$.
Therefore, there are constants $a$, $b$, $c$, $d$ with
\begin{equa}[e:rot2d]
\vp^{\L}&=\psi^1\,a+\psi^2\,b~,\cr
\vp^{\P}&=\psi^1\,c+\psi^2\,d~.
\end{equa}
Since all these vectors  are normalized, one should also have
$a^2+b^2=c^2+d^2=1$.  However, Eq.~\Ref{e:rot2d} is over-determined: four
constants have to satisfy $4N$ equations. Numerically, we use a least-square
method to find the best values for $a$,$b$,$c$ and $d$, which we denote
by $\alpha$, $\beta$, $\gamma$, $\delta$.\footnote{It is also possible
to use a simple projection $\alpha=\vp^{\L}\cdot\psi^1$, etc. This method assumes
that $\vp^{\L}$ and $\vp^{\P}$ are orthogonal vectors, which they only are approximately.}
Thus, the measured modes, $\psi^1$ and $\psi^2$, are decomposed according to
\begin{equa}
\tilde\vp^{\L}&=\psi^1\,\alpha+\psi^2\,\beta~,\cr
\tilde\vp^{\P}&=\psi^1\,\gamma+\psi^2\,\delta~,
\end{equa}
where the vectors $\tilde\vp^{\L}$ and $\tilde\vp^{\P}$ are the
best-possible LP-pair, $\vp^{\L}$ and $\vp^{\P}$, reconstructed from experimental data. \\

%%%%%%%%%%%%%%%%%%%%%%%%%%%%%%%%%%%%%%%%%%%%%%%%%%%%%%%%%%%%%%%%%%%%%%%
\begin{figure}
\begin{center}
\epsfig{file=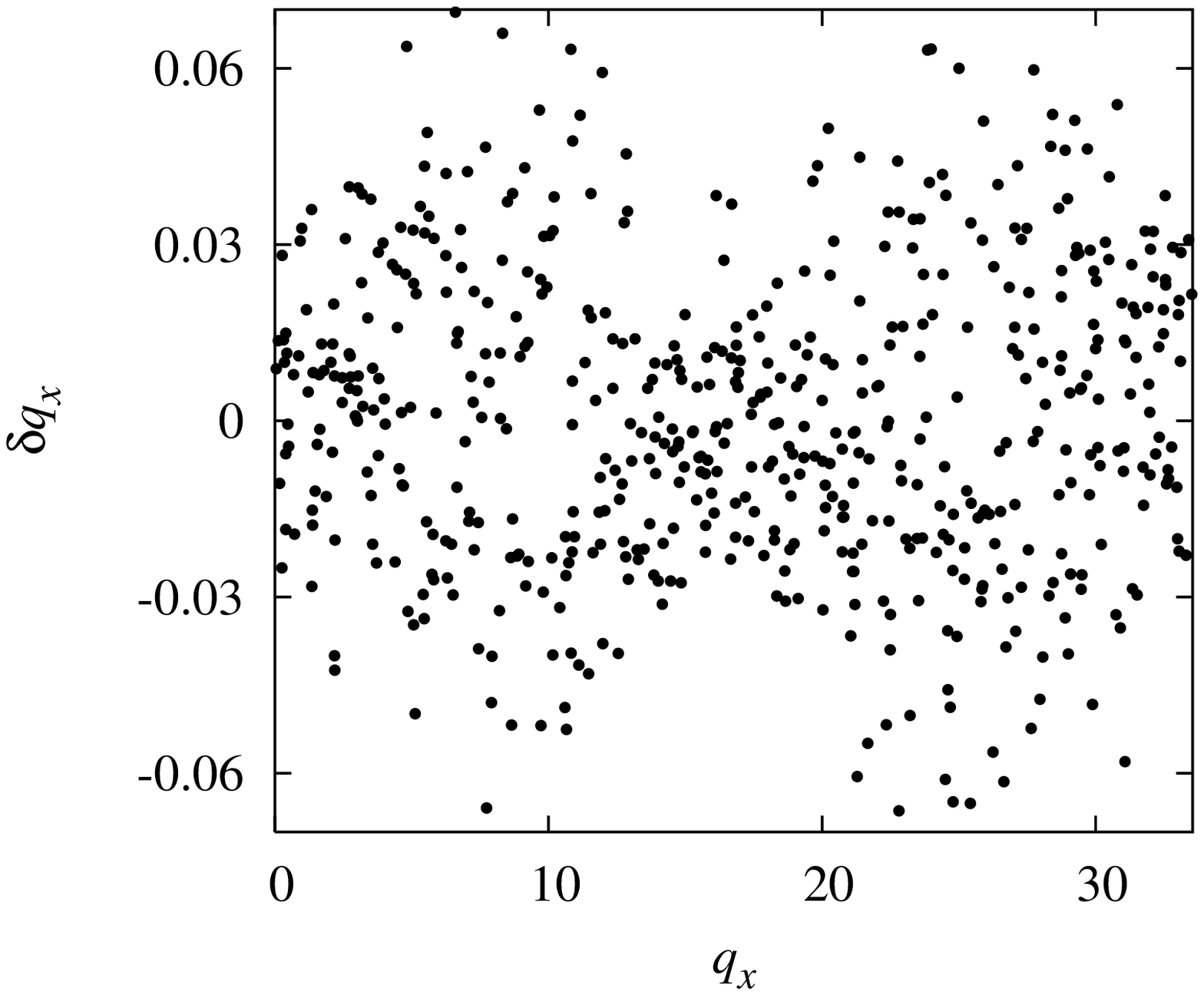, width=5cm, angle=-90}
\epsfig{file=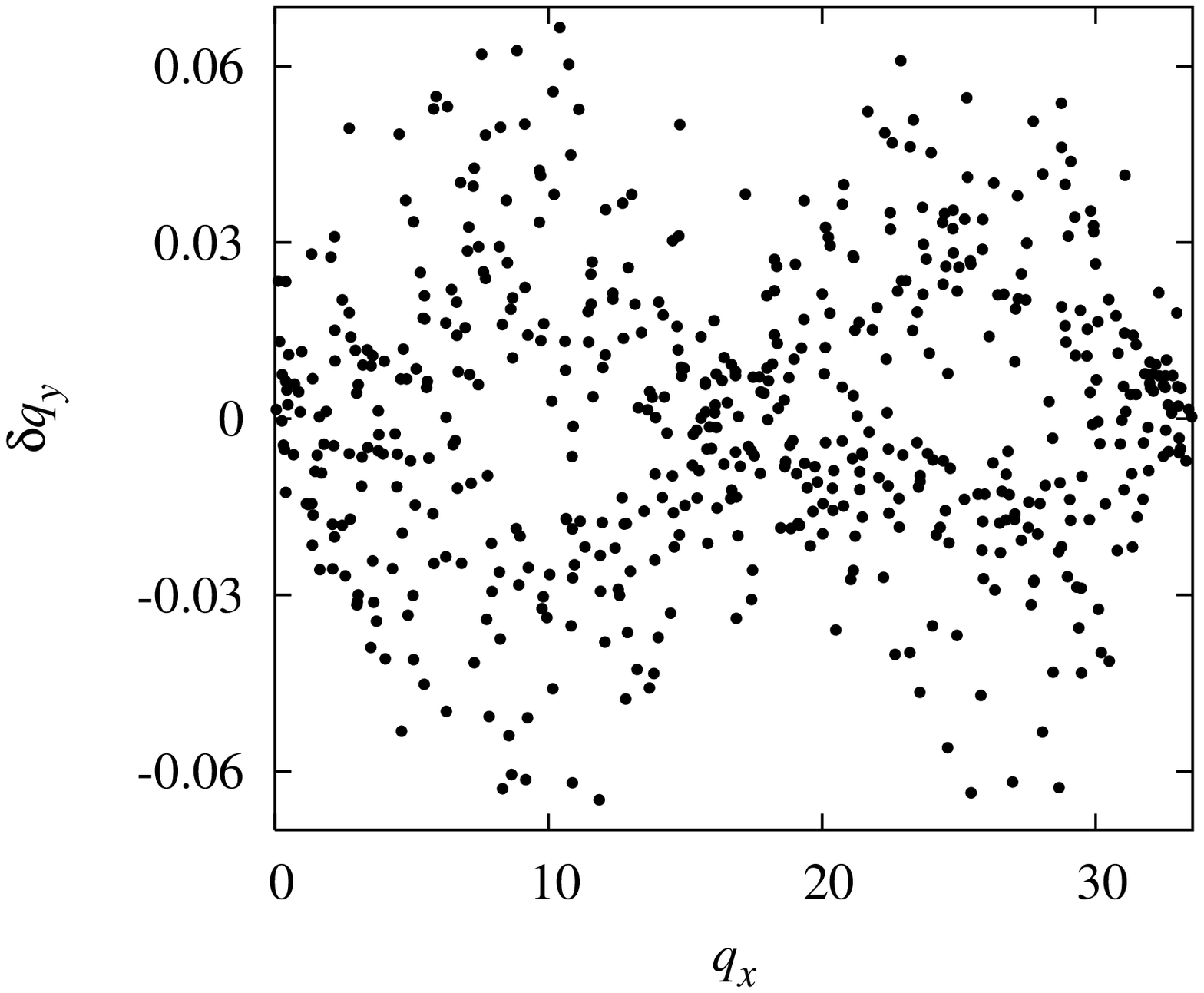, width=5cm, angle=-90}
\epsfig{file=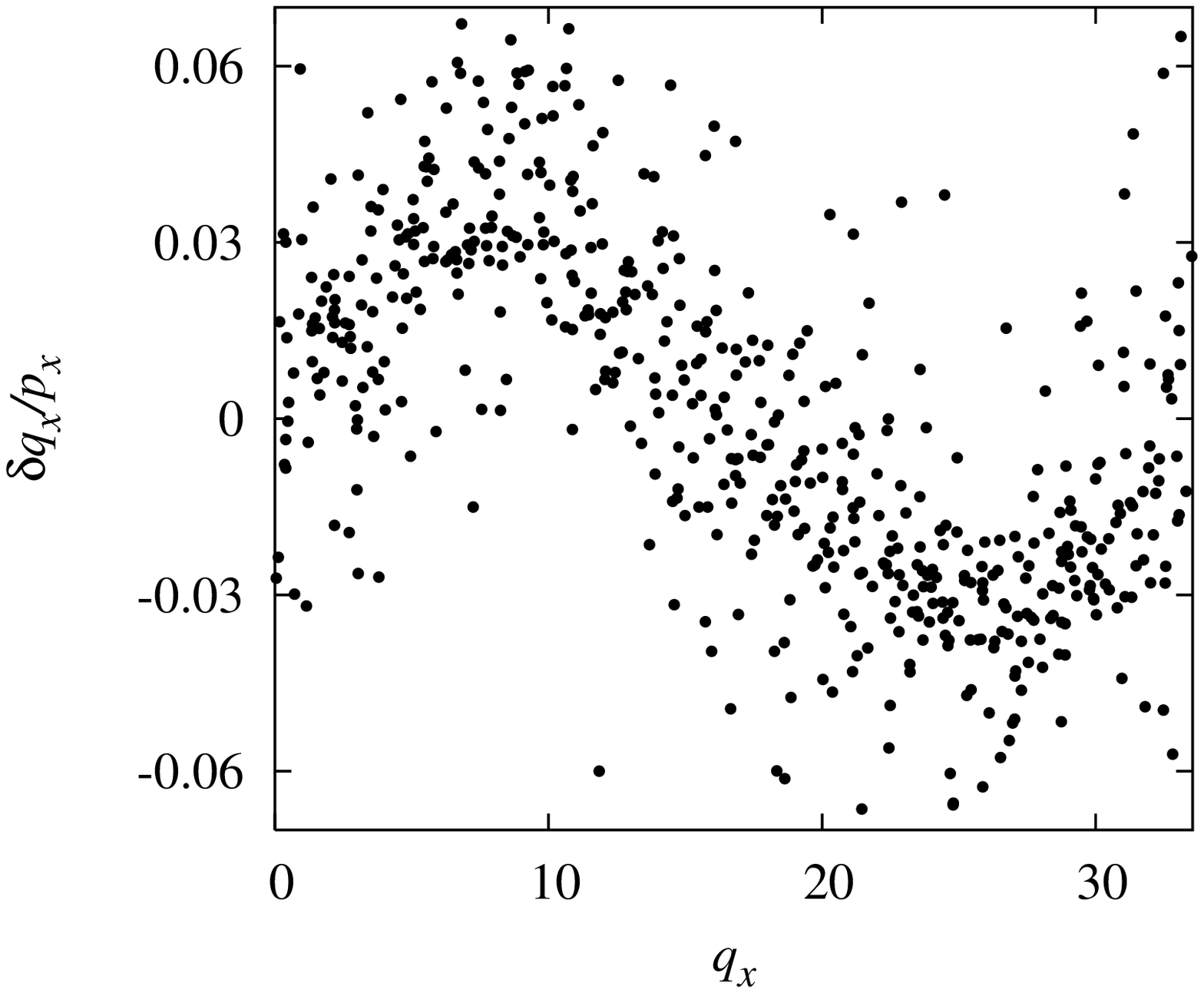, width=5cm, angle=-90}
\epsfig{file=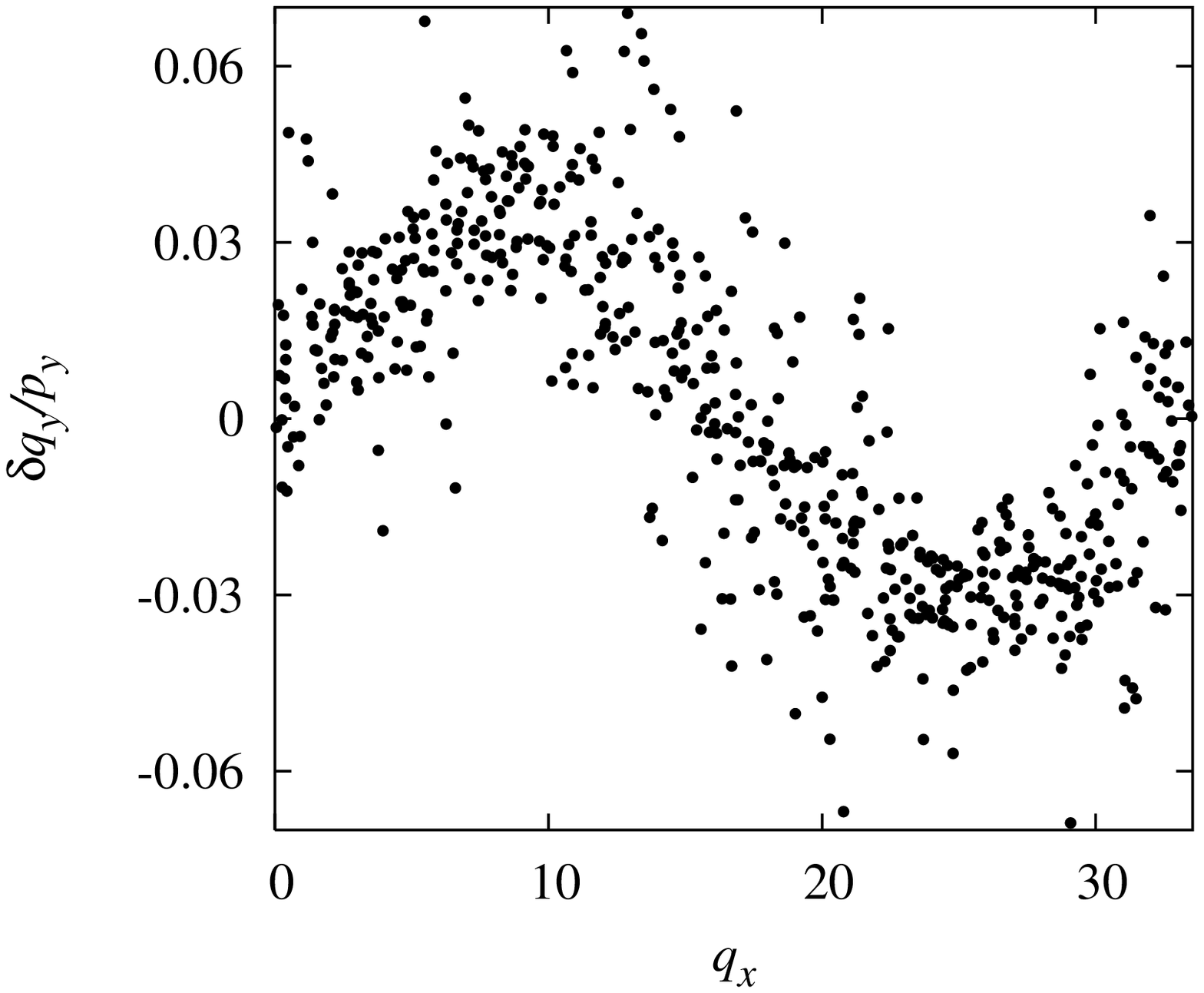, width=5cm, angle=-90}
\caption{Example for the \P-mode reconstruction for the 4-dimensional 
LP(1,0) space
of a system with {\em periodic boundaries.} Only half of the components are
shown, namely those  corresponding to $s_x$. The $c_x$ components are similar.
Top row: $x$ and $y$ components of the reconstructed
 \P-mode $\tilde\vp^{\P} \approx p s_x.$ 
Bottom row: the fields $\tilde\vp^{\P}_x/p_x$ and $\tilde\vp^{\P}_y/p_y$
are wavelike again. Note that in the top row one can recognize the
sinusoidal envelope.}
\label{f:refl_LP_modes}
\end{center}
\end{figure}
%%%%%%%%%%%%%%%%%%%%%%%%%%%%%%%%%%%%%%%%%%%%%%%%%%%%%%%%%%%%%%%%%%%%%%%

Now we can deal with our second problem: whereas the vector $\tilde\vp^{\L}$ is easily
recognized as a vector field, the vector $\tilde\vp^{\P}$ is not.  However, from  
\Ref{e:fldLP} we deduce that 
\begin{equ}
\fld{\vp^{\P}_x/p_x}{\vp^{\P}_y/p_y}=\fld{c_x}{c_x}~\vir 
\end{equ}
or, more precisely, 
\begin{equ}[e:oneoverp]
\fld{\vp^{\P}_{x,j}/p_{x,j}}{\vp^{\P}_{y,j}/p_{y,j}}=
\fld{\cos(k_xq_{x,j})}{\cos(k_xq_{x,j})}~\vir\quad j=1,\ldots,N~.
\end{equ}
Therefore, both $\vp^{\P}_x/p_x$ and $\vp^{\P}_y/p_y$ are smooth functions
and can be easily visualized.\\ 

\noindent
{\bf Remarks}

\begin{mylist}

\item[i)]
This procedure readily extends to higher-dimensional ${\LP}(\nn)$ spaces, 
for instance to the four-dimensional ${\LP}(1,0)$ space of a periodic system.
We take this case as an example to illustrate in Fig.~\ref{f:refl_LP_modes}
our method of mode reconstruction: from the four measured modes
the analogue to Eq.~\Ref{e:rot2d} generates four spanning vectors, namely
two \L-modes (one of which is shown at the
top-right position of Fig.~\ref{s:classif}) and two \P-modes.
The $x$ and $y$ components of one of the \P-modes,
namely  $\tilde\vp^{\P} \approx p s_x,$ are shown in the top row
of Fig.~\ref{f:refl_LP_modes}.  No smooth functions are
recognized. However, after dividing by the momentum components,
$p_{x,j}$ and $p_{y,j},$  as required by (\ref{e:oneoverp}), 
the figures for the fields $\tilde\vp^{\P}_x/p_x$ and $\tilde\vp^{\P}_y/p_y$
in the bottom row of Fig.~\ref{f:refl_LP_modes} clearly display the 
expected $s_x$-dependence.
\item[ii)] 
Division by $p_{x,j}$ or $p_{y,j}$ 
in \Ref{e:oneoverp} is numerically unstable when particle $j$ has a very small 
momentum along a coordinate axis.  We avoid this by multiplying instead
by $p_{x,j}/(p_{x,j}^2+\ve)$, where $\ve\ll1$.  One could also ignore
particle $j$ in this case and only sample the field at those points where 
both $p_{x,j}$ and $p_{y,j}$ are not too small.

\end{mylist}

%%%%%%%%%%%%%%%%%%%%%%%%%%%%%%%%%%%%%%%%%%%%%%%%%%%%%%%%%%%%%%%%%%%%%%%%
%
\section{Dynamics of the modes}
\label{s:dynamics}
%%%%%%%%%%%%%%%%%%%%%%%%%%%%%%%%%%%%%%%%%%%%%%%%%%%%%%%%%%%%%%%%%%%%%%%%

\noindent
{\bf Remark}
The interested reader can look up animated pictures on the web at the address\\
{\small\tt http://theory.physics.unige.ch/modes/}~.\\

In this section we turn to the ``dynamics of the modes'' 
and study what has been called the velocity of the longitudinal modes \cite{PH00,MM03,WB04}. 
This might clarify hydrodynamic theories \cite{MM01,MM03}, which are
mostly based on a partial classification of the modes.
In numerical simulations, the {\em transverse} modes are stationary in space and time:
although the particles move, the vector-field of the mode does not. In other 
words, at any instant of time, the vector field of a \T-mode 
does not move (up to a small jitter due to numerical noise).\\

For {\em longitudinal} modes, however, one seems to observe \cite{FHP03}
a propagation in the direction of the wave vector.\footnote{when either
$n_x$ or $n_y$ vanish, for instance in the $\L(1,0)$ or $\L(0,1)$ space.} 
Using the geometrical 
picture developed above, we can interpret this motion and also explain why no propagation is 
observed  for the {\LP} dynamics in systems with reflecting boundaries as
demonstrated in Fig.~\ref{f:LP_dynamics_refl} below. \\

%%%%%%%%%%%%%%%%%%%%%%%%%%%%%%%%%%%%%%%%%%%%%%%%%%%%%%%%%%%%%%%%%%%%%
\begin{figure}
\includegraphics[width=3.6cm,angle=-90]{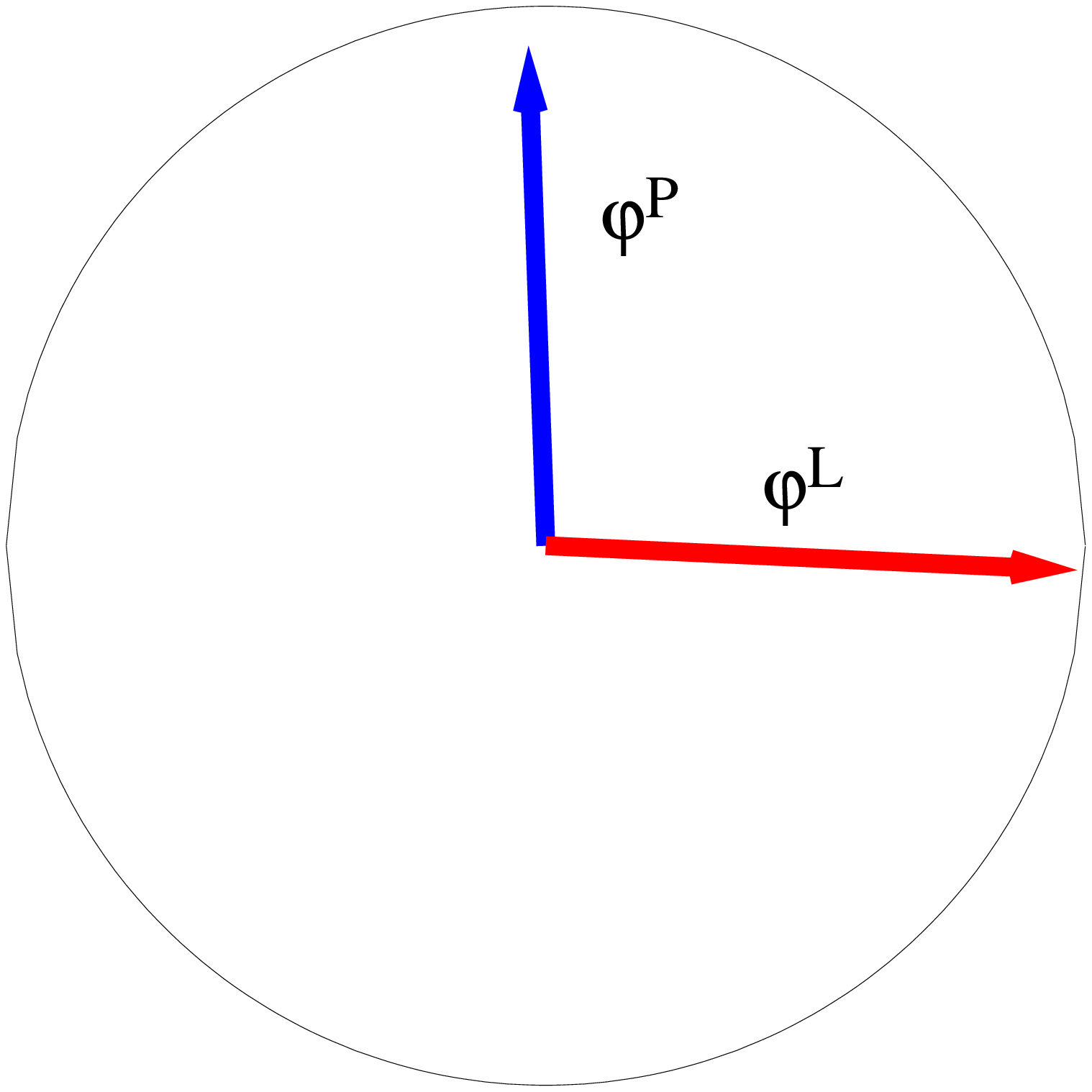}
\hfill
\includegraphics[width=3.6cm,angle=-90]{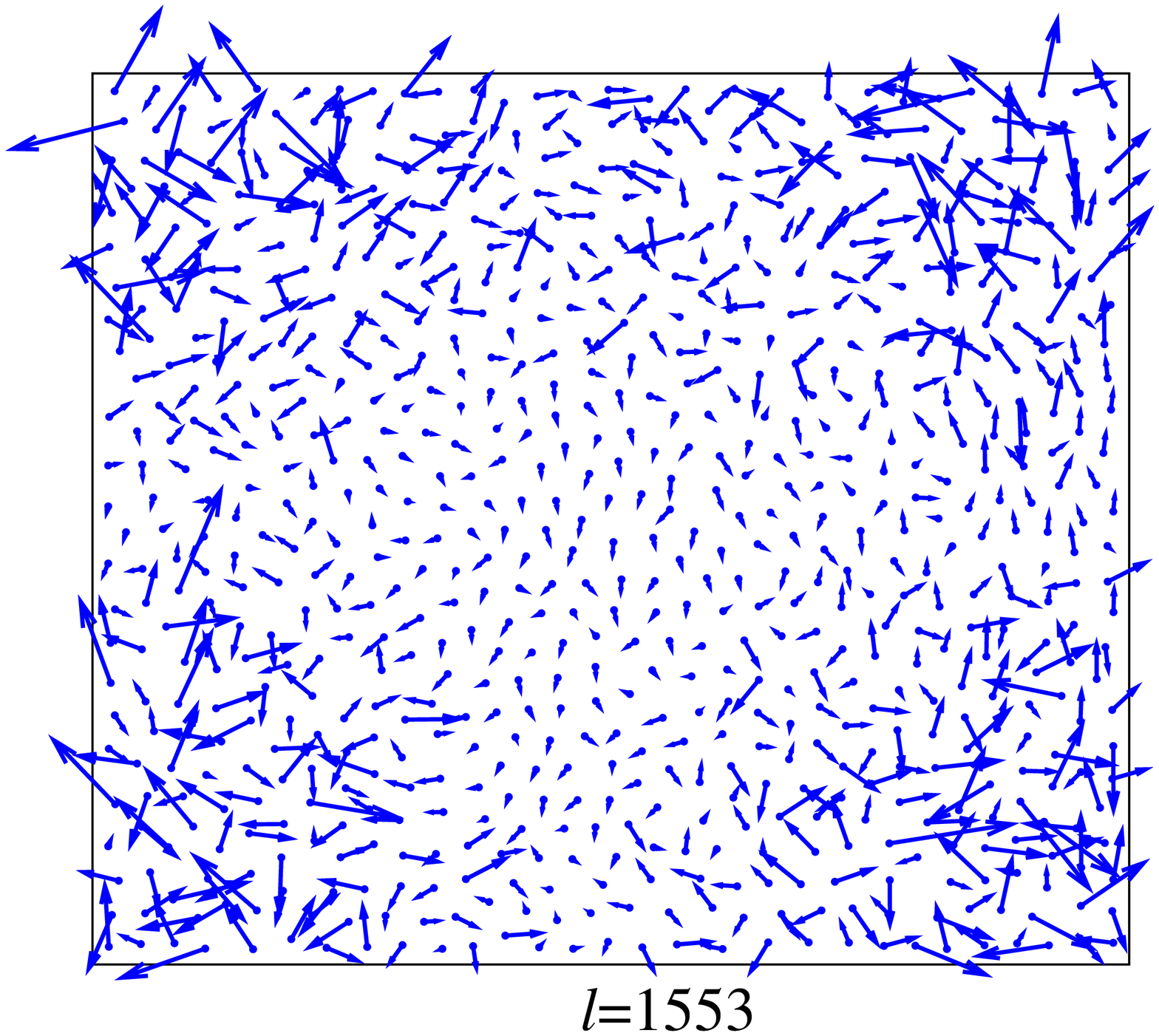}
\hfill
\includegraphics[width=3.6cm,angle=-90]{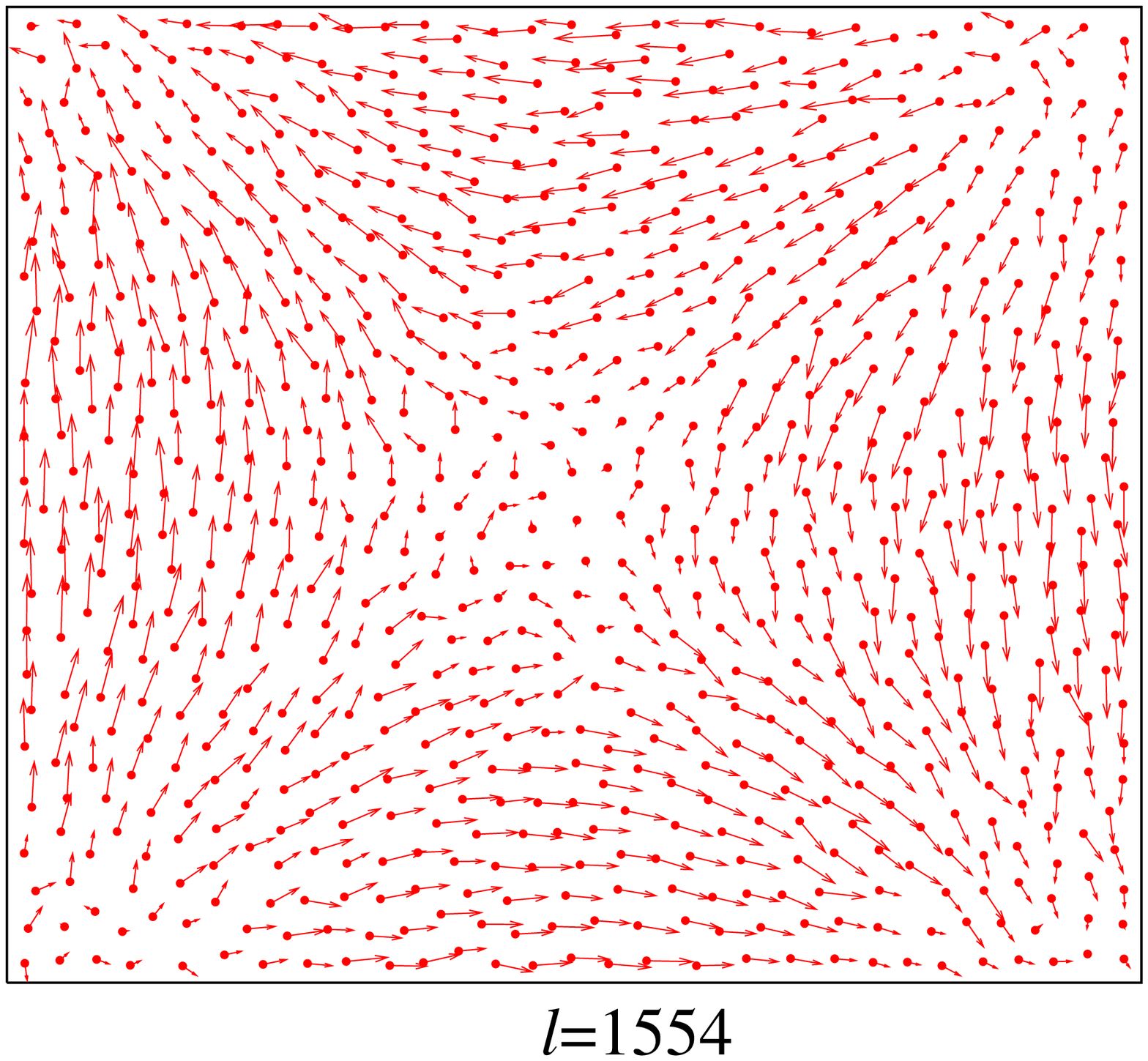}
\includegraphics[width=3.6cm,angle=-90]{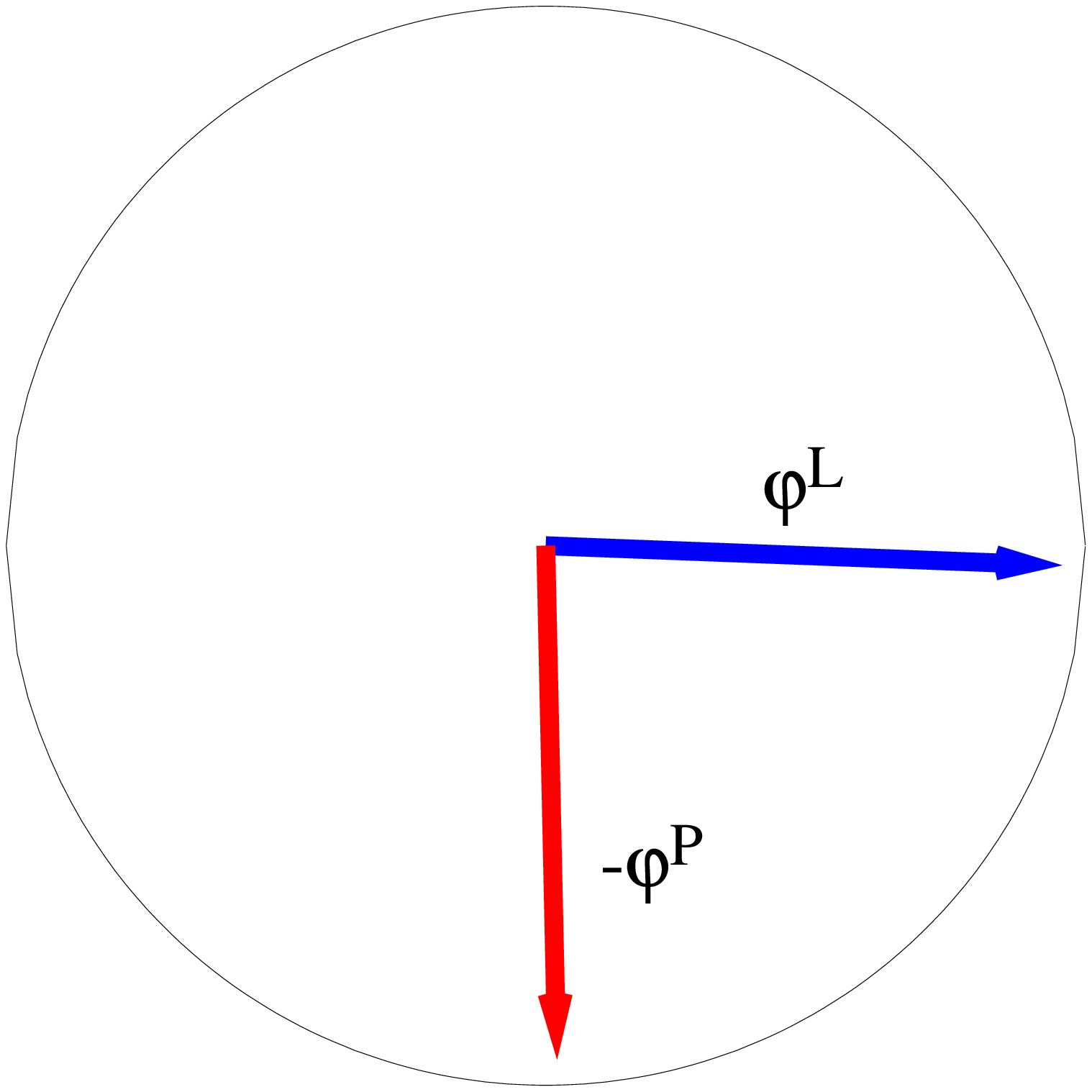}
\hfill
\includegraphics[width=3.6cm,angle=-90]{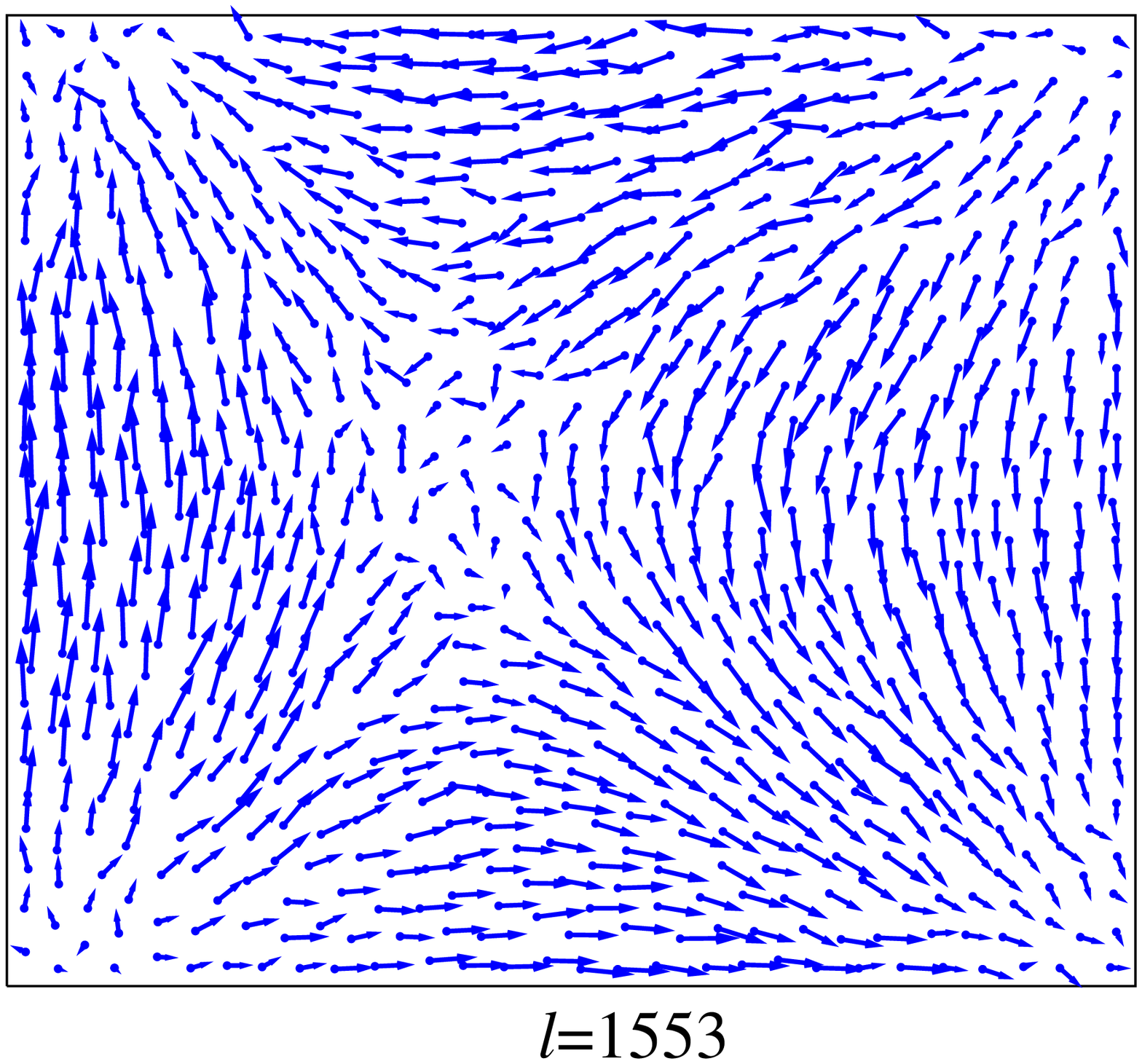}
\hfill
\includegraphics[width=3.6cm,angle=-90]{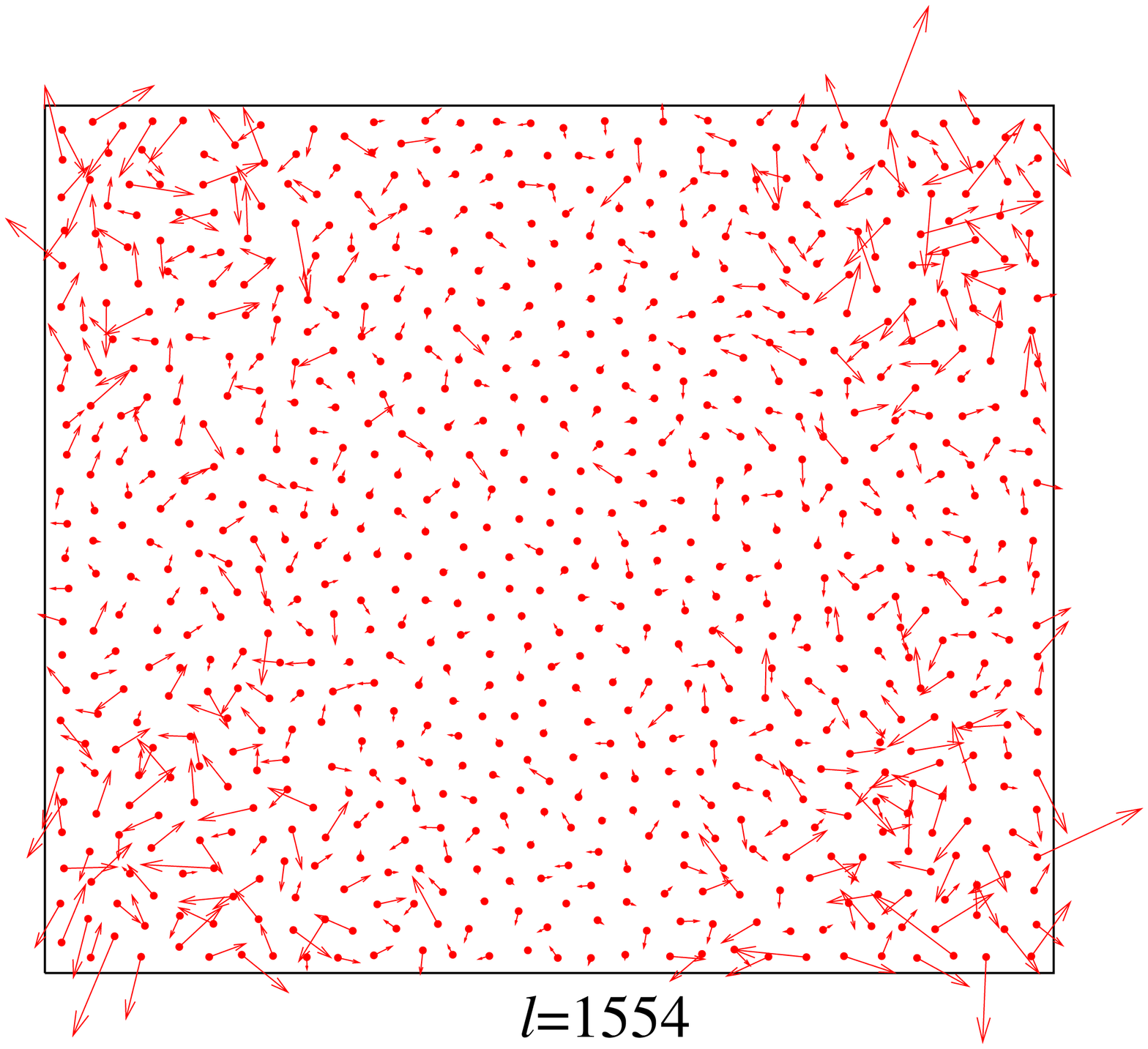}
\includegraphics[width=3.6cm,angle=-90]{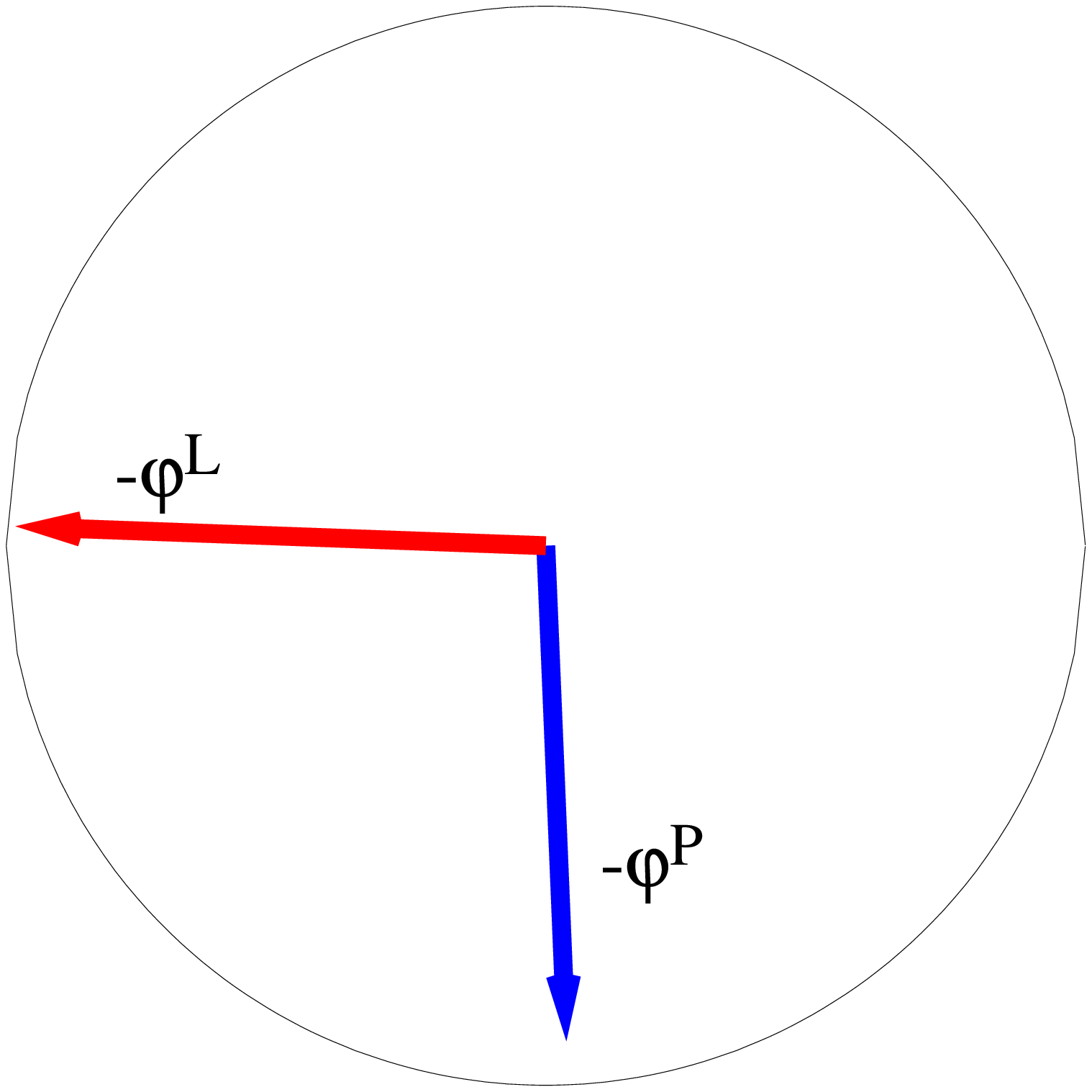}
\hfill
\includegraphics[width=3.6cm,angle=-90]{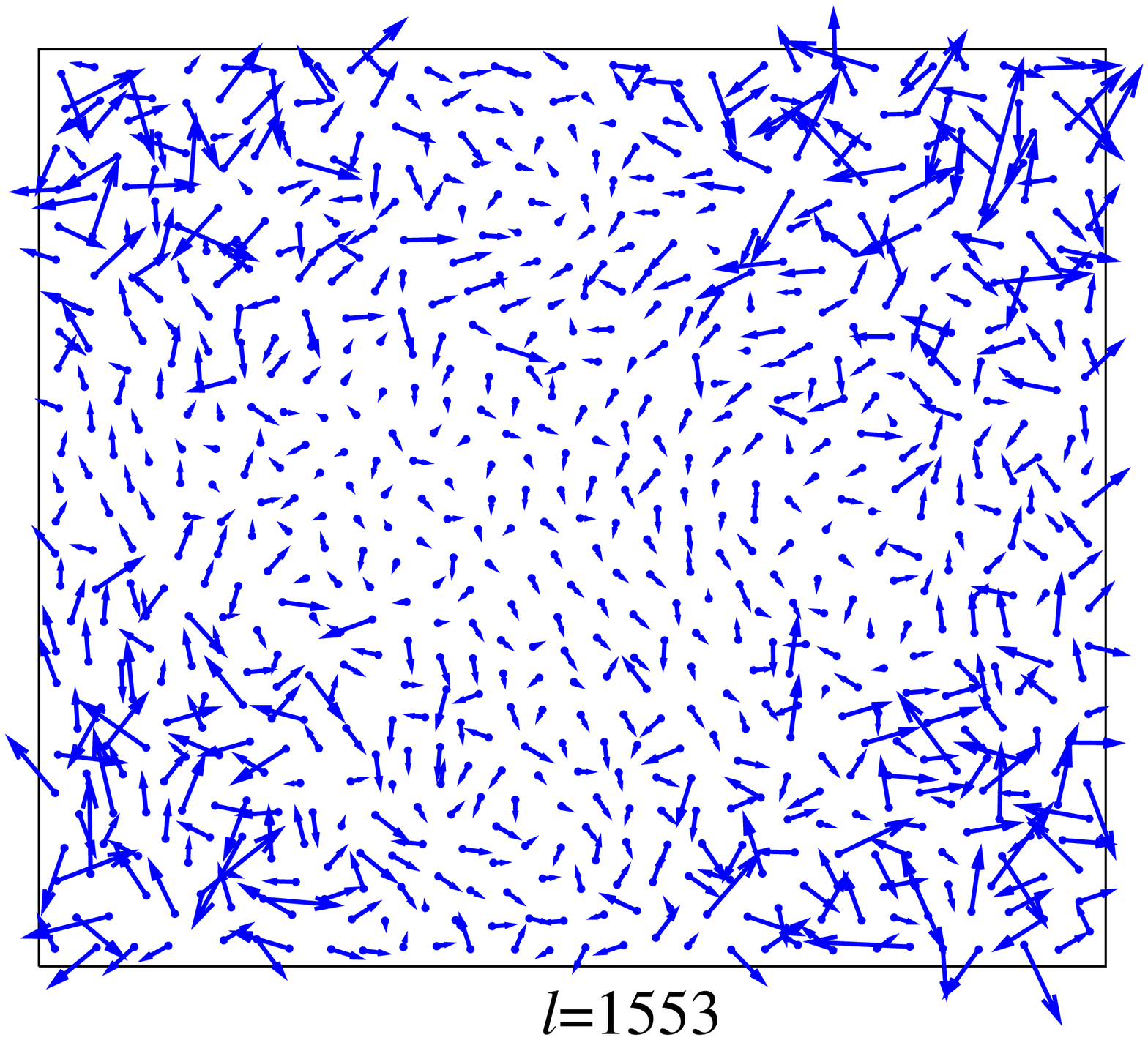}
\hfill
\includegraphics[width=3.6cm,angle=-90]{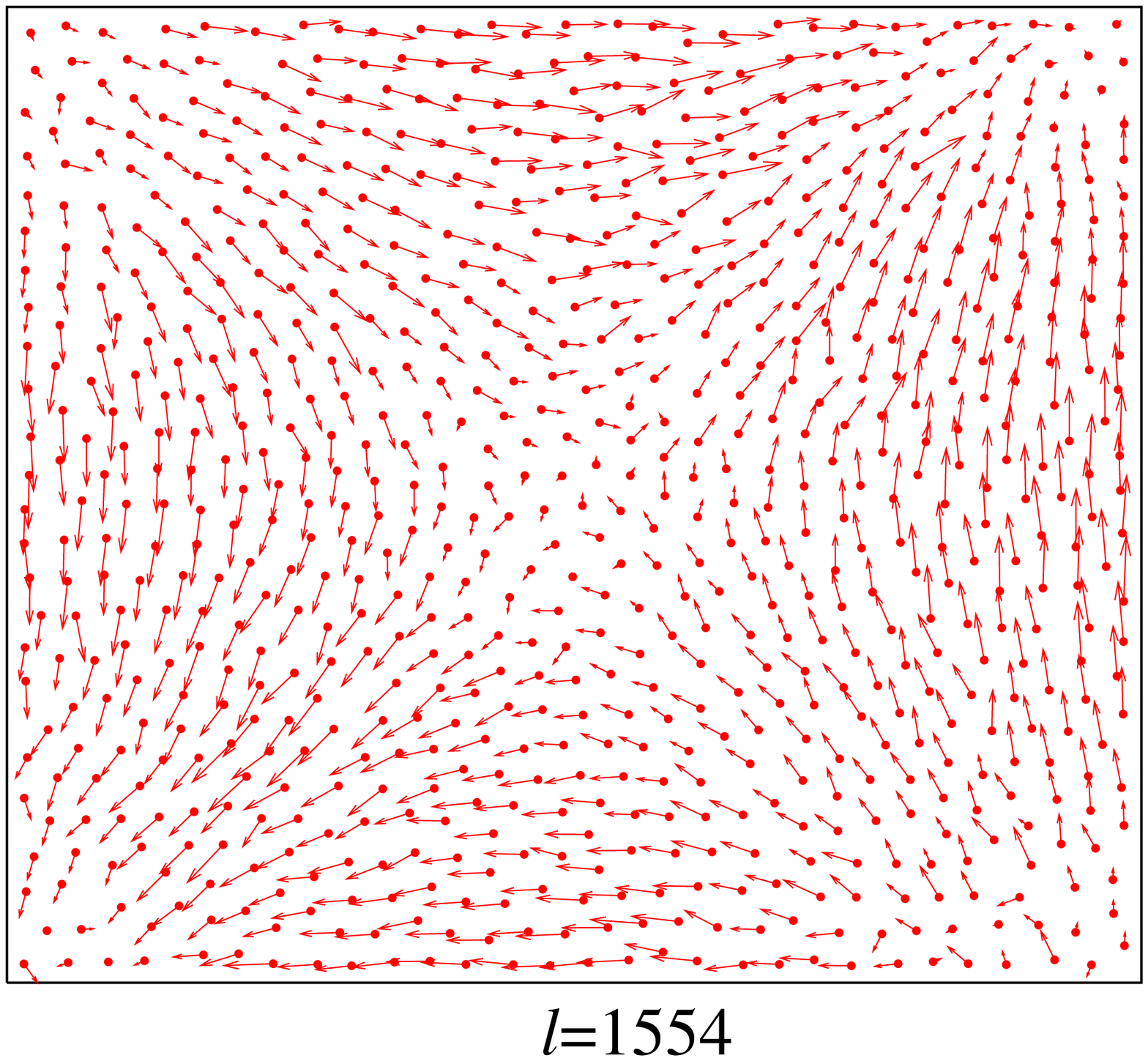}
\includegraphics[width=3.6cm,angle=-90]{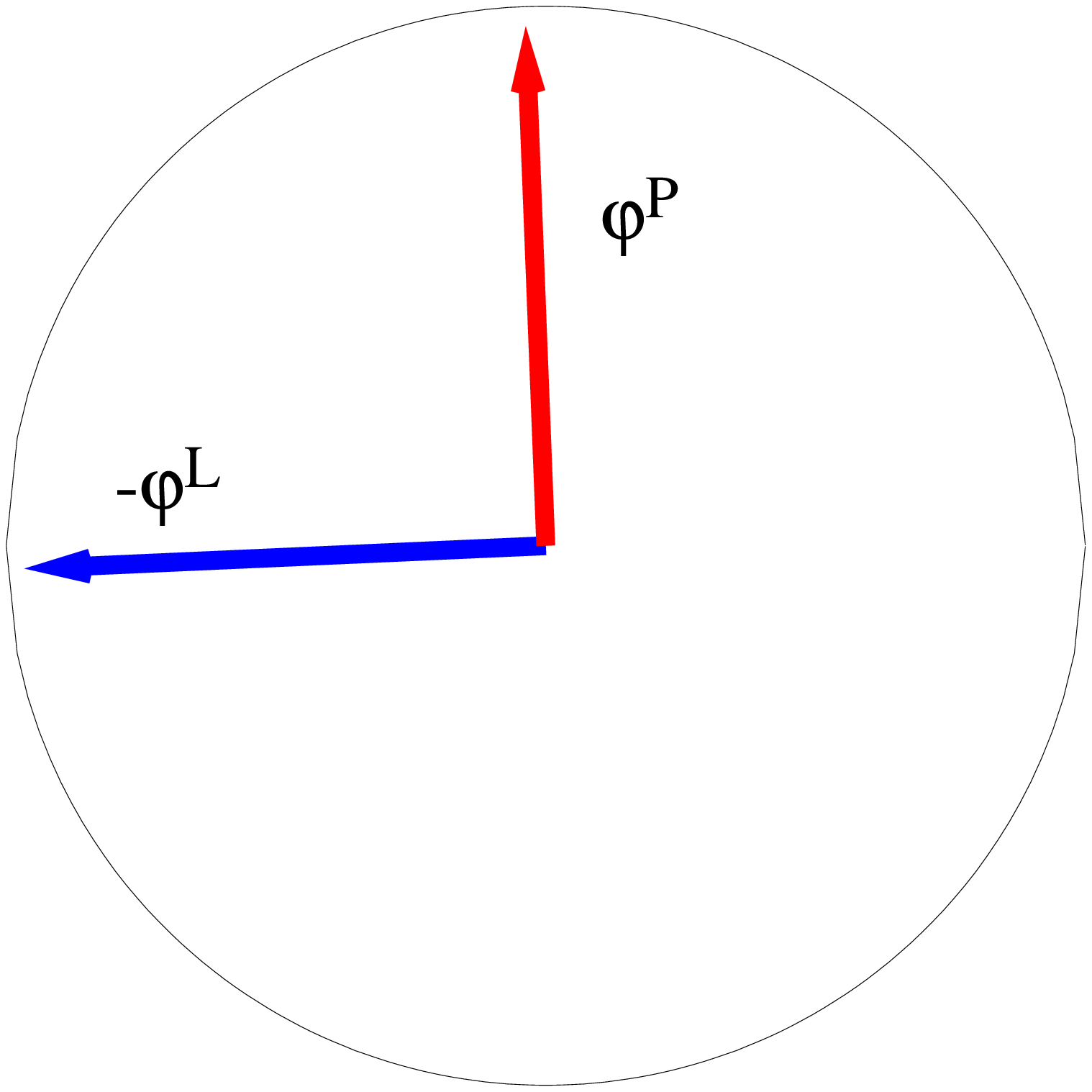}
\hfill
\includegraphics[width=3.6cm,angle=-90]{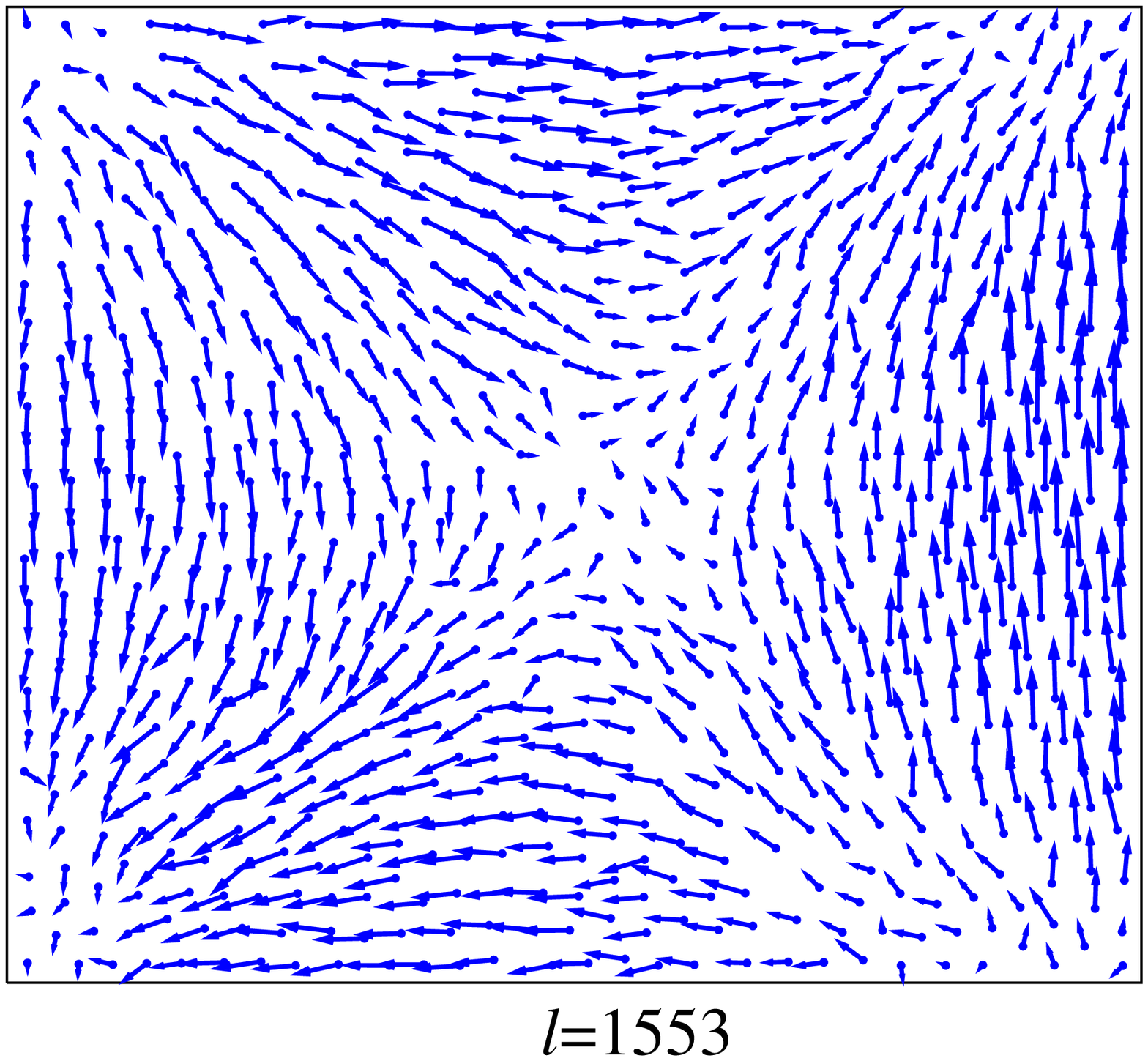}
\hfill
\includegraphics[width=3.6cm,angle=-90]{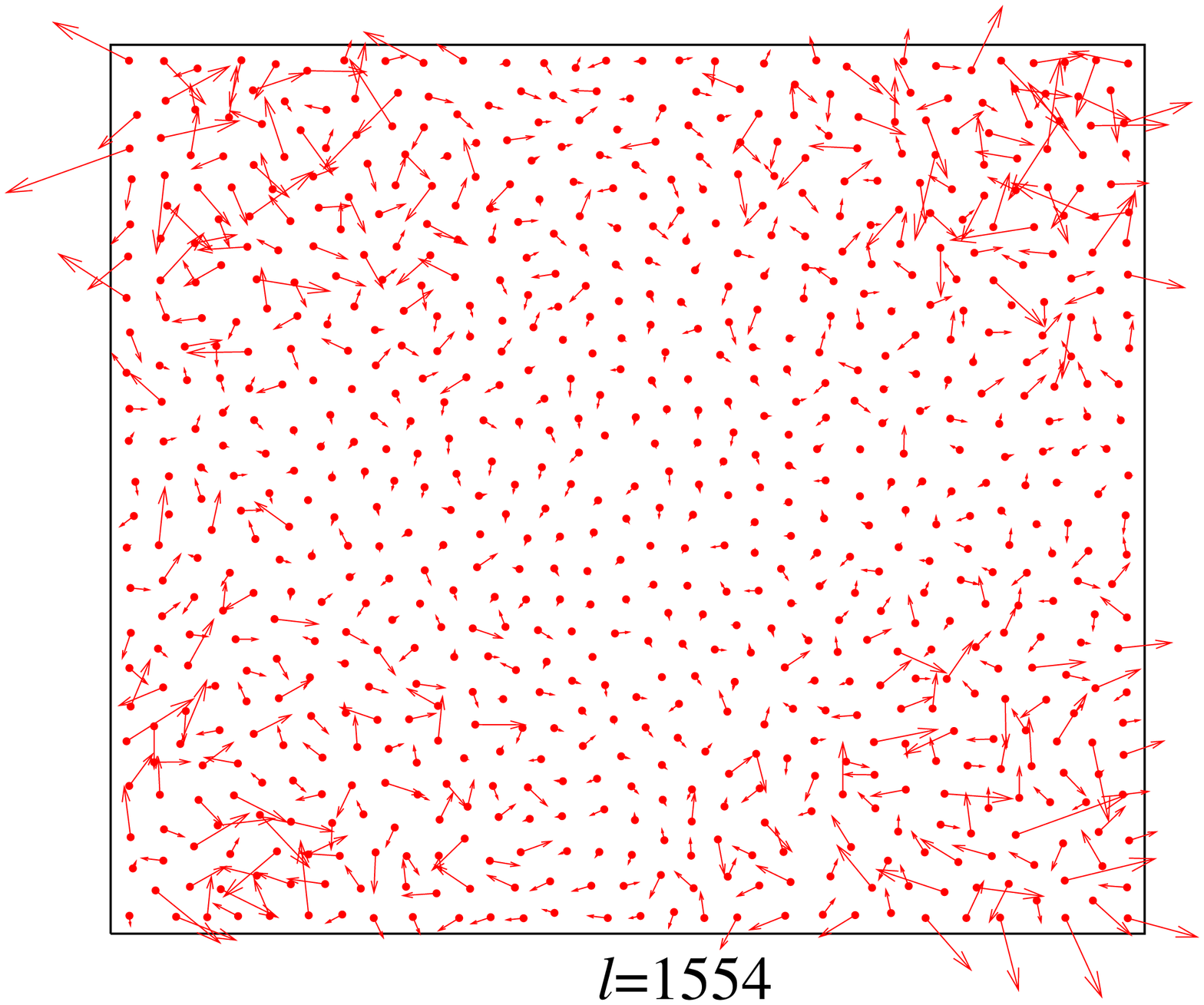}
\caption{{\LP}(1,1) dynamics for a system of 780 disks in a box with an aspect 
ratio 0.867, a density 0.8, and with {\em reflecting} boundaries.
Left: coordinates of the measured fields in the ``standard basis'' of {\LP}(1,1).
The (unit) circle is nearly reached, showing that indeed $\psi^1$ and $\psi^2$
span the same subspace as $\vp^{\L}$ and $\vp^{\P}$.
Center and right: the measured fields $\psi^1(t)$ and $\psi^2(t)$.
The rows from top to bottom are consecutive snapshots separated by time steps of
$\Delta t = 3.20$, 
which corresponds to a phase shift of $\pi/2$ in the {\LP}(1,1) rotation. }
\label{f:LP_dynamics_refl}
\end{figure}

Since multiplicities are not so essential here, we illustrate the
interpretation for the (simpler)
${\LP}(1,0)$ space of a rectangular system with reflecting boundaries (dimension 2).
In Sect.~\ref{s:mesp} we pointed out that at any given time $t$,
where the state is $\xi_t$, the measured modes $\psi^1(t)$ and $\psi^2(t)$ are combinations 
of the two spanning vectors $\vp^{\L}(\xi_t)$ and $\vp^{\P}(\xi_t)$. We re-write
Eq.~\Ref{e:rot2d} in matrix form, but now with explicit time dependence:
\begin{equ}
\big(\vp^{\L}(\xi_t),\vp^{\P}(\xi_t)\big)=\big(\psi^1(t),\psi^2(t)\big)\cdot Q(t)~\vir\quad 
Q(t)=\pmatrix{a(t)&b(t)\cr c(t)&d(t)}~.
\end{equ}
Therefore, the dynamics of the modes reduces to that of the $2\times2$ matrix $Q(t)$.
In our simulations we keep the spanning vectors orthonormal.
Since (in the experiment) the two fields $\vp^{\L}$ and $\vp^{\P}$ are 
also (nearly) orthogonal\footnote{The scalar product 
$\vp^{\L}\cdot\vp^{\P}=\Sigma_{j=1}^N \cos(k_xq_{j,x})\sin(k_xq_{j,x})p_{j,x}$ 
a priori does not vanish. However, as the simulations show, it is of the same
order as $\Sigma_{j=1}^N \cos(k_xq_{j,x})\sin(k_xq_{j,x})$, which is also
small and non-vanishing due to the uneven spacing of the particles.}, the matrix $Q(t)$ 
is close to a {\bf rotation} matrix (that is $c\simeq-b$, $d\simeq a$, and $a^2+b^2\simeq1$). 
Therefore, the dynamics in the two-dimensional subspace is well
described
by a phase 
$\phi(t)=\arctan({b(t)}/{a(t)})$.

\noindent
\proclaim{{LP}-dynamics in 2 dimensions} {The matrix $Q(t)$ is a rotation with
{\bf constant} angular velocity $\omega_\nn$. This velocity is proportional to the 
wave number $k_\nn$, namely
\begin{equ}
\phi(t) = \omega_\nn\,t = v\,k_\nn\,t~.
\end{equ}
Here, $v$ is the product of a frequency and a wavelength and, therefore, is a {\em velocity}. According to the simulations, $v$ depends only on the density of the system.}

\begin{Example}
In Fig.~\ref{f:LP_dynamics_refl} we demonstrate this rotation in the  {\LP}(1,0) space
of a system with reflecting boundaries.
Snapshots of the two measured fields at consecutive times show the rotation  of the two
vectors between the
L and the P direction.
\end{Example}

A similar rotation has been found in narrow systems in \cite{TM03} and
explained in \cite{MM03, WB04} in the low density limit using a Boltzmann-equation 
approach.\\

\begin{Remark}
The velocity $v={\omega_\nn}/{|k_\nn|}$ has been interpreted earlier
\cite{FHP03}
as the 
phase velocity of a traveling wave in physical space. In Appendix \ref{a:LP} we 
demonstrate how the two interpretations can be reconciled.
The definitions given above allow us to apply the same concepts 
also to the ${\LP}$-dynamics of systems with reflecting boundaries,
although they do not show traveling but standing waves.\\
\end{Remark}
%%%%%%%%%%%%%%%%%%%%%%%%%%%%%%%%%%%%%%%%%%%%%%%%%%%%%%%%%%%%%%%%%%%%%

%%%%%%%%%%%%%%%%%%%%%%%%%%%%%%%%%%%%%%%%%%%%%%%%%%%%%%%%%%%%%%%%%%%%%%%%%%%%%%
\begin{figure}
\begin{center}
\epsfig{file=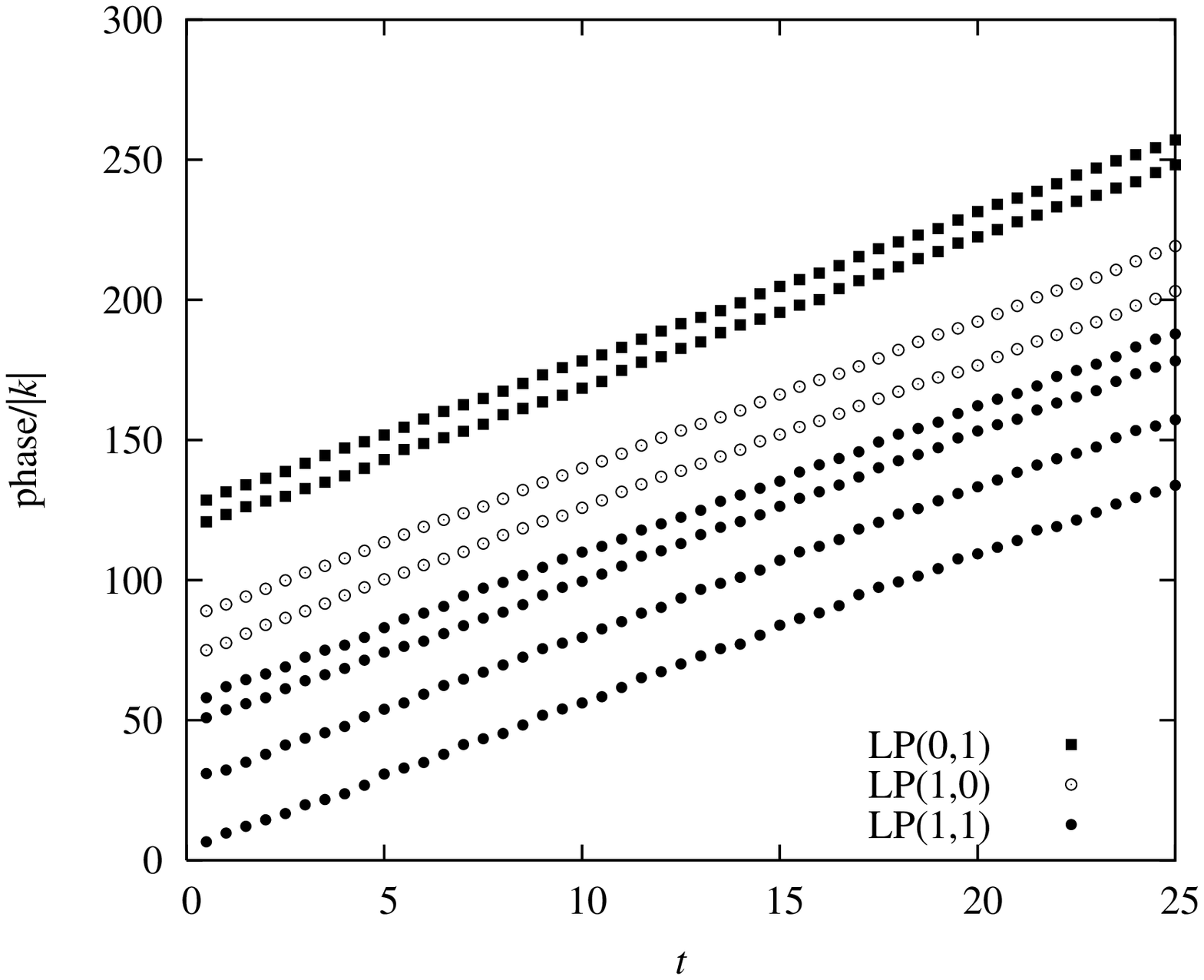, width=6.cm, angle=-90}
\epsfig{file=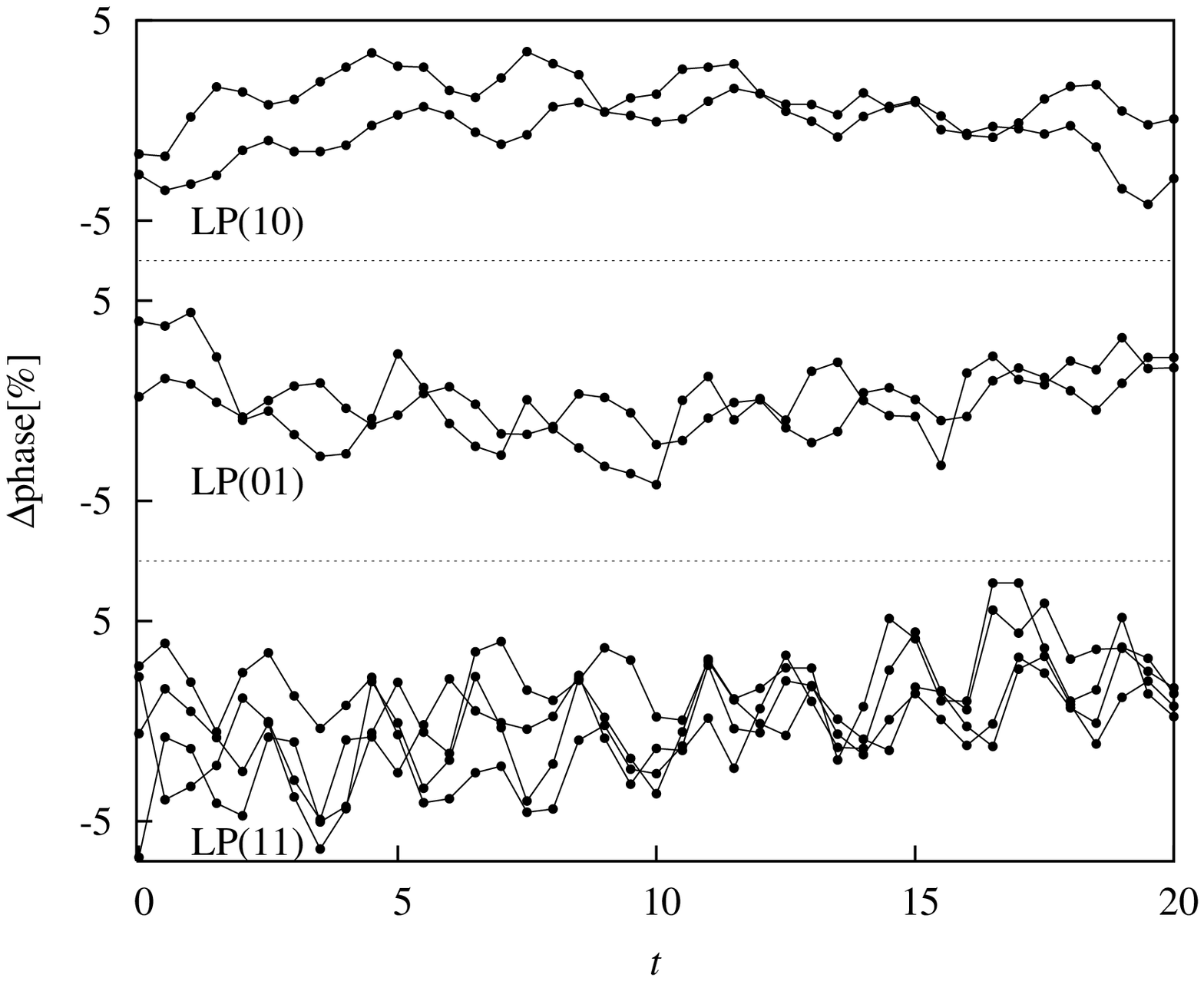, width=6.cm, angle=-90}
\caption{Rotations of various {\LP} pairs for the 780-disk
system of Fig.~\ref{f:lyap_spect}.  Left: Time dependence of the phase. 
For clarity, the phases for different modes are separated by multiples of
$2\pi/|k_\nn|$.  Right: The 
fluctuations around the constant velocity (in percent of $2\pi$). 
For details we refer to the main text.}

\label{f:phase_propagation}
\end{center}
\end{figure}
%%%%%%%%%%%%%%%%%%%%%%%%%%%%%%%%%%%%%%%%%%%%%%%%%%%%%%%%%%%%%%%%%%%%%%%%%%%%%%

Next, we consider the dynamics for the general case of a $2d$-fold degenerate
${\LP}(\nn)$ space. As
explained in Sect.~\ref{s:classif}, an ${\LP}(\nn)$ space is defined by a spanning
set of $d$ longitudinal modes, $\vp^{\L}_1,\ldots,\vp^{\L}_d$, and $d$ \P-modes,
$\vp^{\P}_1,\ldots,\vp^{\P}_d$, where each pair $(\vp^{\L}_k,\vp^{\P}_k)$  is 
an {\LP} pair. The following description is valid for any type of boundary conditions: 

\proclaim{Dynamics in ${\LP}(\nn)$}{The dynamics in ${\LP}(\nn)$, when restricted
to a two-dimensional sub\-space spanned by an {\LP} pair
\begin{equ}
{\rm Span}\{\vp^{\L}_k,\vp^{\P}_k\},
\end{equ} 
is a two-dimensional rotation at a constant angular velocity 
$\pm\omega_\nn$, where $\omega_\nn = v\,k_\nn$, with $v$ independent
of $\nn$. }

\begin{Example}
The {\LP} dynamics for higher-dimensional spaces is illustrated in 
Fig.~\ref{f:phase_propagation}. In the following we concentrate on  {\LP}(1,1). 
This space has dimension 8, that is four {\LP} pairs.  Generically, a measured mode 
has a non-vanishing projection onto all four {\LP} pairs, and four different
phases can be defined.  The four time series of phases for {\LP}(1,1) 
in Fig.~\ref{f:phase_propagation} belong to different projections of 
the {\em same} mode.
\end{Example}

%%%%%%%%%%%%%%%%%%%%%%%%%%%%%%%%%%%%%%%%%%%%%%%%%%%%%%%%%%%%%%%%%%%%%%%%
%
\section{Influence of geometry and system size}
\label{s:density}
%%%%%%%%%%%%%%%%%%%%%%%%%%%%%%%%%%%%%%%%%%%%%%%%%%%%%%%%%%%%%%%%%%%%%%%%

%%%%%%%%%%%%%%%%%%%%%%%%%%%%%%%%%%%%%%%%%%%%%%%%%%%%%%%%%%%%%%%%%%%%%%%%%
\begin{figure}
\begin{center}
\epsfig{file=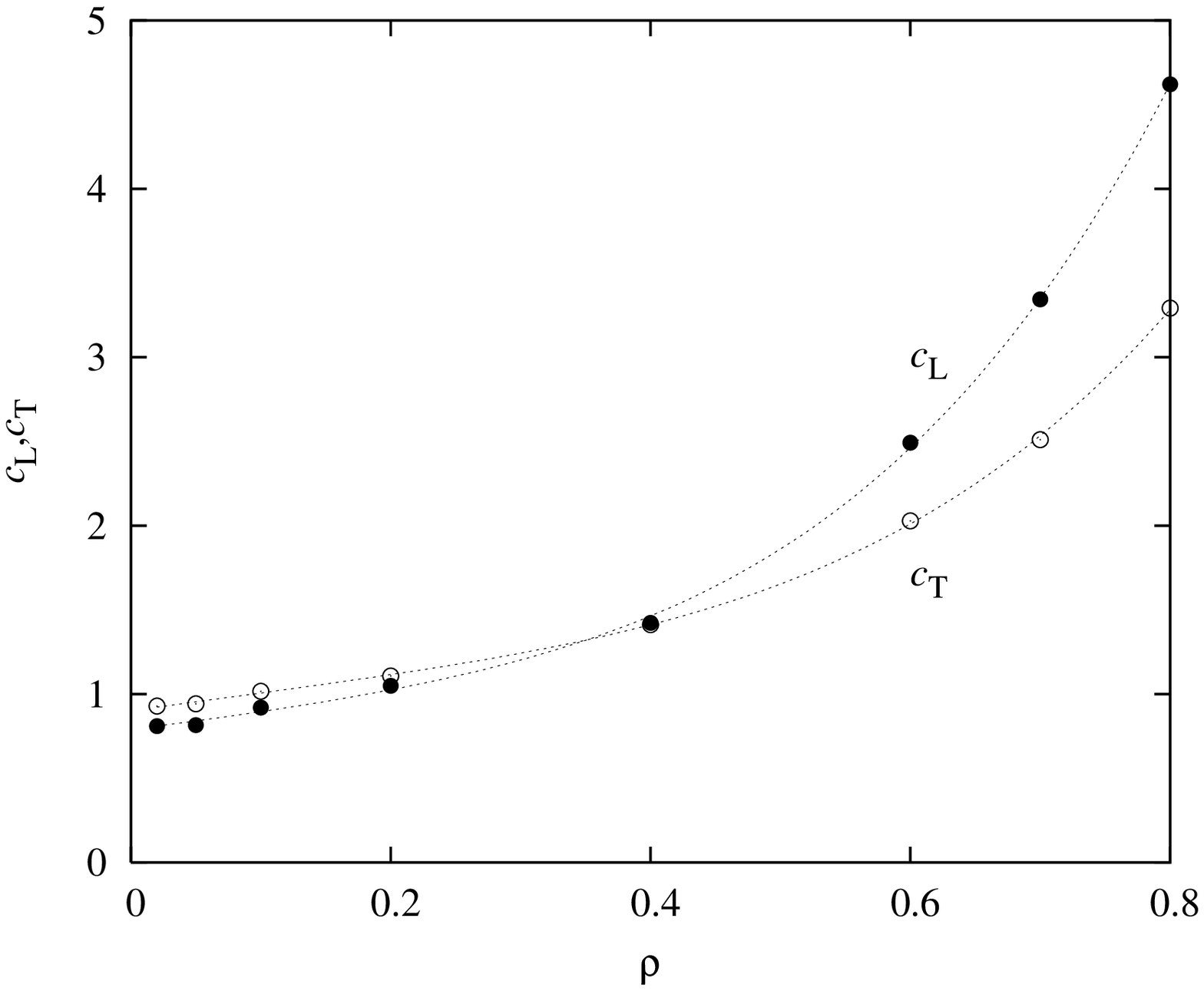, width=6.cm, angle=-90}
\epsfig{file=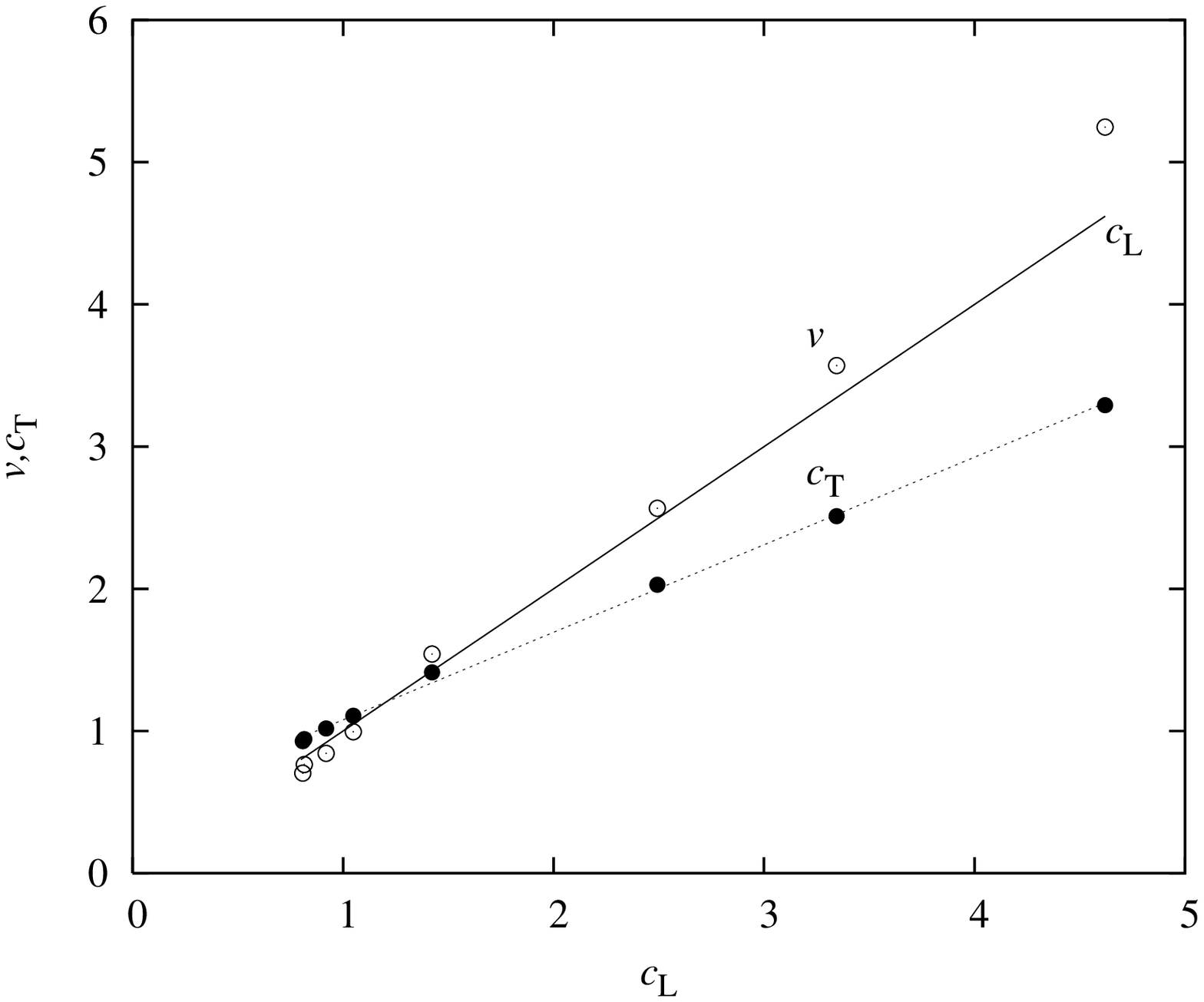, width=6.cm, angle=-90}
\caption{Slopes of the transverse and longitudinal branches $c_{\T}$
and $c_{\L}$ of Fig.~\ref{f:disp_rel}, and of 
phase velocity $v$. The simulations are for a system containing
$N = 780$ particles in a rectangular periodic box with a 
fixed aspect ratio $L_y/L_x = 0.867.$ 
Left: $c_\T$ and $c_\L$ as a function of the particle density $\rho$.
The smooth lines are polynomial fits added to guide the eyes. Although the
fits have no theoretical basis, we provide the fit parameters for convenience:
$c_\L = 0.790 + 0.970 \rho + 0.785 \rho^2 + 6.24 \rho^4$, and 
$c_\T = 0.902 + 1.050 \rho  - 0.053 \rho^2 + 3.85 \rho^4$. Right:
Almost linear relations between all three quantities. }
\label{f:CLCT_rho}
\end{center}
\end{figure}
%%%%%%%%%%%%%%%%%%%%%%%%%%%%%%%%%%%%%%%%%%%%%%%%%%%%%%%%%%%%%%%%%%%%%%%%%
In this section we study what influence the density, aspect ratio, and boundary 
conditions have on $c_\T$, $c_\L$, and $v$.  The density
dependence is significant. Unfortunately, we do not have 
any explanation for this fact. In particular, comparisons with 
the sound velocity \cite{H99}, with the mean free
path, and with similar quantities, do not suggest simple relationships.
Thus, these questions have to await further studies.\\	

The results of our simulations are summarized in Fig.~\ref{f:CLCT_rho}. 
In the left panel, $c_\L$ and $c_\T$, defined in
Sect.~\ref{s:tandyn}, are shown as functions
of $\rho$, and on the right panel $v$ and $c_\T$ are plotted as
functions of $c_\L$. There are two observations which are of interest:
First, as seen in the left panel,  $c_\T$ and
$c_\L$ cross for lower densities. Thus, the intuitively natural conjecture,
$c_\L \ge c_\T$, is not supported by the numerical evidence.
Second, the three ``velocities'' $c_\T$, $c_\L$, and $v$
are almost linearly related. In particular, the phase velocity $v$ agrees
rather well with $c_\L$ for fluid densities $\rho < 0.6$ (see the full line 
in the right panel of Fig.~\ref{f:CLCT_rho}), thus lending support
to referring to the curves $\lambda(k)$ as ``dispersion relations''.\\

%%%%%%%%%%%%%%%%%%%%%%%%%%%%%%%%%%%%%%%%%%%%%%%%%%%%%%%%%%%%%%%%%%%%%%%%%%%%
\begin{figure}[t]
\begin{center}
\epsfig{file=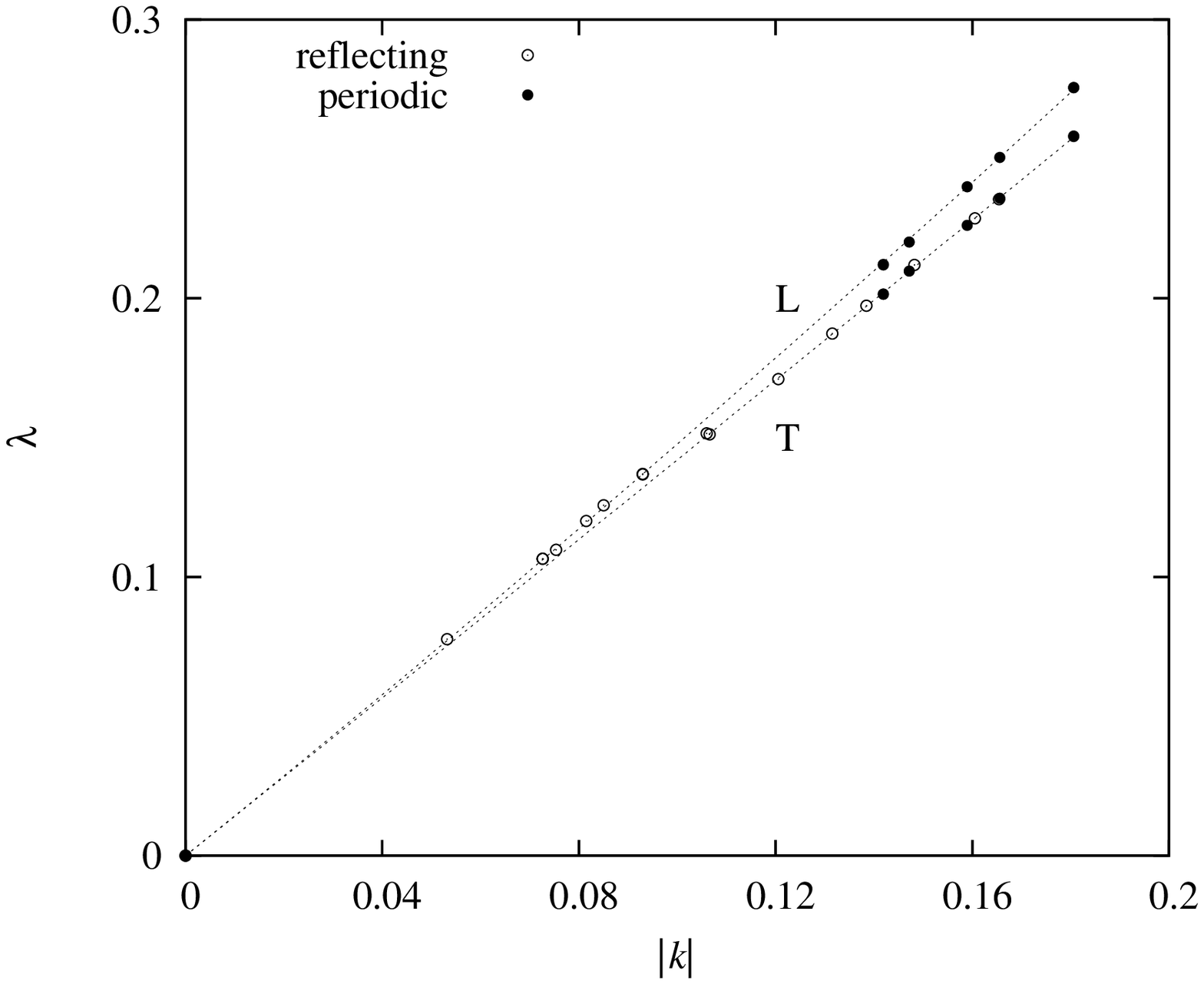, width=6.cm, angle=-90}
\epsfig{file=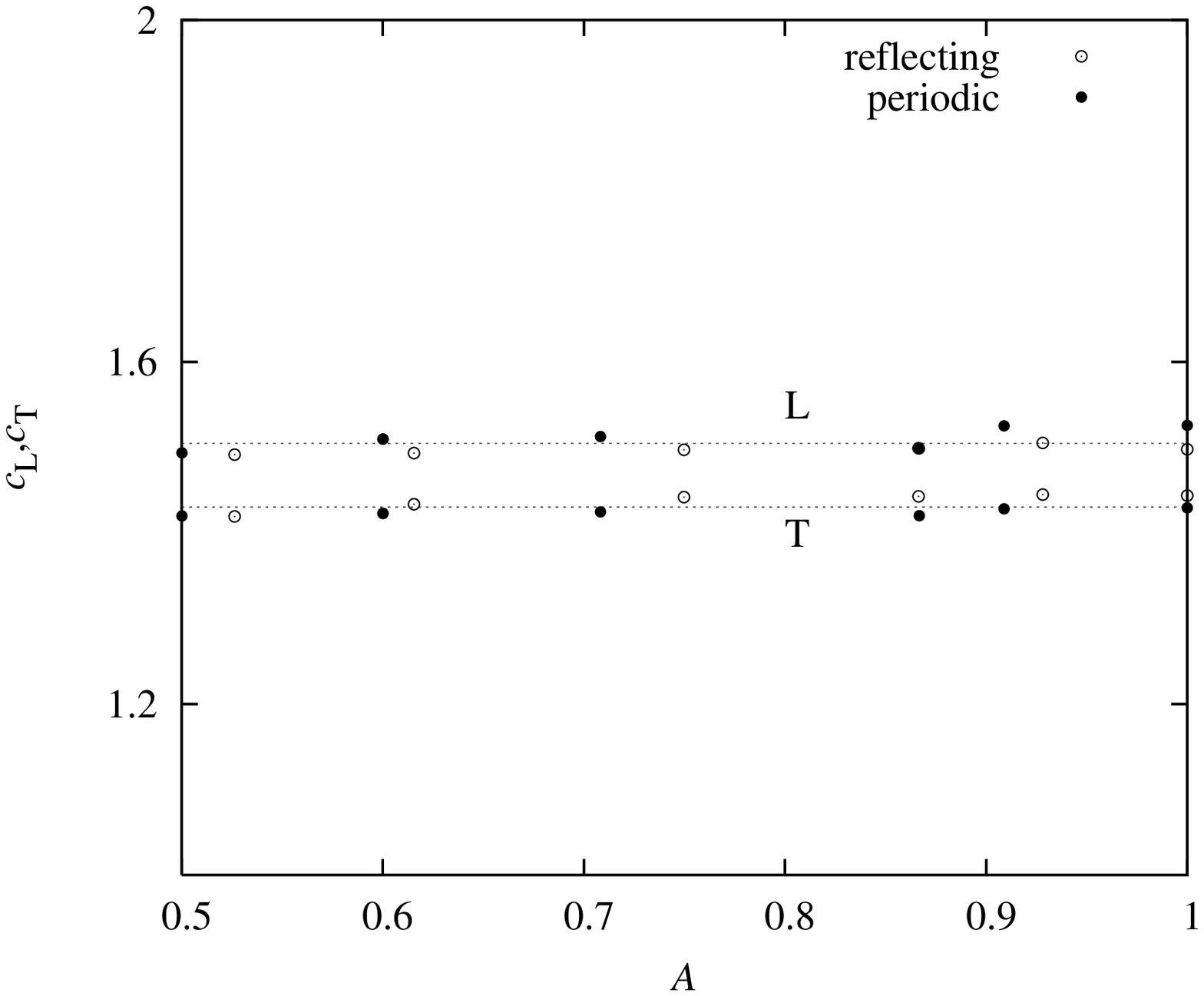, width=6.cm, angle=-90}
\caption{ Simulation results for a hard-disk gas with a density $\rho = 0.4.$
The full and open points refer to periodic and reflecting boundary conditions,
respectively.
Left: The smallest positive Lyapunov exponent for
T and L-modes, respectively, as a function of the wave-number $k$.
The curves for periodic and reflecting boundary conditions agree.
To the second order in $k$, a fit to the data gives
$c_\L = 1.422 k + 0.55 k^2$ and $c_\T = 1.412 k + 0.086 k^2.$
Right: The slopes, $c_{\L}$ and $c_{\T}$, are plotted as a function of the 
aspect ratio $A$ of the simulation box. For a given $A$ we compare points with
the same $k$ to eliminate the influence of the nonlinearity of the
dispersion relations. This means that for periodic boundaries the
linear extensions of the 
simulation box, $L_x$ and $L_y$, are twice those of the respective reflecting
box. The data for periodic and reflecting boundaries agree to within
numerical accuracy.} 
\label{f:PBC_HW}
\end{center}
\end{figure}

%%%%%%%%%%%%%%%%%%%%%%%%%%%%%%%%%%%%%%%%%%%%%%%%%%%%%%%%%%%%%%%%%%%%%%%%%
On the left panel of Fig.~\ref{f:PBC_HW} we plot the longitudinal and 
transverse dispersion relations for a moderately-dense gas with a 
density $\rho  = 0.4$.  The full and open points are for systems with periodic
and reflecting boundaries, respectively. In all cases, $\lambda$
was determined from the lowest step of the Lyapunov spectrum for the system,
for which the aspect ratio varied between 0.5 and 1, and the particle number
between 400 and 800.  The figure demonstrates that periodic and 
reflecting boundary conditions give the same $\lambda(k)$. 
We have experimentally verified (but not shown here) that
the same is true also for larger and lower densities.\\

The situation becomes a little more complicated when we consider the 
dependence of the slopes, $c_{\L}$ and $c_{\T}$, on the aspect ratio $A$,
as we do in the right panel of Fig.~\ref{f:PBC_HW}. The dispersion curves are not strictly 
linear in $k$,  as we have pointed out already in a footnote in
Sect.~\ref{s:tandyn}. A fit to the points in the left panel of 
Fig.~\ref{f:PBC_HW} reveals that the term proportional to $k^2$ 
is much larger for L than for T. To eliminate this nonlinearity in a 
comparison of  the slopes for reflecting and periodic systems with a given $A$, modes 
with the same values for $k$ \footnote{The smallest
wave number $k$ for periodic boundaries is given by $2 \pi/ L$, 
where $L$ is a dimension of the box. For reflecting boundaries,
$k$ is given by $\pi/L$, where $L$ is an effective box size, namely
the size reduced by a particle diameter $\sigma$. } 
are used in the right panel of Fig.~\ref{f:PBC_HW}.
With this precaution the figure demonstrates that the slopes of the dispersion 
relations do not depend on the boundary conditions in any significant way,
as long as the box does not degenerate to a narrow channel \cite{FMP04}. 
Except for this particular case, the nonlinearity of the dispersion curves 
has no noticeable influence on the classification
and description  of the Lyapunov modes in this paper
and has been accordingly ignored.

%%%%%%%%%%%%%%%%%%%%%%%%%%%%%%%%%%%%%%%%%%%%%%%%%%%%%%%%%%%%%%%%%%%%%%%%%
%
\section{Hydrodynamic equivalent of the modes}
\label{s:hydro}
%
%%%%%%%%%%%%%%%%%%%%%%%%%%%%%%%%%%%%%%%%%%%%%%%%%%%%%%%%%%%%%%%%%%%%%%%%%

There is a general expectation that the Lyapunov modes
should be related to the hydrodynamic behavior of the system, and
several papers point in this direction \cite{EG00, MM01, MM03,
TM03}. It should be noted, however, that none of these studies
has reached a totally convincing interpretation, and, furthermore, it
is obvious from this body of work that the LP-modes are more difficult to
explain than the T-modes. 
Here, we add to this a simple calculation
which might be helpful in the future: we determine how the
modes perturb the hydrodynamic fields or, in other words,
what would be the {\em hydrodynamic equivalent} of the modes we measure. 
The results, given in Table~\ref{t:hydro},
have quite a simple form, but do not reproduce the usual
modes of hydrodynamics.\\

Consider a general transformation $T$ of 
the one-particle phase space $\XX\equiv[0,L_x)\times[0,L_y)\times \R^2$
given by
\begin{equ}
T:\left\{\begin{array}{ll}
r\mapsto &r'=r+\ve\,\delta \xi(r,v)\cr
v\mapsto &v'=v+\ve\,\delta \eta(r,v)
\end{array}\right.~\vir
\end{equ}
and $f$ a probability density over $\XX$. If $\ve$ is infinitesimal,
then the new probability density is $f'=f+\ve\delta f$,
where
\begin{equ}[e:DF0]
\delta f = -f\,\big(\nabla_r\cdot\delta \xi+\nabla_v\cdot\delta \eta\big)-
\nabla_rf\cdot\delta \xi - \nabla_vf\cdot\delta \eta~,
\end{equ}
see Appendix~\ref{ap:f}.
Since we study equilibrium dynamics, we assume, furthermore, that $f$ is
the Boltzmann distribution\footnote{with $k_BT=1$ as in the
simulations}. By Sect.~\ref{s:stable} we also have
$\delta \eta=C\delta \xi$, and, therefore, (\ref{e:DF0}) simplifies to
\begin{equ}
\delta f = -f\,\,\big(\nabla_r\cdot\delta \xi+C\nabla_v\cdot\delta
\xi-C v\cdot\delta \xi\big)~.
\end{equ}
We define the variations of the three hydrodynamic fields
(density $\rho$, momentum $u$, and  energy $E$) by
\begin{equ}
\delta\rho(r)=\int dv\,\delta f(r,v)~\vir\quad
\delta u(r)=\int dv\,\delta f(r,v)\,v~\vir\quad
\delta E(r)=\int dv\,\delta f(r,v)\,|v|^2~.
\end{equ}
\begin{table}[ht]
\begin{equ}
\begin{array}{|c|c|ccc|}
\hline
{\rm mode} & \delta \xi & \delta\rho & \delta u & \delta E\\
\hline
\T & \nabla\wedge\A & 0 & \nabla\wedge\A & 0\\
\L & \nabla\A &  \Delta\A & \nabla\A & \Delta\A\\
\P & v\,\A & 0 & \nabla\A & A\\
\hline
\end{array}
\end{equ}
\caption{Modulations of the hydrodynamic fields (multiplicative constants are omitted)}
\label{t:hydro}
\end{table}

For each of the three types of modes, one can compute these quantities as
a function of the scalar modulation $\A$
introduced in (\ref{e:PLT}). The results are given
in Table~\ref{t:hydro}. Since $\Delta\A=-k_\nn^2\A$, we note that
the scalar fields $\delta\rho$ and $\delta E$
are proportional to the initial scalar modulation $\A$. 
Note also that the energy field is only affected
by the L and P-modes but not by the
T-mode.

%%%%%%%%%%%%%%%%%%%%%%%%%%%%%%%%%%%%%%%%%%%%%%%%%%%%%%%%%%%%%%%%%%%%
%
\section{Conclusions}
\label{s:disc}
%%%%%%%%%%%%%%%%%%%%%%%%%%%%%%%%%%%%%%%%%%%%%%%%%%%%%%%%%%%%%%%%%%%%

The picture developed in this paper is for two-dimensional hard
disks, but the method is sufficiently geometric and general to allow
easy extensions to other systems: \\

For example,
the generalization to three-dimensional hard disks is straightforward.
The existence of \L{}  and \T-modes for this case has been confirmed by computer
simulation \cite{H99}. Recently, Lyapunov modes were also found for
two-dimensional soft-particle systems interacting either with a Weeks-Chandler-Anderson
potential \cite{FP04} or a Lennard-Jones potential
\cite{RY04,YR04}. It would also be interesting to extend the work to,
say, circular geometries. Another extension concerns linear molecules such as 
hard dumbbells in a periodic box \cite{MPH98,MP02}. In this case two qualitatively 
different degrees of freedom play a role, translation and rotation. The existence 
of modes has already been demonstrated in this case.     \\

We have ended our wanderings through the rich landscape of Lyapunov
modes. To summarize, we have carefully identified and analyzed the
modes, giving a beginning of a theoretical
classification. Furthermore, we have seen that the Lyapunov exponents
and the phase velocity of the \LP-modes seem to be functions of the
density alone. In particular, they are practically independent of the aspect
ratio of the box (and, where applicable, also insensitive to the
boundary conditions).

\section{Acknowledgments}

Financial support from the Austrian Science Foundation (FWF),
project P15348, and the Fonds National Suisse
is gratefully acknowledged. 

\appendix

\section{Transformation of the one-particle distribution}
\label{ap:f}

Let $f$ be a given distribution. For $\Gamma\subset\XX$, we have
\begin{equ}
F(\Gamma)\equiv
{\rm Prob}({\rm one\ particle\ }\in\Gamma)=\int_\Gamma dr\,dv\,f(r,v)~.
\end{equ}
The transported probability $F'$ is defined by $F'(\Gamma')=F(\Gamma)$,
where $\Gamma'=T(\Gamma)$. 
We shall compute its density $f'$. In the integral
\begin{equ}[e:F1]
F(\Gamma)=\int_{T^{-1}(\Gamma')}dr\,dv\,f(r,v)~\vir
\end{equ}
we change the variables to $(r',v')=T(r,v)$. To first order in $\ve$, we have
\begin{equ}
T^{-1}:\left\{\begin{array}{ll}
r'\mapsto &r=r'-\ve\,\delta\xi(r',v')\cr
v'\mapsto &v=v'-\ve\,\delta\eta(r',v')
\end{array}\right.~\vir
\end{equ}
so \Ref{e:F1} becomes
\begin{equ}[e:F2]
F(\Gamma)=\int_{\Gamma'}dr'\,dv'\,{\rm det}\big[\DDD{T}{-1}{r',v'}\big]\,
f(r'-\ve\,\delta\xi(r',v'), v'-\ve\,\delta\eta(r',v'))~.
\end{equ}
The determinant is
\begin{equa}[e:F3]
{\rm det}\big[\DDD{T}{-1}{r',v'}\big]&={\rm det}
\left.\pmatrix{{\rm Id}-\ve\DD_r\delta\xi & -\ve\DD_v\delta\xi\cr
-\ve\DD_r\delta\eta&{\rm Id} - \ve\DD_v\delta\eta}\right|_{r',v'}\cr
&=1-\ve\big[\nabla_r\cdot\delta\xi+\nabla_v\cdot\delta\eta\big]_{r',v'}~.
\end{equa}
To first order in $\ve$, we obtain with \Ref{e:F2} and \Ref{e:F3}
\begin{equa}
F(\Gamma) = F(\Gamma') &-\ve \int_{\Gamma'}dr'\,dv'\,f(r',v')\,
\big[\nabla_r\cdot\delta\xi+\nabla_v\cdot\delta\eta\big]_{r',v'}\cr
&-\ve\int_{\Gamma'}dr'\,dv'\,\big[\nabla_rf\cdot\delta\xi + \nabla_vf\cdot\delta\eta\big]_{r',v'}~\vir
\end{equa}
which is equivalent to \Ref{e:DF0}.

\section{Traveling waves of LP dynamics}
\label{a:LP}

\noindent

We consider the ${\LP}(1,0)$ space of a rectangular system with
periodic boundary conditions,
defined by the four spanning vectors

\begin{equa}
\vp^{\L}_{\sin}&=\fld{s_x}{0}~\vir\quad\vp^{\P}_{\cos}=\fld{p_x}{p_y}\,c_x~\vir\quad
\vp^{\L}_{\cos}=\fld{c_x}{0}~\vir\quad\vp^{\P}_{\sin}=\fld{p_x}{p_y}\,s_x~.
\end{equa}
We take an initial tangent vector 
\begin{equ}[e:LP1]
\ddxi_0=a\,\vp^{\L}_{\sin}+b\,\vp^{\P}_{\cos}~.
\end{equ}
After a time $\tau={2\pi}/(4{\omega_\nn})$, the vector is
transformed to
\begin{equ}[e:LP2]
\ddxi_\tau=a\,\vp^{\P}_{\cos}+b\,\vp^{\L}_{\sin}~.
\end{equ}
Assume that $a$ and $b$ are more or less equal. Then \Ref{e:LP1} will resemble
a sinus, with much ``noise'' due to the P component, while \Ref{e:LP2} will 
look more like a cosine. In the dynamics leading from \Ref{e:LP1} to \Ref{e:LP2}
a kind of ``traveling wave'' is therefore visible, which seems to cover a distance
$2\pi|k_\nn|^{-1}$ in a time $2\pi\omega_\nn^{-1}$, thus ``moving ''
at velocity $v=\omega _\nn/|k_\nn|$. In actual simulations, we cannot
expect typical vectors to have a 
phase difference of $\fract\pi2$ between their $\vp^{\L}_{\sin}$ and 
$\vp^{\L}_{\cos}$ components, as in our example. Therefore, the
observed wave displacement as seen in \cite{H99,FHP03} has
the shape of ``steps'' in a space-time diagram, with an average slope
equal to $v$.

%%%%%%%%%%%%%%%%%%%%%%%%%%%%%%%%%%%%%%%%%%%%%%%%%%%%%%%%%%%%

\end{document}